\documentclass[12pt]{article}
\usepackage[margin=1in]{geometry}
\usepackage[T1]{fontenc}
\usepackage{erewhon}

\usepackage{setspace}
\usepackage{titlesec}
\titleformat{\subparagraph}[runin]{\normalfont\normalsize\bfseries}{\thesubparagraph}{1em}{}
\titlespacing*{\subparagraph}{0pt}{0.5\baselineskip}{1em}

\usepackage{algorithm}
\usepackage{algpseudocode}
\usepackage{amsfonts}
\usepackage{amsmath}
\usepackage{amsthm}
\usepackage{appendix}
\usepackage{bbm}
\usepackage{booktabs}
\usepackage{caption}
\usepackage{changepage}
\usepackage{comment}
\usepackage{enumitem}
\usepackage{epigraph}
\usepackage{graphicx}
\usepackage{hyperref}
\usepackage{xspace}
\usepackage{pdflscape}
\usepackage{physics}
\usepackage{rotating}
\usepackage{subcaption}
\usepackage{titling}
\usepackage[table,xcdraw]{xcolor}

\usepackage{tikz}
\usetikzlibrary{shapes,arrows,chains,calc}
\tikzset{
  line/.style={draw},
  block/.style={draw}
}

\graphicspath{{./figs/}}

\newtheorem{lemma}{Lemma}
\newtheorem{assumption}{Assumption}

\usepackage[USenglish]{babel}
\usepackage{csquotes}

\usepackage[authordate,backend=biber]{biblatex-chicago}
\DeclareFieldFormat{citehyperref}{%
  \DeclareFieldAlias{bibhyperref}{noformat}%
  \bibhyperref{#1}}

\DeclareFieldFormat{textcitehyperref}{%
  \DeclareFieldAlias{bibhyperref}{noformat}%
  \bibhyperref{%
    #1%
    \ifbool{cbx:parens}
      {\bibcloseparen\global\boolfalse{cbx:parens}}
      {}}}

\savebibmacro{cite}
\savebibmacro{textcite}

\renewbibmacro*{cite}{%
  \printtext[citehyperref]{%
    \restorebibmacro{cite}%
    \usebibmacro{cite}}}

\renewbibmacro*{textcite}{%
  \ifboolexpr{
    ( not test {\iffieldundef{prenote}} and
      test {\ifnumequal{\value{citecount}}{1}} )
    or
    ( not test {\iffieldundef{postnote}} and
      test {\ifnumequal{\value{citecount}}{\value{citetotal}}} )
  }
    {\DeclareFieldAlias{textcitehyperref}{noformat}}
    {}%
  \printtext[textcitehyperref]{%
    \restorebibmacro{textcite}%
    \usebibmacro{textcite}}}

\definecolor{mylinkcolor}{RGB}{0, 102, 204}
\hypersetup{
    colorlinks=true,
    linkcolor=black,
    urlcolor=mylinkcolor,
    citecolor=mylinkcolor
}

\begin{document}

\title{Pricing the Right to Renege in Search Markets: Evidence from Trucking}
\author{Richard Faltings \thanks{The University of California San Diego, 
		\href{mailto:rfaltings@ucsd.edu}{rfaltings@ucsd.edu}.
    Special thanks go to Eugenio Miravete, Dan Ackerberg, Nick Buchholz, and Victoria Marone for their exceptional mentorship and support. I also thank Bob Town, Jorge Balat, Jackson Dorsey, Andrey Ordin, Lauri Kytömaa, Hayden Parsley, Andrey Fradkin, Brett Hollenbeck, Yingkang Xie, and Alexander McKay. This paper benefitted from comments by seminar participants at UT Austin and conference participants at WoPA 2025, and IIOC 2025. This research was made possible through data provided by an anonymous company. I wish to thank JL, AC, FK, and numerous others at the company for their generous support and feedback on the project. The company provided no financial compensation for this research and limited its review to ensuring the protection of proprietary information.
    }}
\date{ First version: July 25, 2024\\ This version: \today \\~\\
 \href{https://rfaltings.com/papers/rfaltings_jmp.pdf}{Click here for the latest version.}
} 

\maketitle

\begin{abstract}
	In many search markets, advance interim contracts include an explicit right to renege, granting one party the option to switch to more attractive matches that emerge later in the search process. This paper studies the design and welfare implications of such interim contracts, leveraging novel data from a brokerage firm in the trucking industry. The broker allocates advance shipment contracts to carriers through a dynamic auction mechanism and penalizes cancellations through a reputational mechanism. I develop a theoretical model linking the carrier’s bidding problem to the firm’s cancellation penalties through a dynamic job-search problem and structurally estimate the model from rich data on bids and cancellations. In counterfactual simulations, I show that the firm is incentivized to lower cancellation penalties as the option value of the right to renege is priced into carrier bids. The results rationalize the large degree of contractual flexibility observed in the trucking industry as an efficient market outcome rather than one constrained by limited enforcement.

\end{abstract}

\newpage

\newcommand{\cfLateClearWelfarePct}{123.5\%\xspace}
\newcommand{\cfLateClearProfitPct}{318.2\%\xspace}
\newcommand{\cfMaxProfitProfitPct}{155.2\%\xspace}
\newcommand{\cfMaxProfitCarrierPct}{41.6\%\xspace}
\newcommand{\cfMaxProfitWelfarePct}{13.3\%\xspace}
\newcommand{\cfZeroProfitPct}{0.1\%\xspace}
\newcommand{\cfZeroWelfarePct}{1.8\%\xspace}
\newcommand{\cfInfWelfarePct}{91.7\%\xspace}
\newcommand{\cfInfMatchesPct}{47.3\%\xspace}
\newcommand{\cfInfBidPct}{15.4\%\xspace}
\newcommand{\cfInfPricePct}{6.0\%\xspace}
\newcommand{\cfLateClearMatchesPct}{74.4\%\xspace}
\newcommand{\cfZeroPenaltyWeight}{64\%\xspace}
\newcommand{\cfMaxTCWeight}{30\%\xspace}

\section{Introduction}

In many decentralized markets, participants search sequentially over alternatives on the other side of the market \parencite{mccall1970economics}. When faced with a potential transaction partner, a market participant must make a decision without knowing future matching opportunities. This creates a timing friction: accepting or rejecting matches without complete information can lead to suboptimal outcomes, reducing market efficiency relative to a centralized clearing mechanism where all potential trading partners are known \parencite{roth1994jumping}. Intuitively, efficiency in such markets can be improved by an explicit \emph{right to renege}, allowing at least one party to continue searching for better opportunities at a predetermined price, or penalty.

Rights to renege against a penalty are a prominent feature of many markets. Travelers are familiar with cancellation terms for hotel and flight bookings, which vary across firms and over time \parencite{nyt2008winter,nyt2015hotel,wapo2024fees}. Job searchers often face exploding offers in the labor market, making it harder for them to compare offers simultaneously \parencite{niederle2009market}. Digital platforms have made it easier than ever for a range of small businesses to charge cancellation fees \parencite{wsj2024noshow}. Despite the widespread prevalence and variety of explicit penalties for reneging, there is scant empirical research on the economic forces that determine firms' optimal choices of penalty terms, or the welfare implications thereof. This paper seeks to fill that gap.

The direct economic trade-offs of a right to renege are straightforward: the party with the option to cancel conserves some of their option value from future opportunities, while the other party sacrifices their option value for the duration of the agreement.\footnote{For example, the federal regulation requiring U.S. airlines to allow free cancellation within 24 hours of booking added an exemption for flights departing within 7 days, after airlines opposed it arguing that it ``takes inventory off the market for the duration of the refund period, blocking it from sale to other customers [\dots].'' \parencite{dot2011enhancing}} The price of this trade is the cancellation penalty, which acts as a foregone deposit on the transaction.\footnote{Some airlines allow passengers to hold a reservation for an upfront fee. On the introduction of United's FareLock program, its president said: ``It's a value to people like a stock option is a value [...] Well, this is an option on a seat'' \parencite{nyt2012cashing}.} Unlike financial options that involve the trade of fungible assets, the right to renege on a match pertains to the right to trade with a specific counterparty. Therefore, we cannot apply the no-arbitrage arguments from standard option pricing theory \parencite{black1973pricing}. Instead, we must consider the pricing problem within the bilateral strategic context of a match. Imposing a cost on reneging also mirrors Williamson's concept of "hostages" enabling trade that requires relationship-specific investments \parencite{williamson1983credible}. The key differences are that the required investments here are in the form of foregone opportunities in a dynamic search process, and that we must also consider the positive re-allocative externality of reneging in favor of a better match.

The equilibrium efficiency of such agreements also depends on the nature of the penalty. While monetary penalties are the simplest, they may be infeasible due to transaction costs or regulatory constraints (such as worker classification rules). As an alternative to monetary penalties, many gig economy platforms use automated reputational penalties \parencite{kandori1992social}, whereby a reneging party is punished through reduced access to future matches or other platform features.\footnote{For example, Uber drivers maintaining a sufficiently low cancellation rate–among other requirements–are eligible for Uber Pro status, offering discounts on gas and other services and geographic match prioritization (\href{https://www.uber.com/cr/en/drive/uber-pro/}{https://www.uber.com/cr/en/drive/uber-pro/}).} Unlike monetary penalties, which transfer utility from one party to the other, reputational penalties only punish the reneging party without compensating their counterparty.\footnote{\textcite{hubbard2001contractual} draws a similar distinction between reputational and contractual enforcement in the context of trucking.} This creates a direct deadweight loss, and may also under-incentivize the provision of flexibility. On the other hand, a monetary penalty turns reneging into a revenue-generating opportunity, which can distort incentives for large firms towards extracting additional rents from workers or consumers.\footnote{Change and cancellation fees accounted for 3.2\% of U.S. airline passenger revenue in 2009 \parencite{wsj2009badluck}} 

To understand firm incentives and characterize equilibrium outcomes and welfare, I study the problem of a firm's optimal pricing of a right to renege in an important empirical context: the U.S. trucking industry, which transports 74.8\% of domestic freight by value \parencite{usdot2021moving} and accounts for 2.3\% of GDP \parencite{usdot2022contribution}. The market is suitable for this research question for several reasons. One, match quality matters because each shipment is unique, with its own pickup and dropoff locations and schedule, which makes it more or less costly for different drivers to haul, depending on their own schedules and locations. Two, finding the ideal match is difficult because the market is fragmented on both the demand and supply side (91\% of carriers operate fleets of ten or fewer trucks \parencite{fmcsa2023pocket}) and there is limited time to search over alternatives (lead times are typically 2 weeks down to less than 24 hours). Consequently, the timing friction described above is a significant issue in the industry. This is evident in the significant flexibility built into advance contracts throughout the market.\footnote{To illustrate, \textcite{caplice2006combinatorial} discuss how this need for flexibility hindered efforts to allocate shipments via combinatorial auctions: ``So, even if another carrier was assigned to the outbound lane in the strategic bidding process, the shipper may choose to tender the load to an alternative carrier for a specific load. Not only is this accepted behavior for shippers, most analysts and transportation management software packages consider such opportunistic continuous move optimization a key capability.'' Consequently, the industry couldn't reliably guarantee bundles to bidders, as would be expected in most combinatorial auctions.} 

The data for the empirical analysis are provided by a large trucking brokerage that matches shipments with freight carriers through an online auction platform. The platform is designed to allow carriers to renege at two stages. Upon winning an auction, the carrier receives a notification and has the option to confirm the match or renege on the bid. After a match is confirmed by the carrier, they can still easily cancel at any time, but at the risk of a reputational penalty on future bids on the platform.\footnote{Auction winners are selected according to the sum of their bids and an algorithmic quality score, which is computed on the basis of several past performance metrics, including cancellation rates.} Whereas most economic data only observe final, realized transactions, these data record every stage of the matching process, making them uniquely suited to studying reneging behavior. Reneging is indeed frequent: over half of winning bids are reneged on, while 15\% of confirmed matches are eventually canceled before pickup. As for the motives behind reneging, two empirical patterns are relevant. First, the cancellation rate is decreasing in the match price, suggesting that carriers cancel opportunistically when they find a better deal. Second, the cancellation rate is increasing in the time interval between match confirmation and pickup. Both these patterns are consistent with carriers continuing a search process in the style of \textcite{mccall1970economics} even after they have won and confirmed a match on the platform.

Guided by the empirical evidence, I build a dynamic model of carrier search and bidding behavior where the option value of reneging is priced into the bidding strategy. In the analysis of the model, I disentangle the direct effects of a penalty on the propensity to cancel from the indirect effects on the bidding behavior. The exercise demonstrates that accounting for the full auction equilibrium effects of the penalty is crucial for understanding the firm incentives to offer flexibility, and the ensuing welfare implications; without the effect on bids, the firm would require full commitment from carriers.\footnote{This incentive extends to firms that can extract part of the option value of reneging by raising posted prices.} I also show that the variance of the carriers' outside offers and the type of penalty---reputational or monetary---affects firm incentives. When the variance of outside offers is low, firm profits increase with higher penalties, even at the cost of overall welfare. Conversely, when the variance is high, both firm profits and overall welfare decrease in the penalty, so that firm incentives are socially aligned. With monetary penalties, the firm can generally profit by raising the penalty level at the cost of overall welfare. This sets up a key tension: while monetary penalties are a more efficient punishment mechanism at any given penalty level, the accompanying rent-extraction motive can lead to a more socially inefficient outcome than reputational penalties, which ``burn'' utility.

To take the model to the data, I enrich the baseline model with additional dynamics, allowing for stochastic arrivals of carriers and shipments over time, and accounting for the multi-round and multi-unit nature of the auctions. In addition, to reconcile the cancellation rates with the much higher bid reneging rates, I introduce an attention mechanism that allows for carriers to miss the auction win notification. I then estimate the primitives governing carrier behavior, including the stochastic process of outside offers, the attention parameters, and the subjective valuation of the platform's reputational cancellation penalty. To support the estimation strategy, I develop a simple partial identification argument based on the additional wedge that the penalty introduces between the carrier's optimal bid and their cost, combined with a non-negativity condition on carrier marginal costs. These effectively bound the estimated penalties from above, as too large a penalty would imply negative marginal costs. Finally, I separately estimate the dynamics of carrier and shipment arrivals, and the on-platform search process by matching directly observable moments in the data.

To establish the profit-maximizing and welfare-maximizing penalties, I use estimates of the model to simulate the profit and welfare consequences of both reputational penalties (which solely decrease carrier utility) and monetary penalties across a wide range of penalty levels. Among reputational penalties, I find that the current near-zero penalties are nearly optimal for both social welfare and firm profits. Reducing penalties to zero would modestly increase both profits (+\cfZeroProfitPct) and welfare (+\cfZeroWelfarePct). A decomposition exercise reveals that higher reputational penalties decrease profits through both of fewer successful matches (despite a decline in the number of cancellations) and higher costs per successful match.

I also find that moving to monetary penalties would allow the platform to increase short-run profits, as predicted by the theoretical exercise. The resulting \cfMaxProfitProfitPct increase in profits, relative to the status quo, is mainly the result of transfers from carriers, whose welfare declines by \cfMaxProfitCarrierPct, for a net welfare loss of \cfMaxProfitWelfarePct. To rationalize why the firm does not use monetary penalties, I explore alternative objective functions which take a weighted average of firm profits and carrier welfare to proxy for long-run motives of carrier acquisition and retention to the platform.\footnote{Platforms in the early stages of growth may value retention more. For example, AirBnB imposed small fees of \$50 to \$100 for host cancellations until August 21, 2022, after which cancellation fees increased dramatically to scale with the reservation amount, up to \$1000 (\href{https://web.archive.org/web/20221127193530/https://www.airbnb.com/help/article/990/}{https://web.archive.org/web/20221127193530/https://www.airbnb.com/help/article/990/}).} I find that a \cfZeroPenaltyWeight weight on carrier welfare is sufficient to explain the lack of monetary penalties. I also explore the effect of transaction or enforcement costs incurred by the firm for every cancellation fee it collects. Under reasonable levels of transaction costs, the firm finds it optimal to avoid monetary fees at lower weights on carrier welfare. However, transaction costs alone are not sufficient to explain the lack of monetary penalties without some weight on carrier welfare---a weight of \cfMaxTCWeight on carrier welfare is required even at transaction costs of \$150 per cancellation fee collected.

While the baseline analysis restricts attention to schedules of cancellation penalties  that follow the same temporal profile as the status quo penalty (which increases sharply for cancellations within the last 48 hours before pickup), the model can be used to analyze alternative temporal profiles. I compare the increasing time profile to a flat one, whereby the cancellation penalty is constant over time. For a given maximal penalty level, carrier welfare is lower with a flat penalty as earlier cancellations are punished more harshly, but this only increases firm profits if the penalty is a monetary fee. Thus, monetary cancellation penalties may not only incentivize harsher cancellation penalties than is socially optimal, but earlier penalties as well.

Additionally, to fully illustrate the value of contractual flexibility in this market, I also investigate the welfare effects of an infinite penalty. Using the same auction format, the full commitment policy reduces total welfare by \cfInfWelfarePct compared to the status quo, driven by a \cfInfMatchesPct decline in the number of final successful matches. The average bid increases by \cfInfBidPct and the transaction price of successful matches increases by \cfInfPricePct, reflecting the increased opportunity cost that carriers price into their bids.

Finally, a reasonable conjecture is that the increased opportunity cost of full commitment is reduced if carriers are matched to shipments \emph{later}. I thus consider another full commitment policy in which auctions are only cleared in the last 24 hours. This indeed leads to a reduction in the transaction price of \emph{successful} matches, but the total number of matches is even lower, with a \cfLateClearMatchesPct reduction relative to the status quo. This can be explained by a greater rate of attrition to offers outside the platform; carriers waiting on the uncertain outcome of a bid are more likely to prefer the sure payoff of an immediate outside offer, as compared to a carrier who has already been matched with a shipment on the platform. Thus, flexible cancellation policies are a more effective \emph{unilateral} policy for mitigating timing frictions when the market cannot be easily coordinated through a centralized matching process.

\paragraph{Literature Review}

Timing frictions were first highlighted in entry-level labor markets \parencite{roth1994jumping,roth1997turnaround,li1998unraveling,li2000risk,kagel2000dynamics}. This literature explores how the strategic timing of binding offers reduces market efficiency. In an experimental setting \textcite{niederle2009market} show that market norms enforcing non-binding agreements can improve efficiency, but that these do not arise in competitive equilibrium under a design where contract flexibility cannot be priced in through transfers. I contribute to this literature, first by extending the theoretical setting to markets with search frictions \parencite{mccall1970economics,pissarides2000equilibrium}, where each market participant experiences a \emph{different} sequence of opportunities. Second, I provide the first empirical application of this largely theoretical and experimental literature in the economically important market of trucking.

Other recent work has explored the strategic incentives to offer non-binding offers in search markets, in the form of both exploding offers of varying deadlines \parencite{lippman2012exploding,lau2014exploding,armstrong2016search,zorc2020deadlines,hu2021size} and cancellation penalties of varying levels \parencite{xie2007service,zhang2021partial,liu2022dynamic,liu2023intraconsumer}. My model nests both forms of non-binding agreements: exploding offers can be modeled as an initial cancellation penalty of zero, jumping to an infinite penalty at some deadline. I replicate and jointly consider several key mechanisms from the theoretical literature, including the acceptance deterrence effect of holding an attractive offer \parencite{zorc2020deadlines}, the interaction between the penalty level and the transaction price \parencite{hu2021size}, and the rent extraction effect of a monetary cancellation penalty \parencite{xie2007service}. The rich model allows for the joint effect of all these mechanisms to be quantified in counterfactual simulations.

For the analogous problem of product returns, \textcite{anderson2009option} show that similar incentives exist for firms to offer flexible return policies when consumers are sufficiently uncertain about their preferences for a product. The key differences in the case of the right to renege is that the relevant uncertainty is over \emph{alternative matches} in the market, and that the cost of reneging is endogenously derived from the foregone matches in the market, rather than from the logistics of returning and restocking a product.

I also add to the literature on auction formats with options to renege or adapt bid amounts. The procurement mechanism for Medicare Durable Medical Equipment (DME) received significant interest \parencite{merlob2012cms,cramton2015designed,ji2022can} for its auction format in which firms may renege on bids after the auction result and clearing price have been revealed. In a different setting, \textcite{bajari2014bidding} emphasize the need to provide flexibility in procurement auctions due to bidders' own uncertainties about construction project costs while \textcite{haberman2023auctions} provide a theoretical justification for withdrawals as part of the price-finding process. I show that in dynamic search settings---similar to \textcite{backus2024dynamic}--- uncertainty over future substitutes also provides an economic rationale for reneging. 

Finally, my paper contributes to the growing literature on the industrial organization of trucking, a market which has garnered significantly more interest from policy-makers since the supply chain disruptions of the COVID-19 pandemic highlighted the dependency of the U.S. economy on the trucking industry. Early contributions most notably include \textcite{hubbard2000demand}, \textcite{hubbard2001contractual}, \textcite{baker2001empirical}, and \textcite{baker2003make}, focusing on issues of vertical relations and contracts under moral hazard, while more recent work by \textcite{yang2022don} explores spatial frictions caused by the need for drivers to return to their home locations. The most closely related work in this area is \textcite{harris2022long}, who examine a different type of contractual flexibility in the trucking market: the option for carriers to reject shipments tendered to them under long term fixed-price contracts, which may diverge significantly from spot market rates. In contrast, I focus on carrier cancellations \emph{after} accepting a shipment under \emph{spot-negotiated} rates. Both papers complement each other in illustrating the value of flexibility at different stages of the contracting process, with implications for different margins of allocative efficiency. The main focus of \textcite{harris2022long} is on the optimal allocation of shipments between long-term contracts, which leverage relationship-specific efficiencies, and the spot markets, which benefit from increased market thickness. I instead focus on the optimal allocation of specific carriers to specific shipments in real time, taking into account timing frictions inherent in the dynamic search process. Another difference is this paper's focus on the incentives of private parties to offer flexibility to their contracting partners, whereas \textcite{harris2022long} take the contract design as given and focus on the resulting equilibrium allocation and welfare implications.

The rest of this paper is structured as follows. In Section \ref{sec:background}, I provide an overview of the trucking industry and the data used in this paper. In Section \ref{sec:toy_model}, I present a stylized model of the carrier search process, which I then extend to a dynamic setting in Section \ref{sec:model}. In Section \ref{sec:estimation}, I describe the estimation of the model, followed by the counterfactual analysis in Section \ref{sec:counterfactuals}. I conclude with Section \ref{sec:conclusion}.

\section{Empirical setting and data}
\label{sec:background}

The trucking industry is a vital sector of the U.S. economy, accounting for 2.3\% of the GDP in 2022 \parencite{usdot2022contribution}, and transporting 74.8\% of domestic freight 2023 \parencite{usdot2021moving}. The market operates on a decentralized basis, through a combination of long-term contracts (typically several months to a year) between shippers and large carriers or brokers, and spot transactions, often negotiated through brokers. The industry is highly fragmented, with over 55\% of carriers operating a single truck, and 91\% operating 10 trucks or fewer \parencite{fmcsa2023pocket}. This fragmentation creates costly search frictions in the spot market, which play a major role in explaining the prevalence of long-term contracts in the industry \parencite{hubbard2001contractual}.

Whereas the industry traditionally relied on labor-intensive matching processes, the firm studied in this paper was an early pioneer of the digital brokerage business model and sought to heavily automate the matching process. The majority of the broker's upstream demand came through long-term  contracts with large shippers (such as major retailers and manufacturers), which would regularly tender shipments to the platform at the contracted rate. Once the firm accepts a shipment, it would seek to procure a carrier at the lowest possible cost, using an online auction platform. Despite the firm's large size in absolute terms, it only accounted for a small fraction of the overall market, with revenues slightly below 1\% of the \$251 billion for-hire trucking market in 2021.

\subsection{Platform design}

The study covers the full two-year period of 2021 and 2022, during which the firm implemented its "timed auction" format. This format consists of several rounds, with the reserve prices non-decreasing across rounds. Additionally, the platform features an option called Accept-Now, allowing carriers to immediately confirm a shipment at a posted price, fixed over time. The majority of shipments are posted to the platform between one and two weeks before their pickup time. Auction rounds begin to clear five days before pickup, with one to two rounds per day. If a shipment is unmatched 24 hours before pickup, the auction format changes to update the reserve price more frequently, with additional intervention from human brokers.

Carriers can discover loads by either searching for specific criteria or browsing a personalized feed. Once a shipment appears on their feed or search results, the carrier can click on it to obtain further details (referred to as a detailed \emph{view}), as well as place a bid, use the Accept-Now feature, or simply move on (referred to as willfully \emph{ignoring} a shipment). When choosing whether to bid or Accept Now on a shipment, the carriers can only see the current round's closing time and the Accept Now price, in addition to the shipment characteristics. No information is provided about other bidders and their bid amounts, and the reserve price is hidden.\footnote{Reserve prices were initially publicly displayed at the launch of the auction format, but were hidden soon after launch, as carrier bids tended to bunch at the reserve price.} Additionally, carriers are never restricted from bidding on additional shipments, even if these conflict with existing confirmed appointments on the platform.

When an auction reaches one of its pre-determined closing times, if all bids exceed the reserve price, the auction proceeds to the next round. If there are bids below the reserve price, the platform accepts the lowest bid and notifies the bidder, who has 10 minutes to confirm. If the bidder does not respond, the process is repeated for the next lowest bid. If no bidders confirm, the auction proceeds to the next round. I use the terminology of the platform \emph{accepting} bids and carriers \emph{confirming} them  throughout the paper.

Based on these rules, it is clear that bidding entails practically no commitment on the side of the carrier. After confirmation, the carriers continue to enjoy some flexibility, as the platform's official policy is that cancellations only incur a reputational penalty when initiated less than 48 hours before the shipment's pickup time.\footnote{Prior to Q1 2021, carriers had to contact a firm representative to cancel. The policy was changed to self-service cancellations with a disclaimer that ``if the request occurs within 48 hours of pickup time, it may negatively affect [the carrier's] scorecard.''} This penalty reduces carrier's probability of winning future bids through a quality adjustment factor which changes the bid ranking, but does not result in any direct monetary loss. The exact mechanics of the reputational penalty are not disclosed to carriers, and \autoref{sec:background_reputational} presents evidence that carriers may overestimate the severity of this penalty. Thus, the structural model takes carriers' subjective valuation of this penalty as a parameter to be estimated rather than recovering it through an explicit model of the repeated interactions.
While it is unclear how widespread explicit cancellation policies are among traditional brokers, for the period of this study the main rival digital platform has a similar policy, penalizing carriers' reputation for cancellations in an even shorter penalized window of 24 hours before pickup.

\subsection{Data and Stylized Facts}

\begin{figure}[htbp]
    \begin{center}
        \caption{Funnel from views to matches}
        \label{fig:data_summary_sankey}
        \includegraphics[width=0.7\textwidth]{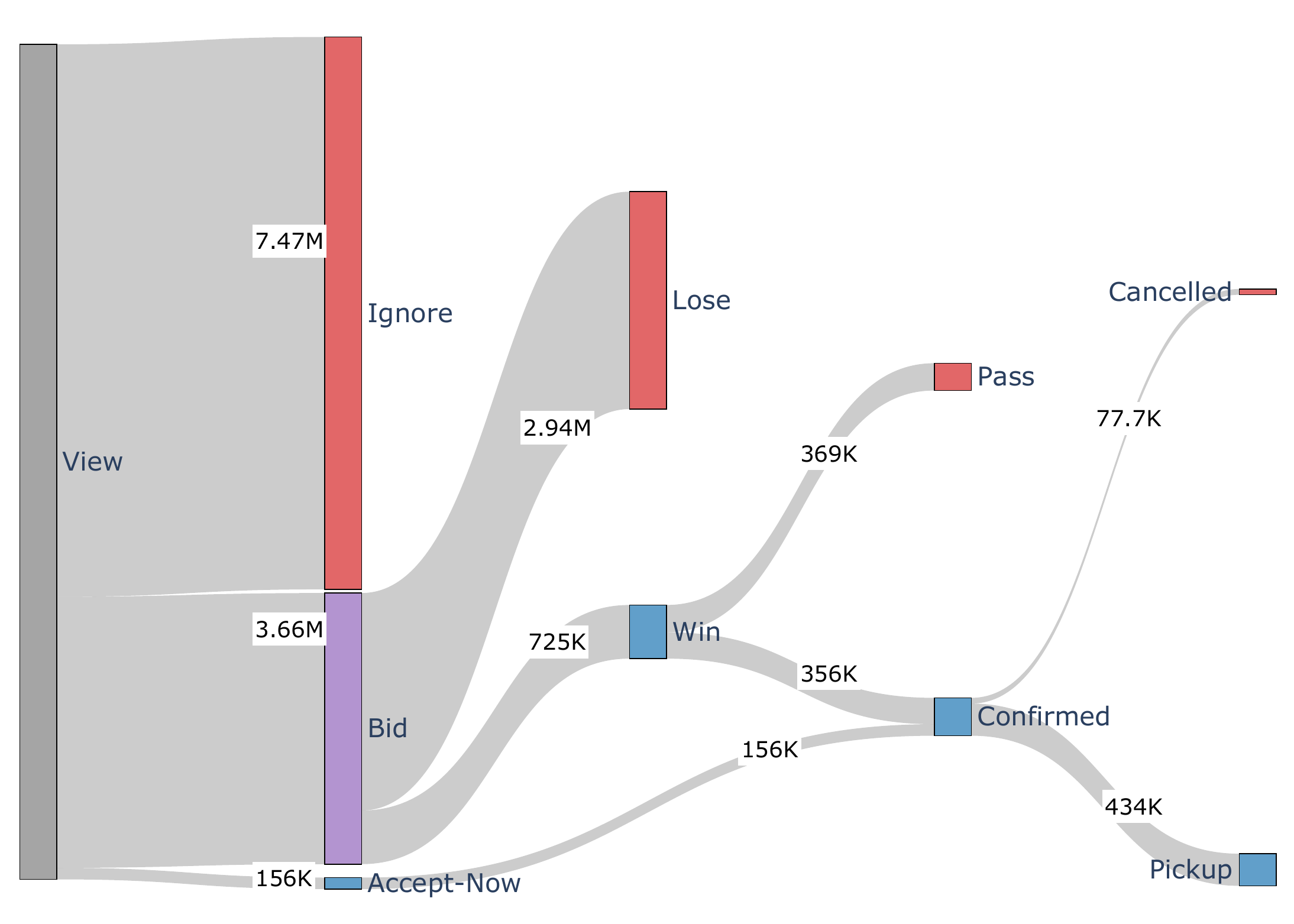}
    \end{center}
    {\footnotesize \textsc{Note}: Data represents the universe of non-local shipments posted to the auction platform in 2021 and 2022. Gray areas and numbers represent mass of flows between states.}
\end{figure}

The design of the platform yields a rich dataset on the endogenous decisions by carriers at the multiple stages of the matching process, allowing me to examine the empirical determinants of attrition on the platform. Although the data was provided directly by the platform, some preprocessing and cleaning steps, detailed in Appendix \ref{sec:data_construction}, were necessary to ensure consistency and accuracy. The process of attrition, converting views to final matches, is shown as a funnel in Figure \ref{fig:data_summary_sankey}. It is immediately apparent that the overwhelming majority of potential matches do not result in a match, while two-thirds of views do not even result in a bid. The attrition of carriers across this funnel is a central focus of the empirical analysis.

While the full dataset covers the universe of auctions run by the platform, I exclude all local shipments (within a metropolitan area) from the analysis, due to their high-frequency nature allowing carriers to match with multiple shipments in a single day. Furthermore, the structural empirical analysis is focused on a single shipping lane---defined as a route between two metropolitan areas---specifically from Seattle, Washington, to San Francisco, California. Table \ref{tab:summary_stats} is a summary and comparison of the two sets of data. The sub-sample is highly representative of the full sample along the measures of attrition and bidding behavior. The main difference between subsets is the mean and variance of shipment distance, which is unsurprising given the focus on a single lane.  %

As my research question focuses on the short-term dynamics of the market for each pickup date, I normalize all measures of prices by the average spot market rate of a shipment's lane (defined as a metro-to-metro) at the shipment pickup time, to remove price variation that is due to aggregate market fluctuations that are caused by macroeconomic factors. This rate was obtained from a third-party data aggregator and is used internally by the firm for various prediction and optimization tasks. Further details on the construction of the estimation sample, including the handling of incomplete records, price normalization, and sample restrictions, are provided in \autoref{sec:data_construction}.

\begin{table}[htbp]
    \begin{center}
        \caption{Summary statistics}
        \begin{tabular}{lcc}
\toprule
{} &   Full data & Seattle-San Francisco \\
\midrule
Unique carriers           &      85,913 &                 4,592 \\
Unique shipments          &     563,620 &                 9,096 \\
Number of views           &  11,286,561 &                87,893 \\
Number of accept now      &     155,772 &                 2,079 \\
Number of bids            &   3,664,334 &                48,673 \\
Bid acceptance rate       &      19.79\% &                21.23\% \\
Bid confirmation rate     &      49.10\% &                48.58\% \\
Cancellation rate         &      15.17\% &                13.45\% \\
Average bid amount        &        1.16 &                  1.16 \\
(Std. deviation)          &      (0.64) &                (0.27) \\
Average shipment distance &      422.96 &                757.78 \\
(Std. deviation)          &    (425.17) &                (6.42) \\
\bottomrule
\end{tabular}

        \label{tab:summary_stats}
    \end{center}
    {\footnotesize \textsc{Note}: Full data refers to the universe of non-local shipments posted to the auction platform in 2021 and 2022. The sub-sample is restricted to the Seattle to San Francisco lane in the same period. Standard deviation of averages are in parentheses. Bid amount is normalized by the contemporaneous average spot rate of the lane at the time of shipment pickup.}
\end{table}

\subsubsection{Cancellation Behavior and Consequences}

In the following, I focus on the determinants of, and consequences of carrier cancellations. Overall, 14.9\% of confirmed shipments are cancelled by the carrier. Among carriers cancelling on a confirmed shipment (reneging on a bid), 27.8\% (25.5\%) ultimately execute a different shipment on the platform. This suggests that carriers use the flexibility to cancel and renege to substitute to other---presumably better---opportunities. The remaining cases likely represent instances of substitution to off-platform opportunities, given the small size of the platform relative to the broader market.\footnote{As additional evidence for carrier multihoming outside the platform, I find that 95\% of carriers drive no more than 10,638 miles on the platform per year (see the full distribution in Figure \ref{fig:CDF_miles_driven} in the Appendix). Given that owner-operators drive over 100,000 miles per year on average, most carriers must find shipments off the platform for a majority of their mileage \parencite{ooida2022survey}. Given this degree of multihoming, a key focus of the empirical strategy will be to recover a distribution of outside offers.}

\begin{table}[htbp]
    \begin{center}
    \caption{Impact of Carrier Cancellations on Shipment Metrics}
    \label{tab:cancellation_impact}
    
  \begin{tabular}{lcc}
    \toprule
    & No Cancellation & $\geq$1 Cancellations \\
    \midrule
    Shipment rescheduled & 10.7\% & 18.1\% \\
    Shipment cancelled & 6.9\% & 8.1\% \\
    Final Margin & 7.7\% & 4.6\% \\
    \midrule
    Overall carrier cancellation rate & \multicolumn{2}{c}{14.9\%} \\
    \bottomrule
  \end{tabular}

    \end{center}
  
  {\footnotesize{\textsc{Note}: Table compares outcomes for shipments with and without at least one carrier cancellation. Shipment cancelled refers to a cancellation from the shipper side.}}
\end{table}

In \autoref{tab:cancellation_impact}, I present the tangible consequences of cancellations for the platform. The table shows that cancellations have a large impact on the rate of shipment rescheduling, with a much more modest effect on shipment cancellations by the shipper. The low impact on shipper cancellation rates suggests minimal disintermediation, unlike findings in \textcite{xie2022platform}, which documented substantial disintermediation on a Chinese intra-city cargo delivery platform. This difference likely stems from U.S. shippers' preference for long-term contracts with brokers or large carriers who can reliably source drivers, rather than engaging with small carriers for one-off shipments at a lower price.

The data also show that in most cases, the platform is able to re-match a shipment following a carrier cancellation, but does so at a higher price, resulting in a lower margin. These financial consequences for the platform raise a natural question: what drives carriers to cancel confirmed shipments?

\begin{figure}[htbp]
    \begin{center}
        \caption{Confirmation/cancellation probability vs. payoff}
        \label{fig:confirm_cancel_prob_price}
        \subfloat[Cancellation]{\includegraphics[width=0.48\textwidth]{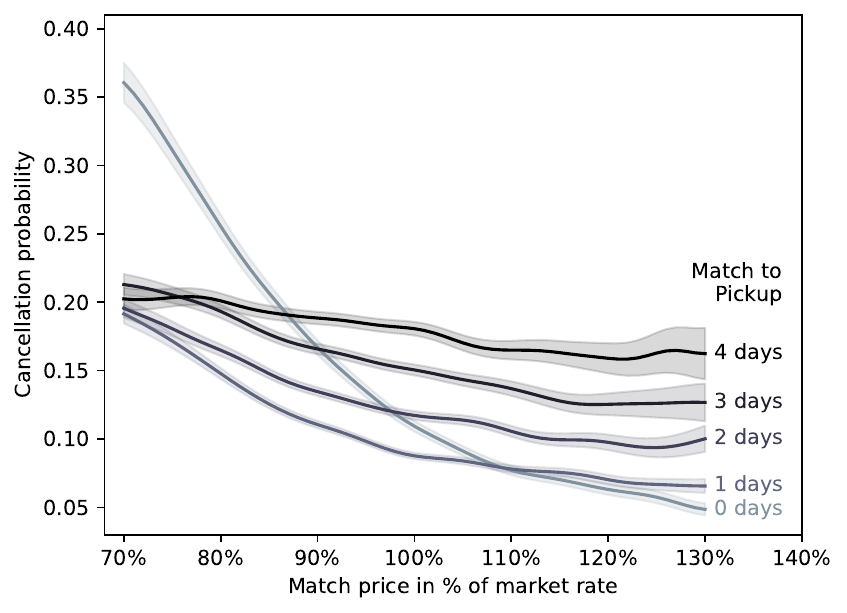}}
        \hfill
        \subfloat[Bid Reneging]{\includegraphics[width=0.48\textwidth]{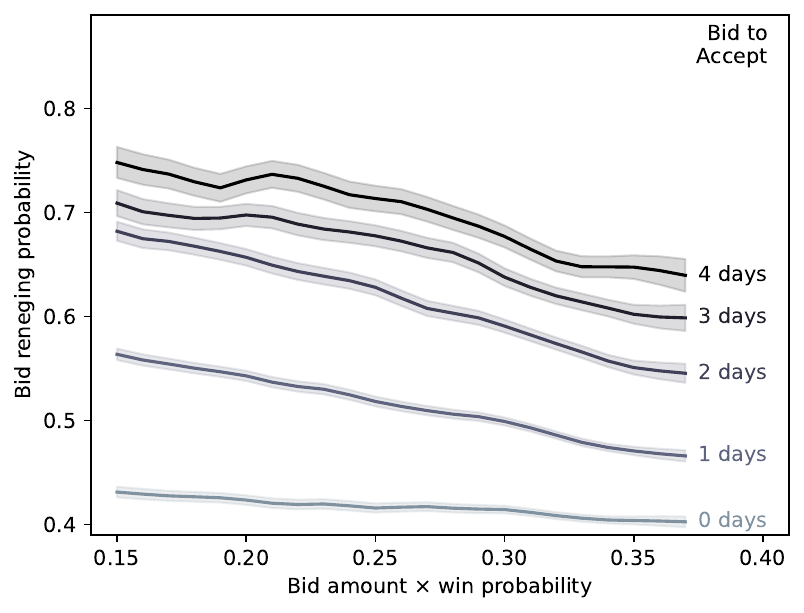}}
    \end{center}
    {\footnotesize{\textsc{Note}: Figures are obtained through a two-dimensional kernel regression, conditioning both on different time horizons for attrition, and measures of payoffs. 95\% confidence intervals are depicted as shaded areas around each respective line. Kernel bandwidth chosen in accordance with Silverman's rule of thumb. Win probability in panel (b) is also computed with a one-dimensional kernel regression of bid acceptance on bid amount.}}
\end{figure}

\autoref{fig:confirm_cancel_prob_price} shows that both cancellations and bid reneging are affected by (i) the payoff of a shipment and (ii) the time interval available for carriers to change their minds. Panel (a) shows that the probability of a carrier cancelling a shipment is decreasing in the price of the shipment, while panel (b) shows that the probability of a carrier reneging on a bid is decreasing in a simple measure of the expected shipment payoff. Both panels also show that the probability of cancelling (reneging) is increasing in the time interval between when the carrier is matched to a shipment (makes a bid) and the pickup time (bid acceptance time).  These patterns are consistent with standard models of search markets \parencite{mccall1970economics} where a stochastic process of outside offers composed of some rate of arrival of outside offers, and some distribution of payoffs conditional on the arrival, rationalize the carrier's decision to exit a match. %

\subsubsection{Reputational Mechanism}
\label{sec:background_reputational}

As described above, the platform's stated policy was to punish cancellations that occurred within 48 hours before the pickup time by penalizing future bids by some quality adjustment factor, which is simply added on to the bid dollar amount. This adjustment factor was not disclosed to carriers and its details are confidential. Nevertheless, in \autoref{tab:penalty_effects}, I present some descriptive evidence of the effect of the reputational penalty on carriers' win probability and their bidding behavior to shed light on the severity of the penalty. 

First, note that a naive regression of a carrier's win probability on past cancellations yields positive coefficients, implying a \emph{benefit} to cancellations. But this regression likely confounds the actual penalty, which occurs via the quality adjustment factor, with behavioral responses to the penalty, in the form of lower bids. To disentangle these effects, I first estimate the effect of the actual bid and the penalty on the win probability, then estimate the effect of past cancellations on the penalty and the bid amount separately. All regressions are specified as linear probability models to allow for high-dimensional fixed effects.

The main win probability regression includes high-dimensional shipment lane and calendar month fixed effects to control for local market conditions. The results show that the quality adjustment factor's impact on the win probability is roughly two-thirds of the impact of the bid amount. This is likely due to the fact that only the raw bid determines whether a carrier beats the reserve price, while the quality adjustment factor only affects the ranking of the bids.

The two regressions of the quality adjustment factor or the bid amount on past cancellations include high-dimensional carrier fixed effects, controlling for time-invariant differences in cancellation behavior and bidding behavior across carriers.

The second regression shows a higher sensitivity of the quality adjustment factor to recent cancellations, but an overall small magnitude---a single cancellation in the past 30 days increases the quality adjustment on a thousand dollar shipment by roughly three and a half cents. As carriers may change their behavior (in other dimensions, such as on-time pickup and delivery) in response to a cancellation to maintain their quality score, which can explain the negative coefficient on the 90 day window, it is possible that the true effect of cancellations is greater than the estimates suggest, but as carriers only receive coarse information about their quality score, it is unlikely that they would be able to precisely adjust their behavior to yield such an exact offsetting effect.

Finally the third regression shows that carriers reduce their bids in response to cancellations. A single cancellation in the past 30 days reduces the bid on a thousand dollar shipment by roughly 80 cents. This effect is an order of magnitude larger than the actual impact on the quality adjustment factor. This suggests that carriers may be overestimating the reputational penalty, which should not occur if carriers are fully informed about how the penalty works (see \autoref{sec:appendix_repeated_model} for a formal analysis of this argument). Ultimately, the policy's effects will largely depend on carriers' \emph{subjective} beliefs about the penalty, which will be recovered through the estimation of the structural model.

\begin{table}[htbp]
    \begin{center}
        \caption{Reputational penalty effects}
        \label{tab:penalty_effects}
        \resizebox{\textwidth}{!}{\begin{tabular}{@{\extracolsep{5pt}}lcccc}
\\[-1.800ex]\hline
\hline \\[-1.800ex]
\\[-1.800ex] & \multicolumn{1}{c}{(1) Win Prob} & \multicolumn{1}{c}{(2) Quality Adjustment} & \multicolumn{1}{c}{(3) Bid} & \multicolumn{1}{c}{(4) Naive Win}  \\
\\[-1.800ex] & (1) & (2) & (3) & (4) \\
\hline \\[-1.800ex]
 Normalized Price & -0.746$^{***}$ & & & \\
& (1.24e-03) & & & \\
 Quality Adjustment & -0.490$^{***}$ & & & \\
& (0.011) & & & \\
 Cancellations (30D) & & 3.51e-05$^{***}$ & -7.79e-04$^{***}$ & 5.81e-03$^{***}$ \\
& & (5.24e-06) & (5.04e-05) & (1.02e-04) \\
 Cancellations (60D) & & 2.24e-05$^{***}$ & -6.85e-05$^{}$ & 6.92e-06$^{}$ \\
& & (5.52e-06) & (5.30e-05) & (1.22e-04) \\
 Cancellations (90D) & & -1.93e-05$^{***}$ & -1.93e-04$^{***}$ & 2.24e-03$^{***}$ \\
& & (5.38e-06) & (5.17e-05) & (1.13e-04) \\
\hline \\[-1.800ex]
Observations & 2244902 & 2244539 & 2244539 & 2249878 \\
 Lane $\cross$ Month FE & Yes & No & No & Yes \\
 Carrier FE & No & Yes & Yes & No \\
\hline
\hline %
\end{tabular}
}
    \end{center}
    
    {\footnotesize
        \textsc{Note}: All prices and quality adjustments are normalized by each shipment's lane-specific spot rate. 
        30D cancellation variable is a rolling sum of cancellations in the 30 days preceding a bid, while 60D and 90D refer to the 30-60 and 60-90 day windows, respectively.
        Standard errors are in parentheses. $^{*}$p$<$0.100; $^{**}$p$<$0.050; $^{***}$p$<$0.010. 
    }
\end{table}

To summarize, the empirical analysis of the platform data has revealed several stylized facts that will guide the development of the structural model. First, the platform's auction format is characterized by a high degree of attrition, with the majority of views not resulting in a bid, and the majority of bids not resulting in a match. Second, the attrition patterns are sensitive to both price and time, suggesting search behavior driven by off-platform opportunities. Third, cancellations have a large impact on the platform's bottom line, yet the reputational penalty imposed by the platform has quantitatively small effects on carriers' win probabilities and bidding behavior. A simplified model capturing these patterns is presented in the next section.

\section{Baseline Model}
\label{sec:toy_model}

I have shown that cancellation and confirmation behaviors are empirically linked to the price and timing of the initial match between a carrier and a shipment. However, there is not enough variation in the data to directly measure the impact of a change in the cancellation policy.\footnote{In general, it would be difficult for any firm to experimentally vary their cancellation penalties, as these are generally less salient than prices, and carriers, or customers, may take time to adapt to a new policy.} The dual objectives of the model are thus to \emph{account} for the empirical patterns in the data and \emph{predict} the counterfactual effects of different cancellation policies.

For the sake of exposition, I begin with a simplified baseline model. This captures the two essential economic mechanisms determining the effect of a change in the cancellation penalty, but abstracts away from the additional complexities of stochastic and heterogeneous arrival times of carriers, and the multi-unit nature of the auction format. The first mechanism is that, holding the match time and price fixed, a higher cancellation penalty should decrease a carrier's propensity to cancel. The second mechanism is that a carrier's forward-looking strategy in the auction depends on the cancellation penalty through their opportunity cost of continuing to search, following well-established models of dynamic auctions \parencite{jofre2003estimation,zeithammer2006forward,hendricks2022sell}. 

Since the auction format allows bidders to renege when they win, solving for the auction equilibrium is a non-trivial challenge, as it must account for the opponent's probability of reneging. 

\subsection{Setup}

In the model, the brokerage platform runs an auction for a single shipment for every pickup date, each of which is treated as a separate market. The auction takes place in the days leading up to the departure date, with the index $d \in \{4,3,2,1,0\}$ counting down the number of days remaining until the pickup time. For each shipment, the platform earns revenue $v \sim N(\mu_v, \sigma_v)$ for a successful match; the payoff for an unsuccessful match is normalized to zero. Two carriers $i \in \{1,2\}$ participate in each auction, with marginal costs of shipping $c_i \sim N(\mu_c, \sigma_c)$. Each of these two carriers also takes independent draws from shipment jobs outside the platform, with net payoffs $\pi - c_i$, with $\pi \sim G(.) \equiv N(\mu_G,\sigma_G)$. These three distributions are the primitives of the model.

The platform's only policy lever is the cancellation penalty $\kappa$\footnote{This reduced-form parameter represents the reduction in the carrier's present discounted value of future work on the platform as a result of the platform's penalty in future auctions. See \autoref{sec:appendix_repeated_model} for the conceptual framework of the long horizon repeated game and why it is abstracted away in the main text.}. In the status quo, the cancellation penalties $\kappa$ are reputational, in that they impact a carrier's utility by reducing their future profits on the platform, but do not result in any direct monetary transfer to the platform. While such reputational mechanisms may also have benefits in terms of \emph{screening out}  low-quality carriers, Appendix \ref{sec:appendix_screening} provides reduced-form evidence that screening has limited benefits for platform profits, as carriers who cancel more frequently also tend to match at lower prices, effectively off-setting the cost of their cancellations. Based on this evidence, I assume the platform simply receives nothing when a carrier cancels, so that $\kappa$ only provides an incentive effect. In counterfactual simulations, I also consider monetary cancellation penalties, which result in a direct monetary transfer of value $\kappa$ from the carrier to the platform when a cancellation occurs.

The timing of the game is as follows:
\begin{itemize}[noitemsep,topsep=0pt]
    \item $d=4$: carriers submit their bid $b_i$
    \item $d=3$: each carrier receives an outside offer $\pi_{i,3}$
    \begin{itemize}[noitemsep,topsep=0pt]
        \item If a carrier takes the outside offer, they exit the market and get payoff $\pi_{i,3}-c_i$
    \end{itemize}
    \item $d=2$: auction clears, awarding shipment to lowest bidder still in the market, subject to bidding less than $v$\footnote{Setting the reserve price equal to the platform's valuation is not generally optimal \parencite{riley1981optimal}. However, the optimal reserve price in this model will also depend on the cancellation penalty. For the sake of tractability, the reserve price is simplified here.}
    \item $d=1$: each carrier receives another outside offer $\pi_{i,1}$
    \begin{itemize}[noitemsep,topsep=0pt]
        \item If an unmatched carrier takes the outside offer, they exit the market and get payoff $\pi_{i,1}-c_i$
        \item If a matched carrier takes the outside offer, they are subject to a cancellation penalty and receive payoff $\pi_{i,1} - c_i - \kappa$
    \end{itemize}
    \item $d=0$: if the auction winner has not cancelled, they receive a payoff $b_i - c_i$, while the platform gets payoff $v - b_i$.
\end{itemize} 

Formally, I model the auction as a Bayesian game between the two carriers. I focus on the symmetric Bayesian Nash equilibrium consisting of conditional win probability beliefs, denoted $\gamma(b) \in [0,1]$, and the carriers' optimal policies, described further below. 

\subsection{Carrier policy}

Carriers are searching for the most profitable shipment they can take given their costs (their type). This may be a shipment from the platform, if they win the auction, or an outside offer they receive. Their behavior can be characterized by a set of type-specific policy functions, consisting of a bidding function $b^*(c)$ and reservation wages $R^p(b,c)$, $R^m(b)$, $R^u(c)$, where the superscripts $p$, $m$ and $u$ correspond to the pending (bid), matched and unmatched states, respectively. Any outside offers $\pi$ that exceed these reservation wages will be accepted. 

I can solve for carrier policies through backward induction. Recall that the time index $d$ is counting \emph{backwards} from the pickup date. In $d=1$, the last period before pickup, a carrier who has not left the market yet is either matched with the platform's shipment or not.

If they are matched, their reservation wage depends on the price agreed-upon with the platform, given by their own bid $b_i$. The carrier then sees the outside offer $\pi_{i,1}$ and decides whether to take it or not, solving $\max\{\pi_{i,1}-c_i-\kappa,b_i-c_i\}$, which results in a reservation wage of $R^m(b_i) = b_i + \kappa$. I can describe the ex-ante value function of a matched carrier in period $d=1$ in terms of their cutoff strategy:
\begin{equation}
    \label{eq:toy_model_match_value}
    V_1(b_i,c_i) = \int_{R^m(b_i)} (\pi_{i,1} - c_i - \kappa) dG(\pi_{i,1}) + G(R^m(b_i))(b_i - c_i)
\end{equation}
An unmatched carrier at $d=1$ compares the outside offer to the payoff of not working at all, solving $\max\{\pi_{i,1}-c_i,0\}$. Their reservation wage is simply $R^u(c_i) = c_i$. The ex-ante value function of an unmatched carrier in period $d=1$ is thus:
\begin{equation}
    U_1(c_i) = \int_{c_i} (\pi_{i,1} - c_i) dG(\pi_{i,1})
\end{equation}
I then move backwards to $d=2$, the period in which the auction is cleared. The lowest bidder who clears the reserve price is awarded the shipment. I assume for now that the equilibrium probability of winning, given bid $b$ is $\gamma(b)$, which I solve for later. The ex-ante value function for this period is thus:
\begin{equation}
    W_2(b_i,c_i) = \gamma(b_i)V_1(b_i,c_i) + (1-\gamma(b_i))U_1(c_i)
\end{equation}
In $d=3$, carriers are waiting for the outcome of the auction, while receiving another outside offer. Their reservation wage at this point is $R^p(b_i,c_i) = W_2(b_i,c_i) + c_i$. The ex-ante value function for this period is then:
\begin{equation}
    W_3(b_i,c_i) = \int_{R^p(b_i,c_i)} (\pi_{i,3} - c_i) dG(\pi_{i,3}) + G(R^p(b_i,c_i))(W_2(b_i,c_i))
\end{equation}
which is similar to Equation \ref{eq:toy_model_match_value}, except that the carrier incurs no penalty for taking the outside offer. Finally, the bidding problem in the first period in $d=4$ is to maximize the continuation value in the next period:
\begin{equation*}
    \max_{b_i} W_3(b_i,c_i)
\end{equation*}
By applying Lemma \ref{lemma:derivative} in the Appendix (which is just a special case of the Envelope theorem), I can show that the FOC is:
\begin{align*}
    &\pdv{W_3(b_i,c_i)}{b_i} = G(R^p(b_i,c_i))\pdv{R^p(b_i,c_i)}{b_i} = 0 \\
    \Rightarrow &\pdv{(W_2(b_i,c_i)+c_i)}{b_i} = \gamma'(b_i)\big(V_1(b_i,c_i)-U_1(c_i)\big) + \gamma(b_i)\pdv{V_1(b_i,c_i)}{b_i} = 0
\end{align*}
Using Lemma \ref{lemma:derivative} again, the derivative of the matched value function is given by $\pdv{V_1(b_i,c_i)}{b_i} = G(R^m(b_i)) = G(b_i+\kappa)$.

The optimal bid satisfies the following equation:
\begin{equation}
    \label{eq:toy_optimal_bid}
    b^*(c_i) = c_i + \underbrace{\frac{(1-G(b^*(c_i)+\kappa))\kappa + \int_{c_i}^{b^*(c_i)+\kappa}(\pi_{i,1}-c_i)dG(\pi_{i,1})}{G(b^*(c_i)+\kappa)}}_{\text{Opportunity cost}} - \underbrace{\frac{\gamma(b^*(c_i))}{\gamma'(b^*(c_i))}}_{\text{Markup}}
\end{equation}
Thus, a carrier's optimal bid consists of their marginal cost, the opportunity cost of foregone outside offers (increasing in the cancellation penalty), and a standard markup term.\footnote{Note that $\gamma'(b) < 0$ in equilibrium, so that the markup is positive.} The bidding function is monotonic in $c$, so it is invertible. %
Thus, although the reservation wages $R^p(b,c)$ and $R^m(b)$ are written in terms of the bid, they can also be expressed solely as a function of the carrier's type.

\subsection{Auction equilibrium}

Carrier strategies are a function of the equilibrium conditional win probabilities $\gamma(b)$. This win probability is given by:\footnote{The terms in parentheses are conditional on $b_i < v$, but since $b_j$ and $v$ are independent, the expression can be simplified.}
\begin{equation}
    \label{eq:toy_win_prob}
    \gamma(b_i) = \underbrace{P(b_i < v)}_{\text{Beat reserve price}}\Big( \underbrace{P(b_i \leq b_j)}_{\text{Beat opposing bid}} + \underbrace{P(b_i > b_j)\mathbb{E}_{b_j}[1-P(confirm|b_j)|b_j < b_i]}_{\text{Opponent wins but reneges}}\big)
\end{equation}

where $P(confirm|b_j) = G(R^p(b_j,b^{*-1}(b_j)))$ is the probability that an opponent with bid $b_j$ doesn't find an outside offer above their reservation wage. Thus, compared to a typical auction, each bidder has an additional chance to win if the opponent reneges. The carrier policies $b^*(c), R^p(b,c), R^m(b), R^u(c)$ jointly influence the above conditional probability, in addition to the three primitive distributions. 

\subsection{Decomposing the effect of an increased penalty}

\begin{figure}[htbp]
    \caption{Cumulative effects of an increased cancellation penalty}
    \label{fig:toy_cf_kappa_effect}
    \begin{center}
    \begin{subfigure}{0.32\textwidth}
        \centering
        \includegraphics[width=\textwidth]{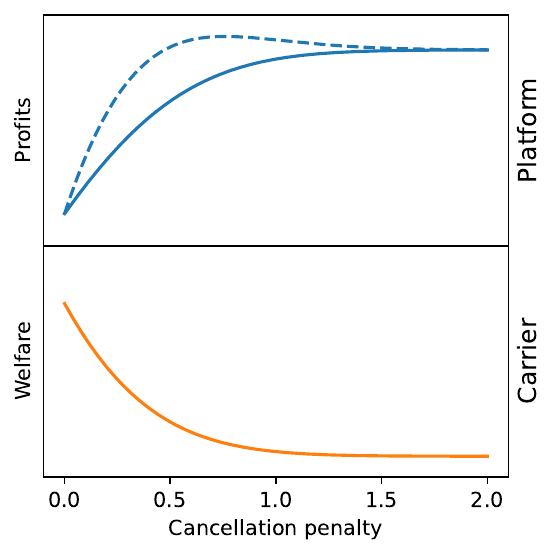}
        \caption{Reduced cancellations}
        \label{fig:toy_cf_kappa_effect_direct}
    \end{subfigure}
    \hfill
    \begin{subfigure}{0.32\textwidth}
        \centering
        \includegraphics[width=\textwidth]{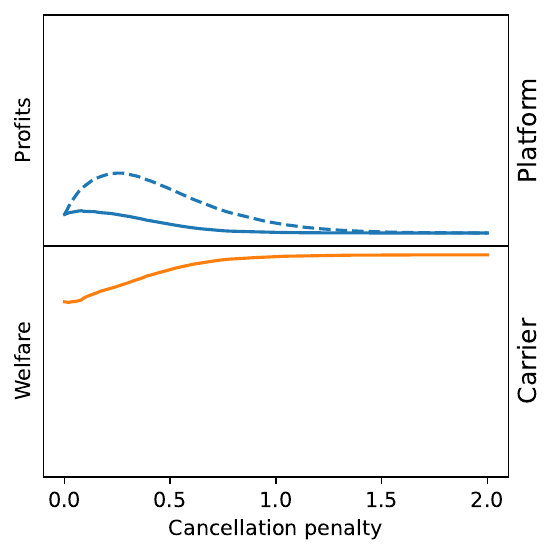}
        \caption{Transaction price}
        \label{fig:toy_cf_kappa_effect_price}
    \end{subfigure}
    \hfill
    \begin{subfigure}{0.32\textwidth}
        \centering
        \includegraphics[width=\textwidth]{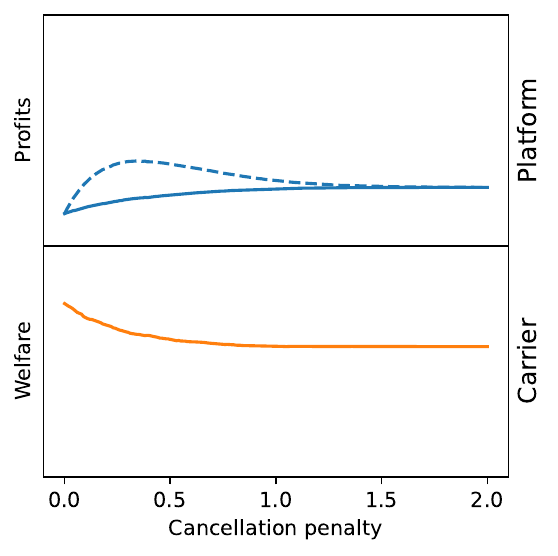}
        \caption{Equilibrium effects}
        \label{fig:toy_cf_kappa_effect_eqm}
    \end{subfigure}
    \end{center}
    {\footnotesize \textsc{Note:} Figures show the cumulative effects of increasing cancellation penalty $\kappa$ from a baseline equilibrium of $\kappa = 0$. Dashed lines represent platform payoffs under monetary penalties. In (a), carrier bids and auction acceptances/confirmations are held fixed, so that only cancellation behavior changes. In (b), the carrier bid is changed, while holding beliefs about win probabilities and auction acceptances/confirmations fixed. In (c), full equilibrium is adjusted, including equilibrium win probabilities and auction acceptances/confirmations. All quantities scaled relative to average shipment value to the platform. Carrier welfare is relative to non-existence of the platform.}
\end{figure}

As alluded to previously, an increase in the firm's cancellation penalty will not only affect the propensity of carriers to cancel, but also their forward-looking bidding behavior, which ultimately changes the equilibrium of the auction. To understand how these effects work in conjunction, I conduct a simple numerical comparative statics exercise, progressively adding each effect. I also show the difference between reputational penalties, and monetary penalties, with the latter generating additional revenue for the firm with each cancellation. I fix $\mu_v = 1$, $\sigma_v = 0.5$, $\mu_c = 0.65$, $\sigma_c = 0.3$, $\mu_G = 0.6$, $\sigma_G=0.6$. As a baseline, I solve for the equilibrium of the model with the cancellation penalty $\kappa$ set to $0$. The results are illustrated in Figure \ref{fig:toy_cf_kappa_effect}.

In panel \subref{fig:toy_cf_kappa_effect_direct}, I hold the carriers' bidding behavior and the auction outcomes fixed, and only adjust their reservation wages $R^m(b) = b+\kappa$. For a given bid $b$, the carriers' propensity to cancel, given by $1-G(R^m(b))$, thus decreases with the penalty, resulting in a welfare loss for carriers, but a profit gain for the platform. When the penalty is reputational, the platform's profits are monotonically increasing in the penalty, but when the penalty is monetary, there is an interior optimum to the platform's profits, which suggests that the penalty is a means for the platform to extract some of the gains that carriers receive from outside offers.

In panel \subref{fig:toy_cf_kappa_effect_price}, I adjust the carrier bids, incorporating the increased penalty into the opportunity cost term in Equation \ref{eq:toy_optimal_bid}, while holding the auction outcomes and win probability beliefs $\gamma(b)$ fixed. The increase in opportunity costs is not uniform across types, since if a carrier with a high bid $b$ is awarded the auction, they are unlikely to find a better opportunity even with no cancellation penalty. What the figure illustrates is that---in expectation---carriers increase their bids enough to more than offset the welfare loss from the increased penalty, when the auction equilibrium is held fixed. Consequently, it is less attractive for the firm to increase the penalty, though they may still collect additional revenue from doing so when the penalty is monetary.

Finally, panel \subref{fig:toy_cf_kappa_effect_eqm} illustrates the full equilibrium effects. This means the auction outcomes are changed, so that some of the previously winning bids now lose, as the bids have increased. It also means solving for a fixed point between carrier policies and the equilibrium win probability defined in Equation \ref{eq:toy_win_prob}. As allocations change, some bids are no longer accepted, and some carriers pre-emptively choose to exit the market, reneging on their winning bids. This alters the equilibrium win probabilities, leading to a new equilibrium. Overall, this restores the pattern observed in panel \subref{fig:toy_cf_kappa_effect_direct}, where the platform benefits from a higher penalty and carriers incur losses, albeit to a lesser extent. This exercise demonstrates the importance of each stage of the analysis in understanding the effects of cancellation penalties on the platform and carriers.

\subsection{Optimal Cancellation Penalties}

\begin{figure}[htbp]
    \caption{Social Welfare vs. Firm Profits}
    \label{fig:toy_cf_plot}
    \begin{center}
    \begin{subfigure}{0.49\textwidth}
        \centering
        \includegraphics[width=\textwidth]{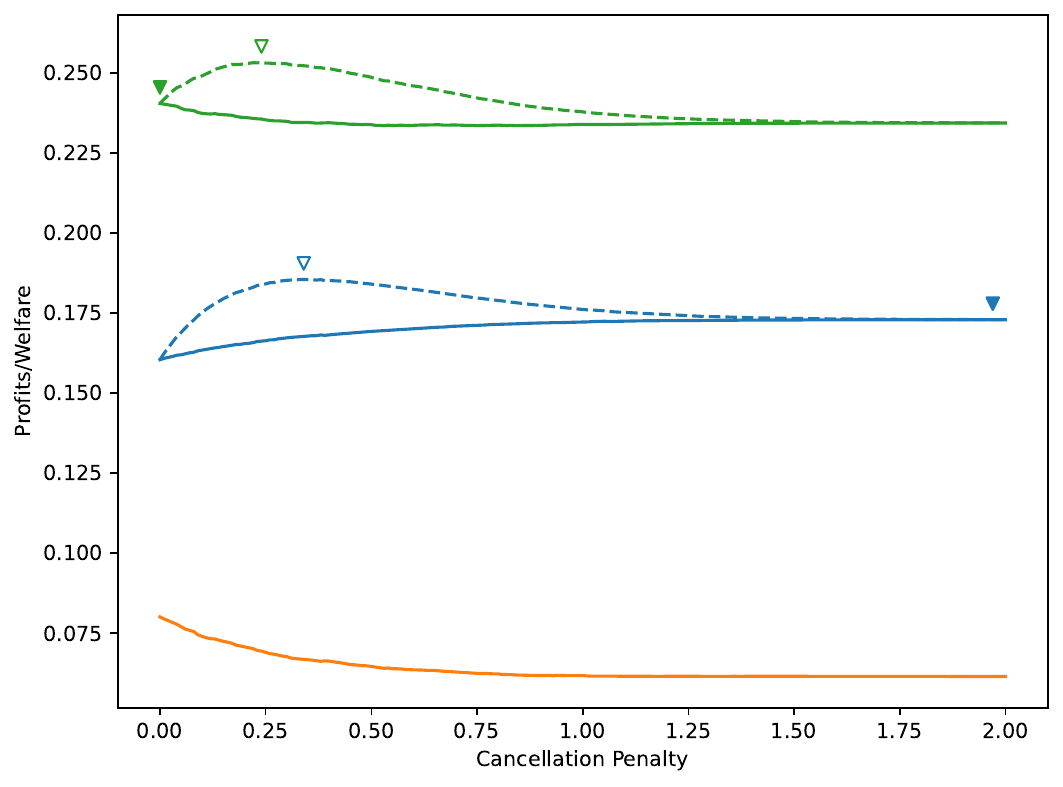}
        \caption{Low variance in outside offers}
        \label{fig:toy_cf_plot_low_sigma}
    \end{subfigure}
    \hfill
    \begin{subfigure}{0.49\textwidth}
        \centering
        \includegraphics[width=\textwidth]{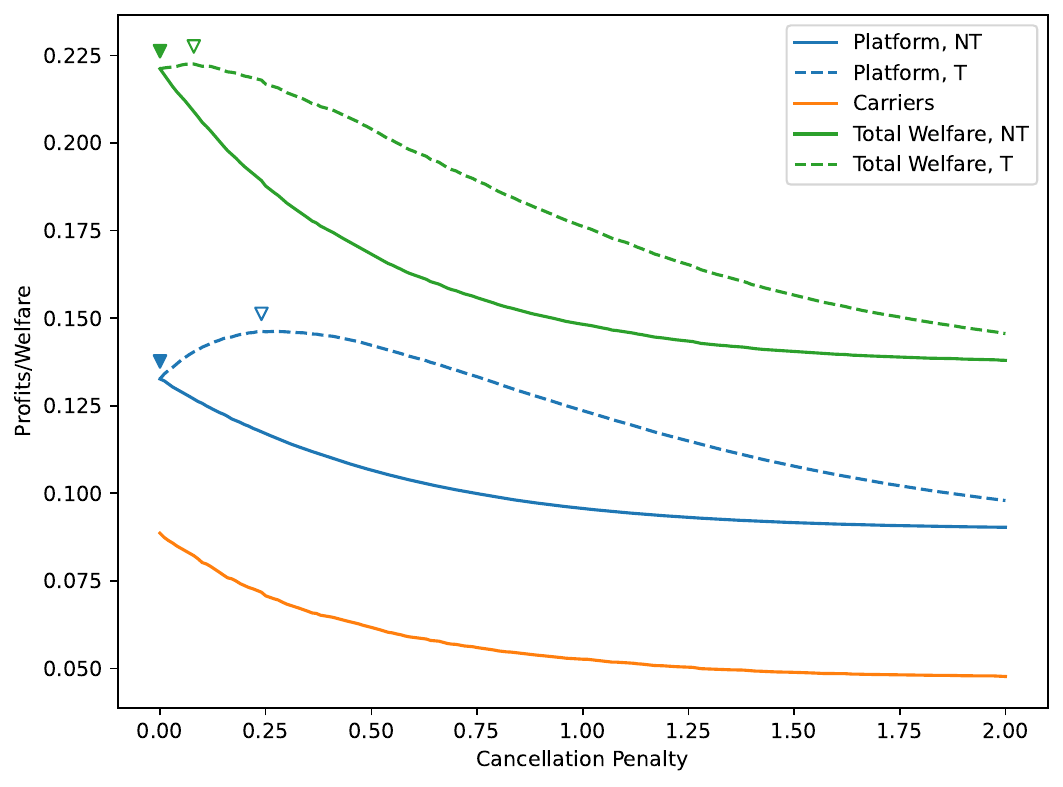}
        \caption{High variance in outside offers}
        \label{fig:toy_cf_plot_high_sigma}
    \end{subfigure}
    \end{center}
    {\footnotesize \textsc{Note:} Dashed lines represent \emph{monetary} penalties (T=Transfer), solid lines represent \emph{reputational} penalties (NT=No Transfer). Triangles denote optimal penalty levels for respective objectives. All quantities are scaled relative to average shipment value to the platform. Carrier and total welfare is relative to non-existence of the platform. Low variance is $\sigma_G=0.6$ and high variance is $\sigma_G=1.2$.}
\end{figure}

In the next exercise I evaluate the social planner's objective and the platform's profits over a range of cancellation penalties, displayed in Figure \ref{fig:toy_cf_plot}. Once again, in the figure I distinguish between reputational and monetary penalties. I repeat the exercise for a scenario with higher variance in outside offers, with $\sigma_G = 1.2$. Intuitively, a higher variance in a carrier's outside option increases the value of the option to cancel.  

I start by considering reputational penalties, given their use in the status quo. Since triggering reputational penalties does not generate any revenue, the firm would---all else equal---prefer to prevent cancellations altogether. But since carriers' bids depend on the penalty level, increasing the penalty could potentially leave the firm worse off because of the pass-through effect. I find that the net effect of these two forces depends crucially on the variance of the carriers' outside offers; when the variance is low (high) increasing the penalty monotonically increases (decreases) firm profits. In both cases, carrier welfare and overall welfare are decreasing in the penalty, although more starkly when the variance is high. The takeaway is that, under reputational penalties, carrier and firm incentives are aligned when the welfare impact of the penalties is strong enough, thanks to the upstream effects on bidding behavior.

In the case of monetary penalties, both objective functions have interior maxima in both scenarios. However, the profit-maximizing penalty is higher than the social welfare-maximizing penalty. This aligns with the idea that penalties allow the platform to capture a portion of the rents that carriers receive from outside offers. Under monetary penalties, the platform and social planner's objectives are generally not aligned, although worst-case divergence in optimal policies is less pronounced than under reputational penalties.

In summary, this illustrative model demonstrates that the welfare implications of different penalty regimes are theoretically ambiguous and depend heavily on the nature of the outside offer process as well as the type of penalty. The model also underscores the importance of considering the equilibrium pass-through of the penalties, rather than just the direct effects on cancellation behavior.

\section{Empirical Model}
\label{sec:model}

The baseline model above captures the fundamental economic impacts of the cancellation penalties on carriers' bidding strategies and reservation wages. However, it does not fully reflect all aspects of the empirical context. To enable a realistic empirical analysis, I enrich the model with additional features, including the multi-unit nature of the auctions, the stochastic arrival of carriers and shipments, the Accept-Now feature, and the multiple auction clearing rounds. With these additions the model better captures the key features of the platform described in Section \ref{sec:background}.

Formally, this model takes place in continuous time, but as all events and decisions arrive according to Poisson processes, there is no reference to a continuous time index in what follows. The dynamics of the model are thus described by a continuous-time Markov chain, as in \textcite{doraszelski2012avoiding}. 

The model is specified on a market-by-market basis, where one market is defined as a combination of a shipping lane (metro to metro) and a departure date. As the model parameters will be estimated on a lane-specific basis, I make no further reference to the lane below. The decision problem takes place in the days leading up to departure, with the state variable $d$ indexing the number of full days until departure. As in the baseline model, the index counts backwards from the pickup time, so that the period from 0 to 24 hours before departure is indexed as $d=0$, the period from 24 to 48 hours before is indexed as $d=1$, and so on. Carriers are indexed by $i$ and shipments are indexed by $j$.

In the first half of this section I describe the individual decision problem of carriers and their optimal policies. In the second half I describe the aggregate dynamics of the platform, which are used to simulate the counterfactuals.

\subsection{Carriers}
\label{sec:model_carrier}

As in the baseline model, the model of carrier behavior includes an optimal bidding function and a set of reservation wages. The empirical model adds a first-stage choice of whether to bid, use the Accept-Now feature, or ignore a shipment, as well as a richer cost structure to allow for match-specific costs in the multi-unit auction setting. In their interactions with a particular shipment, carriers can go through multiple states: unmatched, meaning they are certain not to take the shipment, matched, meaning they are certain to take the shipment, and pending bid, meaning they have placed and are waiting to see if they win the auction.

Carriers vary in their cost of matching with each shipment. Their shipment-specific cost $c_{ij}$ consists of a mean cost $\bar{c}_i \sim N(\mu^c_d, \sigma^c_d)$ and an idiosyncratic term $\epsilon_{ij}\sim N(0,\sigma_\epsilon)$. The mean cost $\bar{c}_i$ is interpreted as a combination of real marginal costs $\tilde{c}_i$ and the opportunity cost of not taking an outside offer with payoff $\bar{u}_i$, so that $\bar{c}_i = \tilde{c}_i + \bar{u}_i$.\footnote{The real marginal costs can include fuel, depreciation and other trip related costs, as well as the change in continuation values from arriving in a different location. This follows the notion of inclusive costs in \textcite{buchholz2020personalized}.} The idiosyncratic component $\epsilon_{ij}$ captures a carrier's cost of taking on a particular shipment. Anecdotal evidence from the firm providing the data suggests that scheduling conflicts and relocation costs are the primary sources of idiosyncratic variation in costs.

In addition to shipments they view on the platform, carriers also receive outside offers off-platform. Whereas in the baseline model, the arrival times of the outside offers is deterministic, in the empirical model, they are stochastic. Each carrier is subject to a Poisson process of outside offers, with rate $\tilde{\lambda}_{d}$. Each offer is characterized by a payoff $\pi_{ij}$, which is distributed according to $N(\mu^\pi_d, \sigma^\pi_d)$. 

Furthermore, each carrier is also subject to a stochastic attention mechanism. For each outside offer that arrives, or each auction win on the platform, carriers have a day-specific probability of paying attention $\alpha_d$. It is also assumed that a carrier must be paying attention in order to confirm a bid. Ultimately, this means that the effective arrival rate of offers is $\lambda_d = \alpha_d \tilde{\lambda}_d$ and the acceptance rate of winning bids (derived in full later) is multiplied by $\alpha_d$. This mechanism is reminiscent of the simple stochastic consideration sets of \textcite{manzini2014stochastic}, and is a reasonable assumption given the large proportion of carriers who are owner-operators and lack dedicated administrative staff to monitor their auctions when they are on the road or off-duty. This model feature also helps explain the high level of attrition at the pending bid stage and the low level of attrition at the matched stage.

Similarly to the baseline model, carriers' behavior can be summarized through a set of type-specific policy functions: the optimal bidding function $b^*_d(c_{ij})$, as well as the reservation wages $R^U_{d}(c_{ij})$, $R^P_d(b_{ij},c_{ij})$, and $R^M_d(b_{ij},c_{ij})$. These reservation wages determine the thresholds for accepting outside offers in the unmatched, pending bid, and matched states, respectively. The empirical model extends carrier policies with the addition of the first-stage choice probabilities $P_{d,ij}(x|c_{ij},b^A_j)$, where $x$ is the choice of whether to ignore, Accept Now, or bid on a shipment after viewing it. Shipment views are treated as exogenous. The model also increases the dimensionality of the carrier policies, with the time indices $d$ denoting the current time and $\bar{d}_i$ denoting the initial time of arrival to the market, which determines the carrier's cost distribution. 

Another complication of the empirical setting is the possibility of bidding on multiple auctions simultaneously. In principle, the optimal policy of a carrier should depend on the entire \emph{set} of shipments available to them. To reduce the dimensionality of the problem, I make a simplifying assumption about the carrier policies:
\begin{assumption}
    \label{assumption:carrier_policy}
    Carriers' first-stage choice probabilities $P_{d,ij}(x|c_{ij},b^A_j)$, bidding function $b^*_d(c_{ij})$, and reservation wages $R^U_{d}(c_{ij}), R^P_d(b_{ij,c_{ij}}), R^M_d(b_{ij},c_{ij})$ on a shipment $j$ only depend on their cost $c_{ij}$ for that shipment. 
\end{assumption}
This does not imply that carriers are myopic about other shipment opportunities on the platform. Instead, I assume that they incorporate their beliefs about future opportunities on the platform into their general belief about future outside options. This is similar to research by \textcite{backus2024dynamic}, in which bidders have beliefs about future auction opportunities described by a Markov chain.

I also make a closely related assumption on carriers' knowledge of their costs:
\begin{assumption}
    \label{assumption:carrier_cost_knowledge}
    In solving for their optimal policy, carriers only know the cost $c_{ij}$ of the focal shipment and not their mean cost $\bar{c}_i$. They form Bayesian beliefs about their mean cost after observing $c_{ij}$, using $N(\mu^c_d,\sigma^c_d)$ as a prior distribution.
\end{assumption}
As will be seen further on, this assumption simplifies the computation of the likelihood of the model. If $\bar{c}_i$ were known to carriers, but unknown to the econometrician, this would create an additional dimension of unobservables to integrate out. The strength of this assumption depends empirically on the relative variance of the mean and idiosyncratic cost components. The estimation results illustrate that the variance of the idiosyncratic component is relatively much smaller, so knowledge of the cost of one shipment is a good approximation to knowledge of the mean cost term. 

However, it is a priori important to specify carriers' rational beliefs about their costs for outside offers in a manner consistent with their cost structure. For example, suppose the variance of idiosyncratic cost component $\epsilon_{ij}$ were zero, then the distribution of profits of outside offers would be fully specified by the distribution of the outside offer payoffs $\pi_{ij}$, independently of their cost for the focal shipment. On the other hand, if the variance of the idiosyncratic component is high, then a carrier that has a particularly low (high) cost on a focal shipment would expect that their costs on outside offers would be higher (lower), creating some dependence between the distribution of profits on outside offers and their cost on the focal shipment on the platform. 

The Bayesian beliefs described in Assumption \ref{assumption:carrier_cost_knowledge} approximate this effect. From this belief, we can readily derive the subjective distribution of the net payoffs of outside offers conditional on the carrier's cost for the focal shipment. To start, note that the net payoff for any outside offer will be $\pi_{ij'} - c_{ij'}$ which can be rewritten as $\pi_{ij'} - (c_{ij'} - c_{ij}) - c_{ij} = \pi_{ij'} - (\epsilon_{ij'} - \epsilon_{ij}) - c_{ij}$. We can focus on the relative payoff of the outside offer, net of the focal shipment's cost, which is $\pi_{ij'} - (\epsilon_{ij'} - \epsilon_{ij})$. Given the normality assumptions, the conditional distribution of this relative payoff, given the focal cost $c_{ij}$, is normal with a shifted mean:
$$\pi_{ij'} - (\epsilon_{ij'} - \epsilon_{ij})|c_{ij} \sim N \Big(\mu_d^\pi + \rho(c_{ij},\bar{d}_i), \sigma^{\tilde{\pi}}_{d,\bar{d}_i} \Big)$$
where $\rho(c_{ij},\bar{d}_i) = \frac{\sigma_\epsilon^2 (c_{ij}-\mu^c_{\bar{d}_i})}{\sigma_\epsilon^2 + (\sigma^c_{\bar{d}_i})^2}$, $\sigma^{\tilde{\pi}}_{d,\bar{d}_i} = \sqrt{(\sigma_d^\pi)^2 + 2\sigma_\epsilon^2 - \frac{\sigma_\epsilon^4}{(\sigma^c_{\bar{d}_i})^2 + \sigma_\epsilon^2}}$, and $\bar{d}_i$ is the carrier's arrival date to the market. Now, define a modified payoff variable $\tilde{\pi}_{ij'}$ such that
$$\tilde{\pi}_{ij'} \sim N \Big(\mu_d^\pi, \sigma^{\tilde{\pi}}_{d,\bar{d}_i} \Big) \sim G_d(.)$$
It is then clear that $\tilde{\pi}_{ij'} + \rho(c_{ij},\bar{d}_i) - c_{ij}$ is distributionally equivalent to $\pi_{ij'} - c_{ij'}$ conditional on $c_{ij}$. Thus, under Assumption \ref{assumption:carrier_cost_knowledge}, whenever a carrier evaluates an outside offer, their realized relative payoff is represented as a draw $\tilde{\pi}_{ij'}$ from the base distribution $G_d$, shifted back up by the adjustment factor $\rho(c_{ij},\bar{d}_i)$. This captures the fact that when a carrier is matched to a shipment at a high cost, their costs on other shipments are likely to be lower, and vice versa.

I derive the optimal carrier policies, which are composed of three parts:
\begin{enumerate}[noitemsep,topsep=0pt]
    \item The choice probabilities of whether to ignore, Accept Now, or bid on a shipment conditional on viewing it.
    \item The bidding function mapping the carrier's cost to their bid.
    \item The state-dependent reservation wage to take on outside offers.
\end{enumerate}
All three policies depend on the carrier's value functions in one of three states (relative to the focal shipment): unmatched, matched at price $b$, and having a pending bid $b$. These can be solved via backward induction, starting with the unmatched state.

\paragraph{Unmatched state}
The value function in the unmatched state is denoted by $U_{d,\bar{d}_i}(c_{ij})$. Whenever an unmatched carrier receives an outside offer in this state, they solve $\max\{U_{d,\bar{d}_i}(c_{ij}),\tilde{\pi}_{ij'} - c_{ij} + \rho(c_{ij},\bar{d}_i) \}$, so that their reservation wage in this state is:
\begin{equation}
    \label{eq:unmatched_reservation}
    R^U_{d,\bar{d}_i}(c_{ij}) = U_{d,\bar{d}_i}(c_{ij}) + c_{ij} - \rho(c_{ij},\bar{d}_i)
\end{equation}

The unmatched value function is thus:
\begin{align}
    \label{eq:unmatched_value}
    U_{d,\bar{d}_i}(c_{ij}) = \frac{1}{\eta + \lambda_d} \Big[ &\lambda_d \Big( \int_{R^U_{d,\bar{d}_i}(c_{ij})} (\tilde{\pi}_{ij'} - c_{ij} + \rho(c_{ij},\bar{d}_i)) dG_d(\tilde{\pi}_{ij'}) \\ \notag & + G_d(R^U_{d,\bar{d}_i}(c_{ij}))U_{d,\bar{d}_i}(c_{ij}) \Big) + \eta U_{d-1}(c_{ij}) \Big] 
\end{align}
with $U_{-1,\bar{d}_i}(c_{ij}) = 0$. Intuitively, the value function in the unmatched state is a weighted average between two events: the arrival of an outside offer and the arrival of a new day. When an outside offer arrives, the carrier only takes it if it exceeds their reservation wage. Otherwise, they remain in the current state.

See Appendix \ref{sec:value_derivation} for the formal derivation of this value function from the continuous time setup. The other value functions that follow below can be derived in a similar manner.

\paragraph{Matched state}

The value function in the matched state is denoted by $V_{d,\bar{d}_i}(b_{ij},c_{ij})$. The value in this state is a function of both the carrier's cost and the agreed upon payment $b_{ij}$, which comes from either the carrier's own bid or the Accept Now feature. 

When a matched carrier receives an outside offer, they solve $\max\{V_{d,\bar{d}_i}(b_{ij},c_{ij}),\tilde{\pi}_{ij'} - c_{ij} + \rho(c_{ij},\bar{d}_i) - \kappa_d \}$. The platform's cancellation penalties $\kappa_d$ enter into this decision in reduced form as a monetary equivalent to the reputational costs of cancelling a shipment. The carrier's reservation wage in this state is
\begin{equation}
    \label{eq:matched_reservation}
    R^M_{d,\bar{d}_i}(b_{ij},c_{ij}) = V_{d,\bar{d}_i}(b_{ij},c_{ij}) + c_{ij} - \rho(c_{ij},\bar{d}_i) + \kappa_d
\end{equation}

The matched value function is thus:
\begin{align}
    \label{eq:matched_value}
    V_{d,\bar{d}_i}(b_{ij},c_{ij}) = \frac{1}{\eta +\lambda_d} \Big[ & \lambda_d \Big(\int_{R^M_{d,\bar{d}_i}(b_{ij},c_{ij})} (\tilde{\pi}_{ij'} - c_{ij} - \kappa_d  + \rho(c_{ij},\bar{d}_i)) dG_d(\tilde{\pi}_{ij'}) \\ \notag & + G_d(R^M_{d,\bar{d}_i}(b_{ij},c_{ij}))V_{d,\bar{d}_i}(b_{ij},c_{ij}) \Big) + \eta V_{d-1}(b_{ij},c_{ij}) \Big]
\end{align}
with $V_{-1,\bar{d}_i}(b_{ij},c_{ij}) = b_{ij}-c_{ij}$. Similarly to the unmatched value function, the matched value function is a weighted average of the payoff from the arrival of an outside offer and the payoff from the arrival of a new day.

I can also derive the overall probability of a matched carrier cancelling a shipment, which will be used in constructing the model's likelihood. The cancellation probability of a matched carrier on day $d$ is jointly determined by the arrival process of outside offers and the carrier's reservation wage, as follows:
\begin{equation}
    \label{eq:cancel_prob}
    P_{\bar{d}_i}(cancel|b_{ij},c_{ij}) = \frac{\lambda_d \big(1-G_d(R^M_{d,\bar{d}_i}(b_{ij},c_{ij})) \big)}{\eta + \lambda_d \big(1-G_d(R^M_{d,\bar{d}_i}(b_{ij},c_{ij})) \big)}
\end{equation}

\paragraph{Bidding}

A carrier with a pending bid $b_{ij}$ has the value function $W_{d,\bar{d}_i}(b_{ij},c_{ij})$. At any point in time, the carrier may win the auction at day-specific rate $\gamma_d(b)$. Upon winning, the carrier may accept or decline the shipment. In the model, they will only decline the shipment if they have already matched with another shipment, or if they are not paying attention.

While waiting on a pending bid, carriers continue to receive outside offers. When one arrives, they solve $\max\{W_{d,\bar{d}_i}(b_{ij},c_{ij}),\tilde{\pi}_{ij'} - c_{ij} + \rho(c_{ij},\bar{d}_i) \}$, so that their reservation wage in this state is
\begin{equation}
    \label{eq:pending_reservation}
    R^P_{d,\bar{d}_i}(b_{ij},c_{ij}) = W_{d,\bar{d}_i}(b_{ij},c_{ij}) + c_{ij} - \rho(c_{ij},\bar{d}_i)
\end{equation}

The pending bid value function is:
\begin{align}
    \label{eq:pending_bid_value}
    W_{d,\bar{d}_i}(b_{ij},c_{ij}) = &\frac{1}{\eta + \gamma_d(b_{ij}) + \lambda_d} \Big[ \lambda_d \Big(\int_{R^P_{d,\bar{d}_i}(b_{ij},c_{ij})} (\tilde{\pi}_{ij'} - c_{ij} + \rho(c_{ij},\bar{d}_i)) dG_d(\tilde{\pi}_{ij'}) \\ \notag &+ G_d(R^P_{d,\bar{d}_i}(b_{ij},c_{ij}))W_{d,\bar{d}_i}(b_{ij},c_{ij}) \Big) \\  
    \notag &+ \gamma_d(b_{ij})(\alpha_d V_{d,\bar{d}_i}(b_{ij},c_{ij}) +(1-\alpha_d) U_{d,\bar{d}_i}(c_{ij})) + \eta W_{d-1}(b_{ij},c_{ij}) \Big] \nonumber
\end{align}

with $W_{-1,\bar{d}_i}(b_{ij},c_{ij}) = 0$. The value function is now a weighted sum of three events: the arrival of an outside offer, a new day, and the possibility of winning the auction. When a carrier wins an auction, they have the option to confirm the bid if they are paying attention, which occurs with probability $\alpha_d$. If the bid is confirmed, the carrier enters the matched state. Otherwise, they remain unmatched.

I use this to construct the likelihood of bid confirmations in the data, using only knowledge of the day the carrier placed the bid, denoted $d^b_{ij}$, and the day the bid was accepted, denoted $d^a_{ij}$. The confirmation probability of a bid is given by:
\begin{align}
    \label{eq:confirm_prob}
    P_{\bar{d}_i}(confirm|b_{ij},c_{ij},d^a_{ij},d^b_{ij}) = \alpha_{d^a_{ij}} \prod_{k=d^a_{ij}}^{d^b_{ij}} \Big(1 - \underbrace{\frac{\lambda_k(1-G_k(R^P_{k,\bar{d}_i}(b_{ij},c_{ij})))}{\eta + \gamma_k(b_{ij}) + \lambda_k(1-G_k(R^P_{k,\bar{d}_i}(b_{ij},c_{ij})))}}_{\text{Probability of taking an outside offer on day $k$}} \Big)
\end{align}

Intuitively, the confirmation probability is simply the cumulative probability of the carrier not taking any outside offer between the bid being placed and being accepted, multiplied by the probability of paying attention.

I turn to the carrier's problem of choosing their optimal bid. I assume that bidders cannot change their bids as time goes on.\footnote{In reality, they can freely change their bids, but this is empirically rare, which may imply a high cognitive cost of re-optimizing the bid.} The bidding problem at the time of first viewing the shipment can then be written as:
\begin{equation}
    \max_{b_{ij}} W_{d^b_{ij},\bar{d}_i}(b_{ij},c_{ij})
\end{equation}
The first order condition---shown here in Equation \ref{eq:bid_foc}---takes the form of a weighted sum of day-specific marginal utilities, each of which accounts for the benefit of a higher bid for a win on each particular day. The full derivation is given in Appendix \ref{sec:bid_derivative}.
\begin{multline}
    \label{eq:bid_foc}
    \pdv{W_{d^b_{ij},\bar{d}_i}(b_{ij},c_{ij})}{b_{ij}} = \sum_{k=0}^{d^b_{ij}} \overbrace{\Big[ \prod_{\ell=k}^{d^b_{ij}} \frac{\eta}{\eta + \gamma_\ell(b_{ij}) +\lambda_\ell\big(1 - G_\ell(R^P_{\ell,\bar{d}_i}(b_{ij},c_{ij})) \big)} \Big] \eta^{-1}}^{\text{Probability of bid surviving from $d$ to $k$}} \\ 
    \gamma_k'(b_{ij}) \Big( \underbrace{\alpha_k V_{k,\bar{d}_i}(b_{ij},c_{ij}) + (1-\alpha_k)U_{k,\bar{d}_i}(c_{ij}) - W_{k,\bar{d}_i}(b_{ij},c_{ij})}_{\text{Benefit of winning net of opportunity costs}} + \underbrace{\frac{\alpha_k\gamma_k(b_{ij})}{\gamma_k'(b_{ij})} P_{k,\bar{d}_i}(\text{no cancel}|b_{ij},c_{ij})}_{\text{Markup}} \Big) \\
     = 0
\end{multline}
Let $b_d^*(c_{ij},\bar{d}_i)$ denote the optimal bidding function implicitly defined by the first-order condition.

\paragraph{First-Stage choice}

When first viewing a shipment, a carrier can choose between three options: ignoring the shipments, bidding on the shipment, or using the Accept-Now feature, to immediately match with the shipment at the price set by the platform, denoted $b_j^A$, which is observed by the carrier. I can formulate the continuation value associated with each of these choices. Let $u^k_{ij}, k \in \{B, A, I\}$ be the continuation value of bidding, using the Accept-Now feature, and ignoring the shipment. 

To increase the flexibility of the model and fit the empirical first-stage choices, I add additional non-structural shocks (which do not enter welfare calculations) to rationalize the observed behavior. I assume that carriers make their first stage choices based on a set of mean pseudo-values---combining the continuation value of a choice with a ``hassle'' cost---and a choice-specific Type-1 Extreme Value shock with scale parameter $\sigma^{choice}$.\footnote{Hassle costs and choice-specific shocks are omitted from welfare computations as they would mechanically increase carrier utilities for every shipment view. In addition, the estimates of these parameters are implausibly large, so their economic interpretation is unclear. However, the counterfactual results are qualitatively robust to alternative values of these parameters, so they are not the main focus of the empirical strategy.} The mean pseudo-value of bidding 
is given by $u^{B}_{ij} = W_{d,\bar{d}_i}(b^*(c_{ij}),c_{ij}) - c_{bid}$. The mean pseudo-value of the Accept-Now feature is $u^{A}_{ij} = V_{d,\bar{d}_i}(b_j^A,c_{ij}) - c_{AN}$, with the value function of a matched shipment given by Equation \ref{eq:matched_value}. The mean pseudo-value of ignoring the shipment is $u^{I}_{ij} = U_{d,\bar{d}_i}(c_{ij})$. This formulation leads to the familiar multinomial logistic choice probabilities:

\begin{equation}
    \label{eq:first_stage_prob}
    P_{ij}(X=k) = \frac{exp(u^k_{ij}/\sigma^{choice})}{\sum_{\ell \in \{B, A, I\}} exp(u^\ell_{ij}/\sigma^{choice})}, k \in \{B, A, I\}
\end{equation}

To summarize, I have solved for the first-stage choice probabilities in Equation \ref{eq:first_stage_prob}, the bidding function, defined implicitly through Equation \ref{eq:bid_foc}, and the reservation wages in the unmatched, matched, and pending bid states in Equations \ref{eq:unmatched_reservation}, \ref{eq:matched_reservation}, and \ref{eq:pending_reservation}, respectively. Together, these fully specify the single-agent behavior of carriers on the platform. Next, I deal with the aggregate dynamics and equilibrium of the platform.

\subsection{Platform}

In the counterfactual simulations, the model of carrier behavior is combined with an aggregate dynamic model of the auction platform for each shipment pickup date. This model is used to simulate the arrival of carriers and shipments, the matching process, and the auction clearing process. As before, the model dynamics are described by a continuous-time Markov chain. At any point in time, one of the following events can occur:
\begin{enumerate}[noitemsep,topsep=0pt]
    \item A carrier $i$ arrives to the market with mean cost $\bar{c}_i$
    \item A shipment $j$ arrives to the market with value $v_j$
    \item An existing carrier $i$ views an existing shipment $j$ (drawing matching cost $\bar{c}_i+\epsilon_{ij}$), choosing to ignore, bid, or take the shipment at the Accept-Now price $p^{AN}(v_j)$
    \item A carrier accepts an outside offer and exits the market, potentially cancelling their matched shipment on the platform
    \item A shipment is cancelled and exits the market
    \item An auction on the platform attempts to clear at reserve price $r_{d}(v_j)$
    \item The days remaining until pickup count down from $d$ to $d-1$
\end{enumerate}
These events occur at state-dependent \emph{rates}, which determine the transition matrix of the Markov chain. The carrier and shipment arrivals are assumed to occur exogenously at rates $\lambda^{carrier}_d$ and $\lambda^{shipment}_d$. Each carrier-shipment pair that has not previously met can generate a view at independent rate $\lambda^{view}_d$. Following a view, carriers decide whether to ignore the shipment, bid, or take the shipment at the Accept-Now price based on the value functions derived above in section \ref{sec:model_carrier}. 

Carriers compute their outside offer acceptance threshold as the maximum of all thresholds on the set of pending bids or their currently matched shipment. As above, they receive outside offers at rate $\lambda_d$ with payoff distribution $G(.)$. When they accept an outside offer or a different offer within the platform, they take the payoff $\pi_{ij'} - c_{ij'}$ and exit the market, at the additional cost of a cancellation penalty $\kappa_d$ if they were currently matched on the platform.

Shipment cancellations occur exogenously at rate $\lambda^{shipcancel}_d$. Because the focus of the paper is on the carrier side of the market, any potential welfare gains from a shipment cancellation are not modeled.

Every currently unmatched shipment reaches an auction clearing round at rate $\lambda^{clear}_d$. In a clearing round, all bidders below the reserve price $r_d(v_j)$ are offered the shipment in sequence, starting from the lowest bid, until a carrier accepts. If no bids below the reserve price confirm their bids, nothing happens. Bids that are rejected by carriers are deleted. Bids that have yet to be accepted by the platform remain pending.

As before, the rate of transition between days $\eta$ is normalized to 1. Given the properties of the Poisson process, all other rates can thus be interpreted as the mean number of the corresponding event per day.

Finally, the equilibrium of the model is once again a fixed point between carrier policies and the conditional auction win probabilities. However, unlike the baseline model, these win probabilities do not have a closed form solution. They are solved for numerically through simulation of the platform dynamics.

\newcommand{\eoMaxKappaPct}{1.4\%\xspace}
\newcommand{\eoAvgVarianceRatio}{15.61\%\xspace}
\newcommand{\eoDataConfirmPct}{51.94\%\xspace}
\newcommand{\eoModelConfirmPct}{52.10\%\xspace}
\newcommand{\eoDataCancelPct}{10.19\%\xspace}
\newcommand{\eoModelCancelPct}{11.22\%\xspace}
\newcommand{\eoViewProb}{13\%\xspace}
\newcommand{\eoViewDerivation}{$\frac{\hat{\lambda}_v}{1 + \hat{\lambda}_v} = \frac{0.149}{1 + 0.149} \approx 13\%$\xspace}
\newcommand{\eoInitShipmentRate}{11.92\xspace}
\newcommand{\eoMeanShipmentValue}{1.002\xspace}

\section{Estimation}
\label{sec:estimation}

The estimation of the structural model is split into two parts. The carrier model above is used to derive likelihood functions in order to estimate the parameters of carrier behavior. The parameters governing arrivals to the platform, the distribution of shipment values, and the conditional reserve price and Accept-Now price models are estimated through the method of moments.

\subsection{Carrier Model Estimation}

The carrier likelihood combines four types of observations: bids (which are inverted to costs using the bidding first-order condition), first-stage participation choices, bid confirmations, and cancellation decisions. The main technical challenge arises from the panel structure of the data: because the carrier-level mean cost $\bar{c}_i$ is unobserved, it must be integrated out \emph{jointly} over all of a carrier's decisions. This integration is simplified by the fact that, for shipments where a bid is observed, the carrier's cost is pinned down by the bid inversion, so the confirmation and cancellation terms can be factored out of the integral. However, for shipments where a carrier used the Accept-Now option or ignored the shipment altogether, no bid is observed, and the idiosyncratic cost component must also be integrated out. The full derivation of the likelihood is provided in Appendix \ref{sec:appendix_likelihood}.

While the full likelihood could in principle be used to estimate all parameters jointly, I instead adopt a block-wise alternating approach (similar to two-step estimators but alternating back and forth). The parameters are split into two groups: the outside offer process and cancellation penalties ($\theta_1$), and the cost distribution and first-stage choice parameters ($\theta_2$). In the first step, $\theta_1$ is estimated using only confirmation and cancellation decisions which do not require integration over the cost distribution but do depend on the cost parameters via the carrier's Bayesian updating on the cost distribution of outside offers. In the second step, $\theta_2$ is chosen to maximize the full likelihood conditional on the current $\theta_1$. This alternation is repeated until convergence. Formally, this estimator is interpreted as a GMM estimator with stacked score conditions \parencite{newey1994large}. 

The key motivation for this separation is robustness to model misspecification. The first-stage choice model is not the primary focus of this paper and may be misspecified given the simplified distributional assumptions imposed on the cost distribution (and necessary for tractability of the carrier model solution). Because first-stage observations vastly outnumber confirmation and cancellation observations, joint estimation would allow any misspecification in the first-stage model to dominate the objective and bias the estimates of the outside offer distribution and cancellation penalties, which are the parameters of primary interest. Further details on the formal estimation procedure, including the objective functions and additional parametric restrictions, are provided in Appendix \ref{sec:appendix_likelihood}. The offline estimation of the conditional win probabilities is detailed in Appendix \ref{sec:win_rate}. For details on the estimation of the asymptotic variances of the estimates, the reader is referred to Appendix \ref{sec:struct_est_var}.

Furthermore, a penalty on negative inverted costs is also added for economic plausibility and to support identification as discussed in the next section. 

\subsection{Carrier Model Identification}
\label{sec:identification}

I begin by discussing the main identification challenge that is unique to this paper, which concerns the distribution of outside offers. It is not directly observed, but can be inferred from the conditional cancellation probabilities in the raw data, presented in Figure \ref{fig:confirm_cancel_prob_price}. However, as the cancellation decision depends linearly on the cancellation penalty, the \emph{location} of the outside offer distribution is not separately identified from the penalty level based on the conditional cancellation probability alone. 

To deal with this, I develop a partial identification argument to separately identify these objects based on the non-negativity condition on inverted costs. To enable a tractable analysis, I develop the argument  using the baseline model described in Section \ref{sec:toy_model}. Note that the following argues that the parameters are \emph{at least} partially identified, without making use of any  parametric restrictions imposed in the estimation, as discussed above. While not shown formally here, it is likely that the additional parametric restrictions on the shape of the outside offer distribution and its evolution over time provide additional constraints on the model that make it point-identified.

Using the threshold $R^m_1(b) = b + \kappa$, the cancellation probability can be written as:
\begin{equation}
    P(cancel|b_i) = 1 - G({b + \kappa})
\end{equation}
Let $\tilde{G}(x) = G(x + \kappa)$ be the location normalized distribution, which is directly identified from the cancellation probability above. Taking the bid $b$ as given, the first-order condition for optimal bidding from Equation \ref{eq:bid_foc} can be re-written in terms of the normalized distribution as follows:
\begin{equation}
    b - c - \frac{(1-\tilde{G}(b))\kappa + \int_{c-\kappa}^{b}(\tilde{\pi}-c+\kappa)d\tilde{G}(\tilde{\pi})}{\tilde{G}(b)} + \frac{\gamma(b)}{\gamma'(b)} = 0
\end{equation}
The derivative of the implicit function of $c$ with respect to $\kappa$ is:
\begin{equation}
    \frac{dc}{d\kappa} = -\frac{1 - \tilde{G}(b) + \int_{c - \kappa}^{b} \tilde{G}'(\tilde{\pi}) \, d\tilde{\pi}}{\tilde{G}(c - \kappa)} < 0, \quad \forall b \quad s.t. \quad 1 - \tilde{G}(b) > 0
\end{equation}
Now, let $\underline{b}$ be the lowest bid in the data. If $1 - \tilde{G}(\underline{b}) > 0$, that is, the (data-derived) conditional cancellation probability at the lowest bid is greater than 0, then the inverted cost $b^{*-1}(\underline{b})$ is strictly decreasing in the penalty $\kappa$. Given the lower bound $c \geq 0$, this provides an upper bound on the penalty $\kappa$. The lower bound $\kappa \geq 0$ is assumed. For any $\kappa$ in the identified set, there is a corresponding $G^\kappa$. Thus, $\kappa$ and the location of $G(.)$ are jointly set-identified.

The Monte Carlo exercise in Appendix \ref{sec:monte_carlo} provides further evidence for the identification of the parameters of interest and examines the identification of the attention probability $\alpha$ from the addition of the confirmation probabilities. 

This identification argument also supports the estimation of the outside offer distribution parameters and cancellation penalties from the confirmation and cancellation behavior alone. Conditional on these parameters, the identification of carrier's cost distribution and the parameters governing their first-stage choice probabilities are straightforward.

The cost distribution is identified from a combination of the costs obtained from the bid inversion and the first-stage choice probabilities. Carriers with very low costs for a shipment will generally find it more attractive to use the Accept-Now option, as the posted price will be closer to their optimal bid, while carriers with very high costs for a shipment will generally prefer to ignore it, as they stand very low chances of winning. Thus, the first-stage choice probabilities ``fill in'' the truncated regions of the cost distribution. The use of the joint likelihood helps to separately identify the variance of the idiosyncratic cost term $\epsilon_{ij}$ from the variance of the mean cost term $\bar{c}_i$.

Finally, the parameters governing the first-stage choice probabilities are identified from the first-stage choices alone and effectively act as a residual. The variance of the idiosyncratic cost term $\epsilon_{ij}$ is identified from the elasticity of choices to the value functions associated with each choice, while the hassle costs help to fit the overall average choice probabilities.

With these identification arguments in place, I now turn to the estimation results.

\subsection{Carrier Model Estimates}

\begin{figure}[htbp]
    \begin{center}
    \caption{Features of the carrier search process}
    \label{fig:carrier_search}
    \subfloat[Arrivals and attention]{\includegraphics[width=0.48\textwidth]{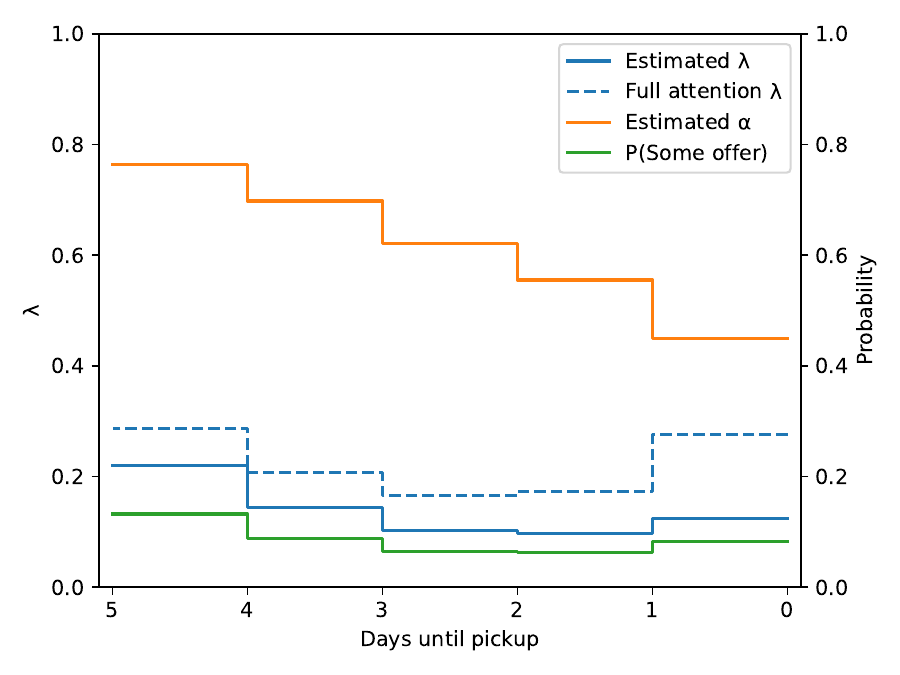}\label{fig:arrivals_attention}}
    \hfill
    \subfloat[Effective distribution of max outside offer]{\includegraphics[width=0.48\textwidth]{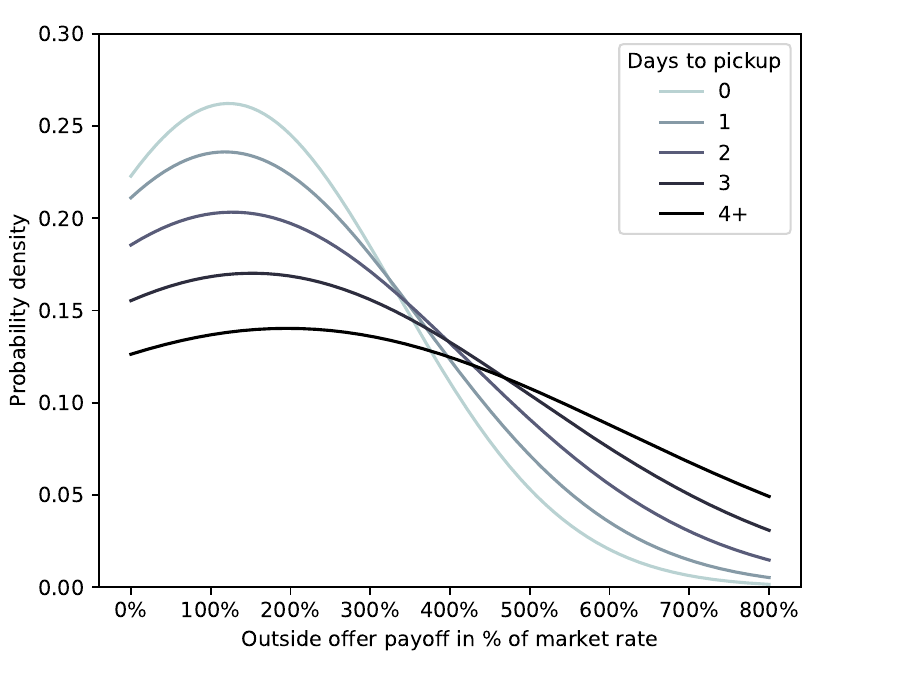}\label{fig:outside_dist_amalgam}}
    \\
    \subfloat[Cancellation penalty]{\includegraphics[width=0.48\textwidth]{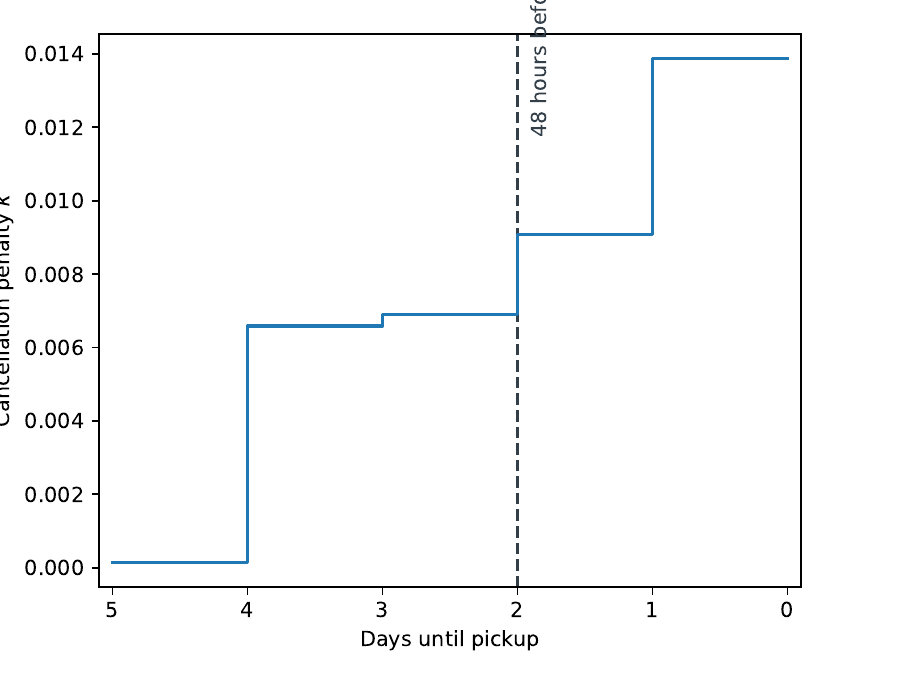}\label{fig:kappa}}
    \hfill
    \subfloat[Carrier cost distributions]{\includegraphics[width=0.48\textwidth]{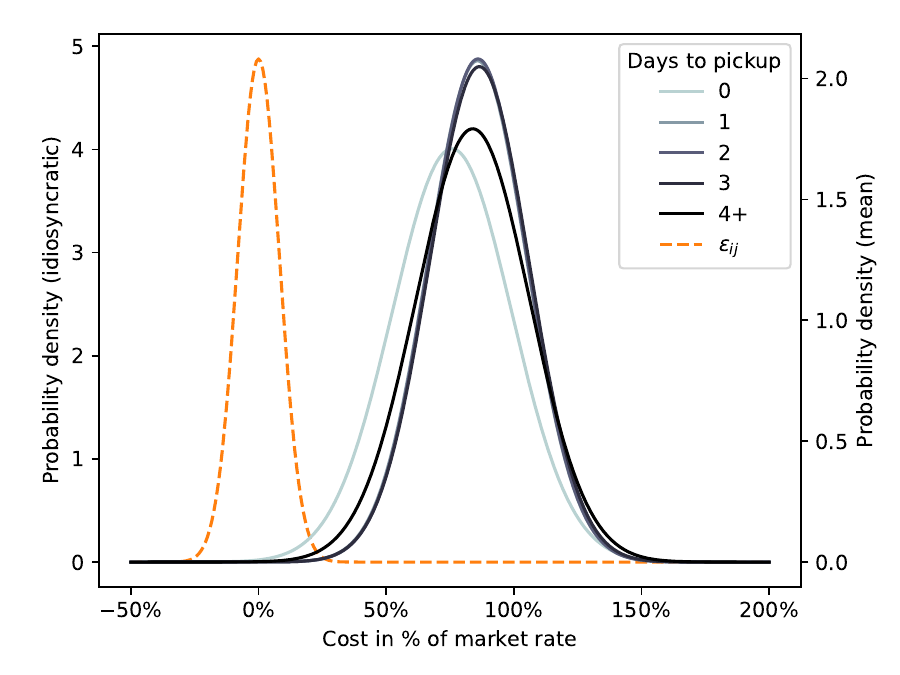}\label{fig:carrier_cost_dist}}
    \end{center}
    {\footnotesize \textsc{Note:} In (a), P(Some offer) is the probability of an offer arriving and being non-negative. In (b), the effective distribution is integrated over the number of offer arrivals from the Poisson process, and is conditioned on the offer being non-negative. In (c), the cancellation penalty is expressed as a fraction of the market rate. In (d), the day-specific distributions (right axis) represent the distribution of mean carrier costs $\bar{c}_i$; the dashed line (left axis) shows the idiosyncratic component $\epsilon_{ij}$ centered at zero.}
\end{figure}

Recall that the carrier model is estimated in two alternating blocks, with the first block estimating the parameters of the outside offer process and the penalty schedule, and the second block estimating the carrier cost distribution and the parameters governing the first-stage choice probabilities. Appendix Table \ref{tab:raw_params_block1} reports the raw structural parameter estimates from the first block and their asymptotic standard errors. Appendix Table \ref{tab:raw_params_block2} reports the raw structural parameter estimates from the second block and their asymptotic standard errors. In the following, I present the main parameters of interest in the carrier model in \autoref{fig:carrier_search} and discuss their economic implications.

Formally, the stochastic process of outside offers consists of four elements: the arrival rates of offers $\lambda_d$, the mean and standard deviation of the distribution of outside offers $\mu_d^\pi, \sigma_d^\pi$, and the attention probability $\alpha$. However, what is effectively identified under the arguments in the previous section is the amalgamated distribution of the maximal outside offer received by a carrier within a day. 

Panel (a) of Figure \ref{fig:carrier_search} presents the offer arrival rates and attention probabilities, along with the \emph{overall probability of receiving a non-negative offer}. The estimates imply that arrival rates are generally declining over time, particularly once inattention is accounted for, except for an uptick in the last 24 hours. Panel (b) presents the effective distribution of the maximal outside offer received by a carrier, which shows that as the pickup time nears, the distribution of outside offers tightens up, potentially reflecting selection on the part of other carriers.

The cancellation penalties are presented in Panel (c) of Figure \ref{fig:carrier_search}. While the estimated penalties do grow higher as the pickup time approaches, they already reach half the maximum level even four days ahead of pickup, which contradicts the platform's stated policy that penalties only apply in the last 48 hours. Regardless, the estimated penalties are fairly modest, reaching a maximum of \eoMaxKappaPct of the market rate, which is consistent with the notion that relational penalties are weak in this spot market.

I now turn to the estimates from the second step of the estimation, presented in Panel (d) of Figure \ref{fig:carrier_search}, which include the parameters of the carrier cost distribution and those governing the first-stage choices. The mean carrier cost is steady over nearly all days, but is substantially lower on the very last day, which reflects that carriers, like shippers, have a greater urgency to match as the pickup date nears, though the greater variance on the last day indicates that there is substantial heterogeneity in this respect. The variance of the carrier-shipment idiosyncratic cost shocks is substantially smaller than the variance across carriers, with an average variance ratio of \eoAvgVarianceRatio.

\begin{table}[htbp]
    \begin{center}
        \caption{Model fit: reneging behavior}
        \label{tab:model_fit}
        \begin{tabular}{l|c|c}
    \toprule
    Type & Data & Model \\
    \midrule
    Confirmation & 51.94\% & 52.10\% \\
    Cancellations (bids) & 10.19\% & 11.22\% \\
    \bottomrule
\end{tabular}

    \end{center}
    {\footnotesize \textsc{Note:} Confirmation and cancellation rates computed as averages across all accepted bids in the sample.}
\end{table}

\paragraph{Model Fit}

\begin{figure}[htbp]
    \caption{Confirmation and cancellation probabilities in model vs. data}
    \label{fig:confirmation_cancellation_fit}
    \centering
    \includegraphics[width=0.6\textwidth]{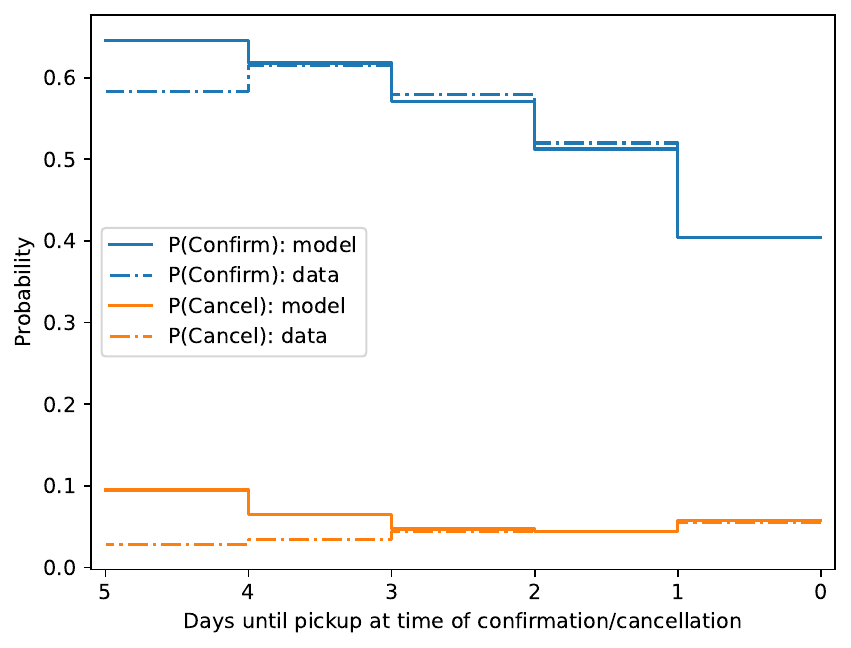}
\end{figure}

Next, I evaluate the fit of the model to the data. Table \ref{tab:model_fit} shows that the overall fit of the model in terms of confirmation and cancellation rates is high, with the model closely matching the data. Figure \ref{fig:confirmation_cancellation_fit} is presented to check the fit along the temporal dimension of the model. The fit of cancellation rates diverges somewhat in the two earliest periods, as it tracks the pattern of outside offer arrivals. The fit of confirmation rates also diverges in the earliest period. Note that since there are few accepted and matched bids early on, there is relatively less data in these earlier periods. 

Figure \ref{fig:confirmation_cancellation_fit} illustrates this performance over time. While the predictions are generally accurate, cancellation rates diverge slightly in the first two periods as they track the pattern of outside offer arrivals. Confirmation rates exhibit a similar early divergence. Both divergences can be further explained by the paucity of data in the early stages, when few shipments have undergone auction clearing rounds and been matched.

\subsection{Platform Parameters}
\label{sec:platform_params_est}

The additional parameters to be estimated are the arrival rates of carriers, shipments, and views (a proxy for the search process on the platform), the distribution of shipment values to the platform, and conditional reserve price and Accept-Now price models. These parameters are all estimated directly in reduced form from their empirical analogues as they all involve observable events. More details on the estimation method, the parameter estimates and their standard errors can be found in Appendix \ref{sec:platform_params_est}. In addition to the estimated parameters, the first-stage choice parameters of carriers are calibrated using the platform model to match the empirical first-stage choice probabilities. This calibration will eventually be replaced by the full likelihood estimation of the carrier model described above.

\begin{figure}[htbp]
    \begin{center}
    \caption{Features of the platform matching process}
    \label{fig:platform_features}
    \subfloat[Arrival rates of events]{\includegraphics[width=0.48\textwidth]{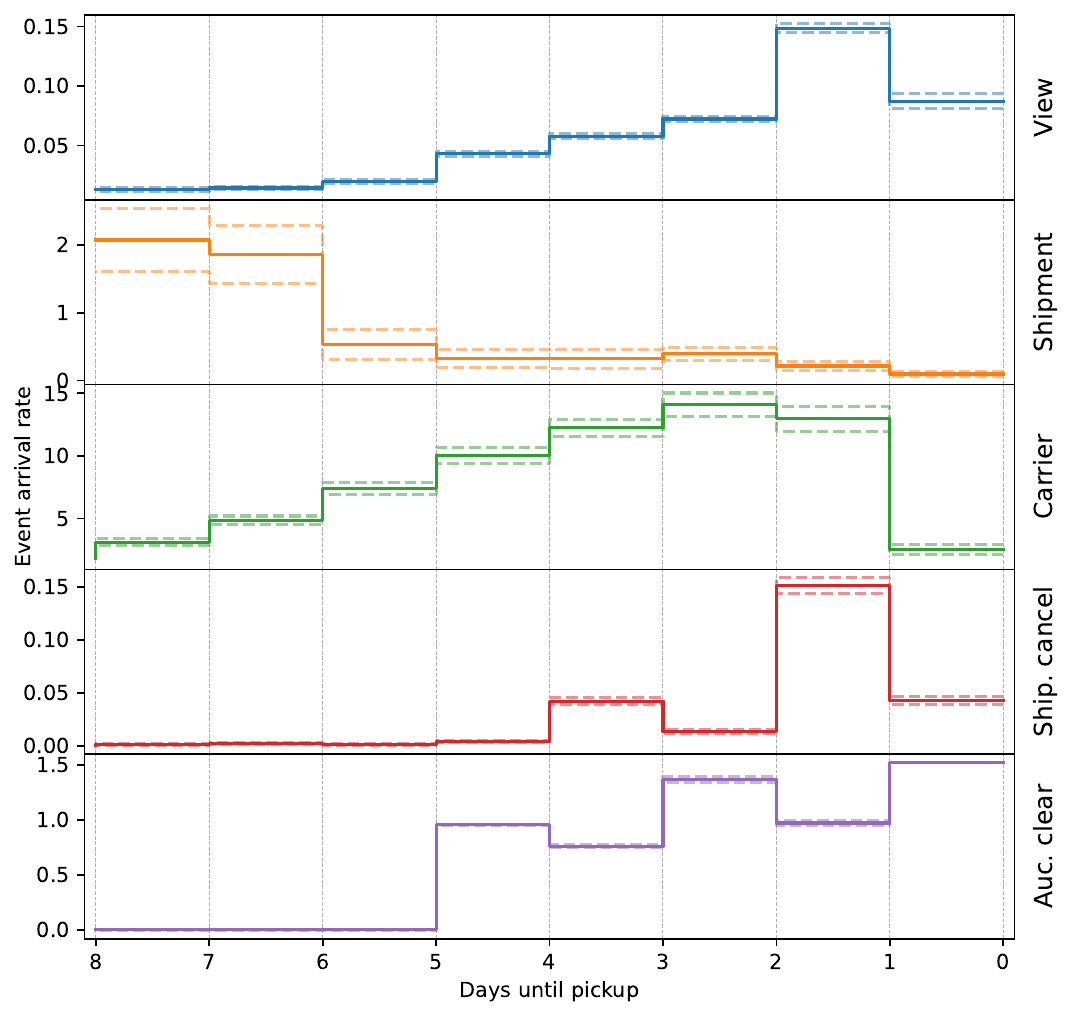}\label{fig:lambda_platform}}
    \hfill
    \subfloat[Reserve and Accept-Now prices]{\includegraphics[width=0.48\textwidth]{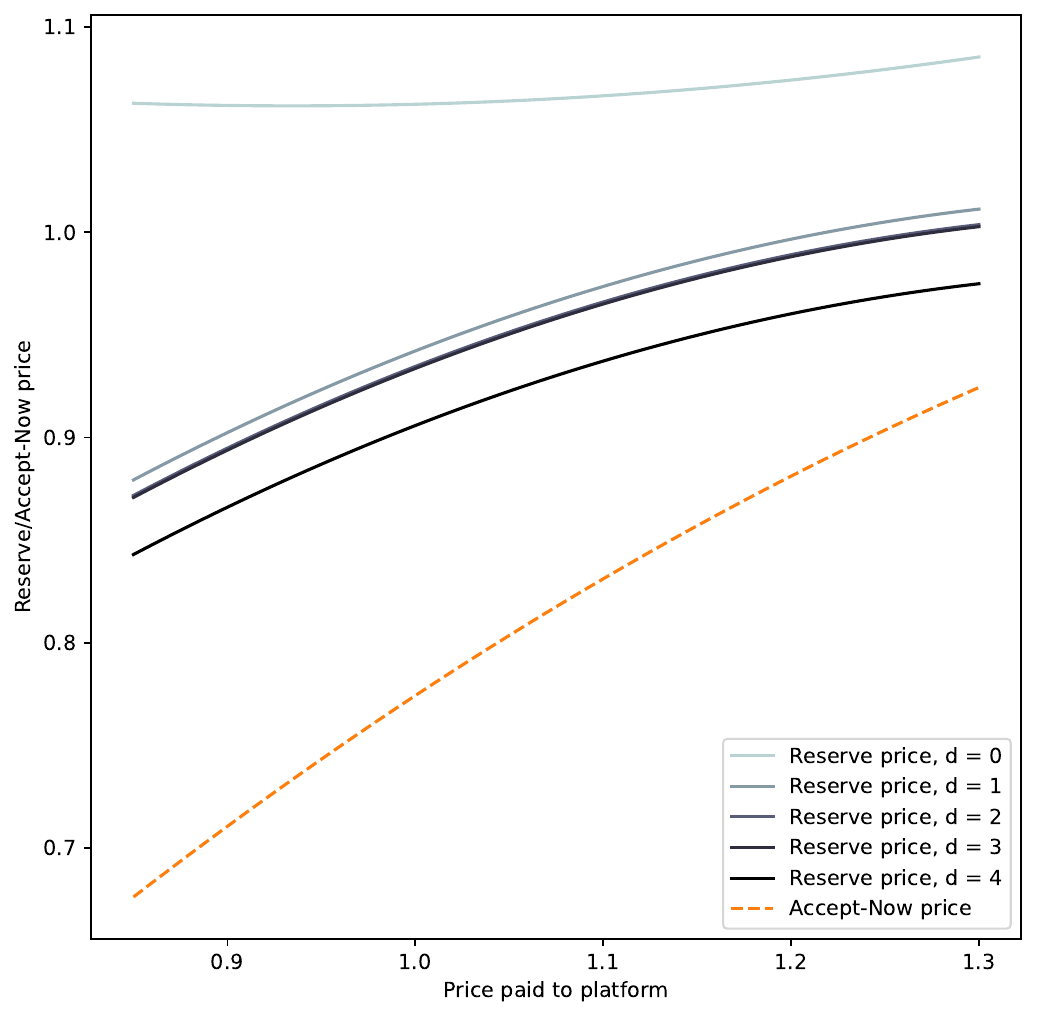}\label{fig:reserve_accept_now_prices}}
    \end{center}
    {\footnotesize \textsc{Note:} In (a), estimated Poisson arrival rates of carriers, shipments, views, shipper cancellations, and auction clearing rounds on the platform; dashed lines represent the 95\% confidence intervals. In (b), reserve prices for $d\geq 1$ and Accept-Now prices estimated via linear regression on second-order polynomials of the shipment price paid to the platform; reserve price regression is pooled, with a separate intercept for each $d$; reserve price on $d=0$ uses data from the fallback auction model with manual intervention, using inequality conditions on highest accepted bids and lowest unaccepted bids for each shipment.}
\end{figure}

Figure \ref{fig:platform_features} shows the estimated parameters of the platform matching process. The left panel shows the estimated arrival rates of various events on the platform. The rate of view events is defined at the carrier-shipment level; for example, between 48 and 24 hours before pickup, each carrier-shipment pair that has not previously been viewed has a \eoViewProb chance of being viewed by the end of the period.\footnote{This comes from the properties of the Poisson process and the normalization that the arrival rate of days $\eta=1$. Thus, the probability of an event with arrival rate $\lambda$ arriving within a day is $\tfrac{\lambda}{1 + \lambda}$.}

The shipment and carrier arrival rates reveal an important asymmetry: while shipments are mostly posted to the platform with at least a week of lead time,
the arrival rate of carriers is at its highest in the last few days, before dropping substantially in the last 24 hours before pickup.\footnote{Most shipments are posted to the platform before the arrivals of the first carriers. To account for this, I draw an initial set of shipments from a Poisson distribution with rate $\lambda=\eoInitShipmentRate$, equal to the mean number of shipments posted prior to the observation window.} This asymmetry has implications for the design of the penalty schedule, as it gives the platform many options in the last few days before pickup.

Shipment cancellations are also estimated directly from the data and are part of the counterfactual simulations. They are, however, abstracted away from the carrier's decision model, because the existing industry norm is to pay pecuniary cancellation penalties to carriers when cancelling shipments close to the pickup time.\footnote{Referred to as ``TONU'' (Truck Order Not Used) in the industry. These are typically applied for cancellations within 24 hours of pickup, potentially explaining the bunching in shipper cancellations in the 48-24 hour window prior to pickup.}

Finally the auction clearing rounds are also modelled stochastically. In general, the platform would schedule at least one clearing round per auction per day, and sometimes more, beginning when there were fewer than 5 days remaining until pickup. In the last 24 hours, however, timed auction format is replaced by a continuous auction format with escalating prices and frequent manual broker intervention. The clearing rate for this phase is approximated by taking the count of unique hours in which bids were accepted on the last day.

The other feature of auctions is their reserve prices and Accept-Now prices, which I model as second-order polynomials of the rate paid \emph{to the platform} from the shipper. As before, the estimation details of these parameters and their standard errors can be found in Appendix \ref{sec:platform_params_est}. These are plotted in the right panel of Figure \ref{fig:platform_features}.

All pricing functions are increasing in the rate paid to the platform, which is consistent with the platform's profit-maximizing behavior, while the Accept-Now price is substantially lower than the reserve prices on any day. In the last 24 hours, the true reserve price is not observed, due to the change in formats. Instead, I use the highest accepted bid and the lowest unaccepted bid as bounds to estimate the reserve price. The much higher curve may imply that the platform is more desperate to match very high-paying shipments in the last 24 hours. 

Finally, the distribution of shipment values to the platform is fitted to a log-normal distribution resulting in a mean shipment value of \eoMeanShipmentValue. See Appendix \ref{sec:platform_params_est} for additional details. The cost of not matching a shipment is calibrated at 10\% of the shipment value, which is the same heuristic used by the platform in its development of the newer auction format introduced in 2023.

The estimated structural model is then used to simulate counterfactuals, discussed in the following section.

\section{Counterfactuals}
\label{sec:counterfactuals}

The counterfactual analysis explores the effect of alternative cancellation policy designs on platform profits and overall welfare. Several features of the design are of interest. One, what is the effect of changing the overall level of the penalties? Two, how does the type---reputational or monetary--of the penalty affect outcomes and incentives? 

To evaluate the optimal level and type of the penalties, I consider counterfactual policies that are linear transformations of the status quo penalty schedule. That is, given the estimated penalty schedule $\hat{\kappa}_d$ for day $d$, I consider a counterfactual penalty schedule $\kappa_d = s \hat{\kappa}_d$, where $s$ is a scaling factor. This allows me to explore the effect of changing the overall level of penalties, while keeping the shape of the penalty schedule constant. I consider other variations in the shape of the penalty below.

Simulating outcomes under each counterfactual policy involves solving for the fixed point of win probabilities and carrier policies, which is why I restrict attention to a linear grid.\footnote{With five possible days of cancellation and the restriction that penalties are non-decreasing, exploring a grid of $K$ values for each day would require ${5 + K -1 \choose K-1}$ simulations. Simulating a single counterfactual policy requires approximately 15 hours of computation time on a single thread.} The grid is composed of two sub-grids that are merged together: a fine grid of 96 points covering roughly one tenth of the penalty range, and a coarser grid of 96 points spanning the full range up to approximately the market rate for a single shipment (a penalty of 100\%). This nested structure is meant to capture potentially fine-grained patterns at low penalty values. To reduce simulation variance across the grid, all counterfactual scenarios share the same underlying sequence of random draws, so that differences in outcomes reflect only the change in penalty policy.

The main outcomes of interest are average platform profits, carrier welfare, and total welfare per market (defined as a lane and pickup date combination). Because a large portion of carrier welfare stems from offers \emph{outside} the platform, I subtract the carrier's welfare under a no-platform counterfactual to obtain welfare numbers in platform value-added terms. As in the stylized model in Section \ref{sec:toy_model}, I also distinguish profits and overall welfare under reputational penalties (as employed in the status quo) and monetary penalties, which involve a transfer from cancelling carriers to the platform, providing the latter with an additional source of revenue. In the following sections, I discuss the optimal reputational penalty, and then compare the trade-offs between monetary and reputational penalties.

\subsection{Optimal Reputational Penalty}

\begin{figure}[htbp]
    \begin{center}
    \caption{Visualization of counterfactual penalty policies.}
    \label{fig:full_cf}
    \includegraphics[width=0.7\textwidth]{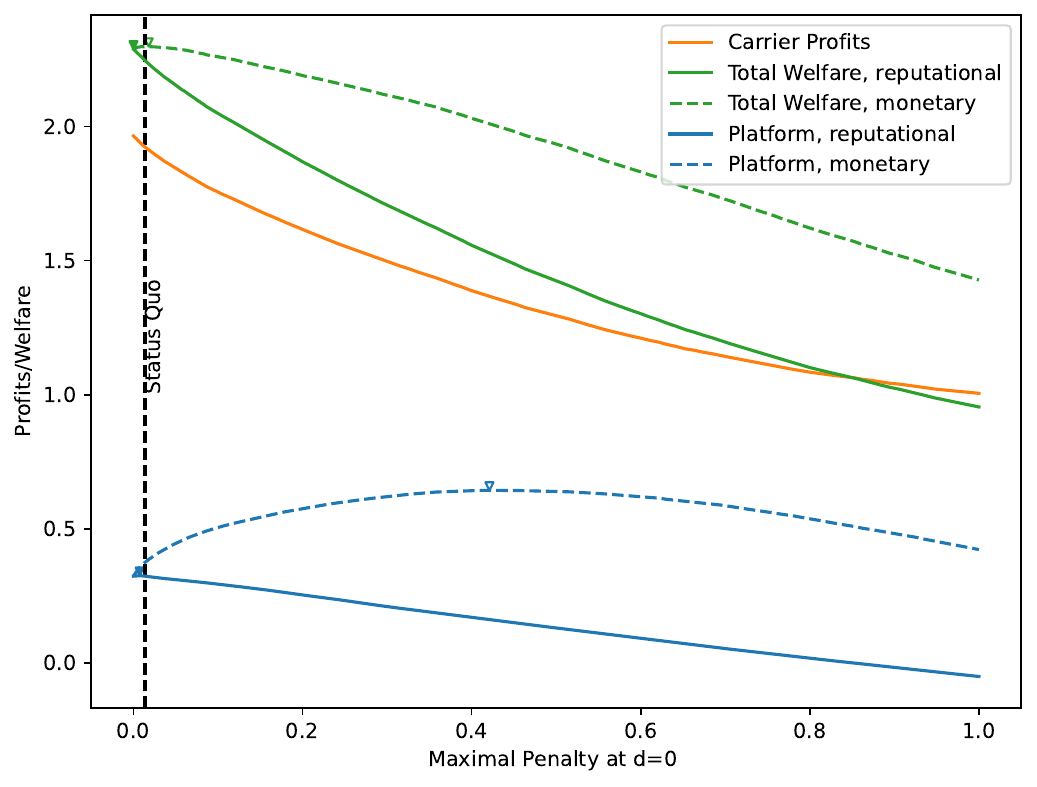}
    \end{center}
    {\footnotesize \textsc{Note}: All quantities scaled by average lane market price of a shipment. Carrier welfare is relative to platform not existing. Maxima of platform profits and total welfare are indicated by triangles in the corresponding color.}
\end{figure}

Figure \ref{fig:full_cf} presents the evolution of profits and welfare over the range of counterfactual policies. Is is immediately noticeable that the optimal reputational penalties for both social welfare and platform profits are at zero, so that fully flexible agreements are optimal. This is consistent with the stylized model under high variance of the outside offers, depicted in Figure \ref{fig:toy_cf_plot_high_sigma}. The status quo reputational penalties, depicted in Figure \ref{fig:kappa}, are very close to this optimum.

To further understand why the profits are decreasing in the level of the reputational penalty, I decompose the change in profits into three different margins: a change in the number of successfully completed shipments (the extensive margin), the firm's gross profit margin on each shipment (the intensive margin), and the indirect losses suffered by the firm as a consequence of shipments that failed to be matched in time. 

\begin{figure}[htbp]
    \begin{center}
    \caption{Decomposition of profits under reputational penalties.}
    \label{fig:profit_decomp}

    \begin{subfigure}{0.48\textwidth}
        \centering
        \includegraphics[width=\textwidth]{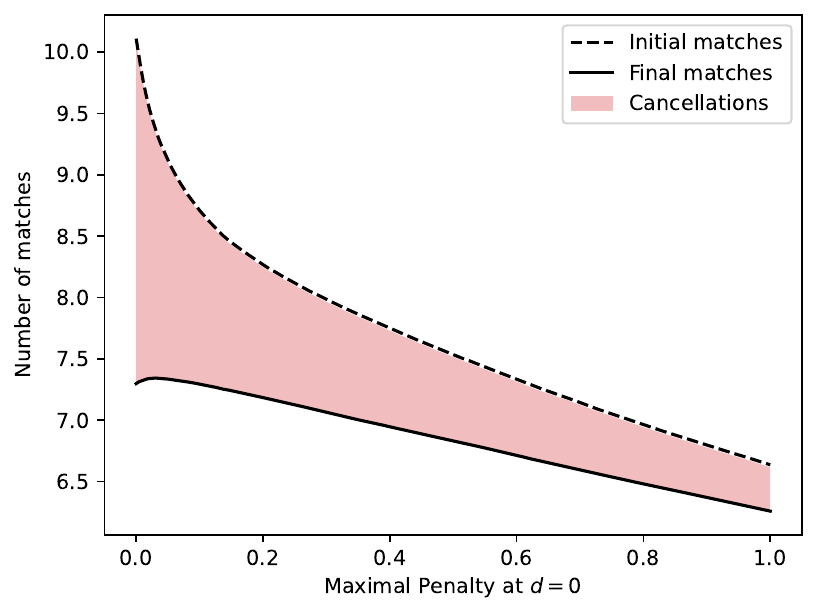}
        \caption{Extensive margin}
        \label{fig:sub1}
    \end{subfigure}
    \hfill
    \begin{subfigure}{0.48\textwidth}
        \centering
        \includegraphics[width=\textwidth]{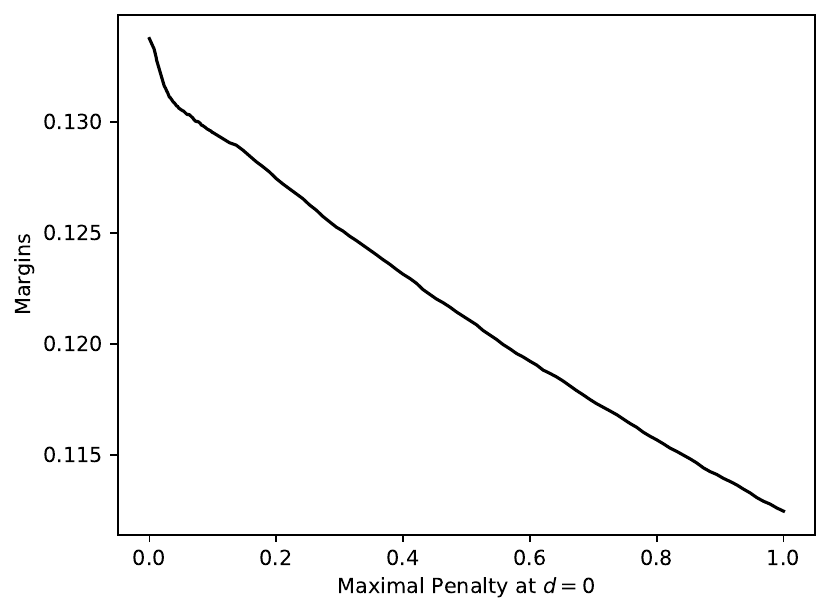}
        \caption{Intensive margin}
        \label{fig:sub2}
    \end{subfigure}
    
    \vspace{0.5cm} %

    \begin{subfigure}{0.48\textwidth}
        \centering
        \includegraphics[width=\textwidth]{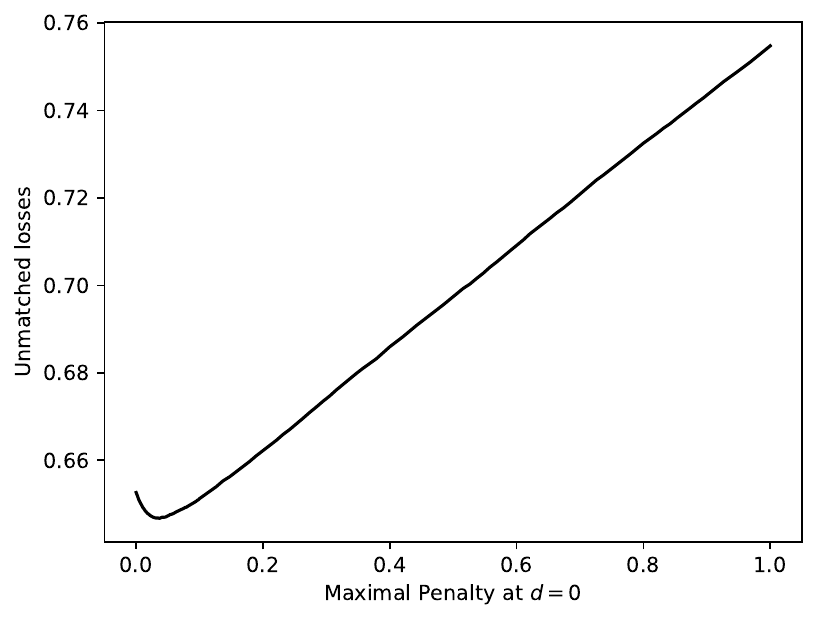}
        \caption{Indirect losses}
        \label{fig:sub3}
    \end{subfigure}
    \hfill
    \begin{subfigure}{0.48\textwidth}
        \centering
        \includegraphics[width=\textwidth]{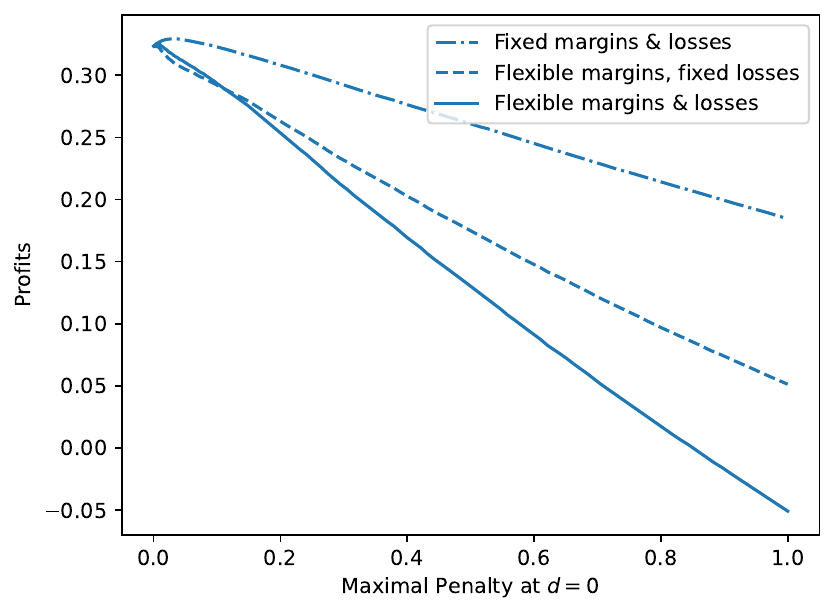}
        \caption{Cumulative effects on profits}
        \label{fig:sub4}
    \end{subfigure}

    \end{center}
    {\footnotesize \textsc{Note}: Profits, losses, and penalties are scaled by average lane market price of a shipment. Panel \subref{fig:sub4} presents the cumulative effect of the three margins on platform profits.}
\end{figure}

The results of the decomposition exercise are presented in Figure \ref{fig:profit_decomp} with Panel \subref{fig:sub4} summarizing the cumulative effect of the three margins on platform profits. All three margins are generally monotonically declining as the penalty level increases. Breaking down the extensive margin further, we see that although both initial matches and cancellations fall with stricter penalties, the loss of initial matches dominates at all but the most lenient levels. The intensive margin shrinks as well, since higher penalties trigger increased bids. Finally, indirect losses grow due to the diminished match volume. 

The decomposition exercise highlights the importance of accounting for the strategic response of bids to changes in the cancellation penalty, which drives both the higher transaction prices and lower matching rates.

\subsection{Trade-offs between Monetary and Reputational Penalties}

Figure \ref{fig:full_cf} also presents the results of the counterfactual analysis under monetary penalties. For a given intermediate penalty schedule, monetary penalties are more efficient than reputaional penalties because they do not "burn money." Nevertheless, the counterfactual analysis shows that, under penalty schedules of the same shape as in the status-quo, the welfare-maximizing monetary penalties are still at zero. In contrast, the profit-maximizing monetary penalties are much higher. The increase in profit is accompanied by a large decline in both carrier and total welfare, suggesting that these penalties mainly transfer rents from carriers to the platform.

\begin{figure}
    \begin{center}
    \caption{Monetary penalties with weighted objective and transaction costs.}
    \label{fig:monetary_caveats}

    \begin{subfigure}{0.48\textwidth}
        \centering
        \includegraphics[width=\textwidth]{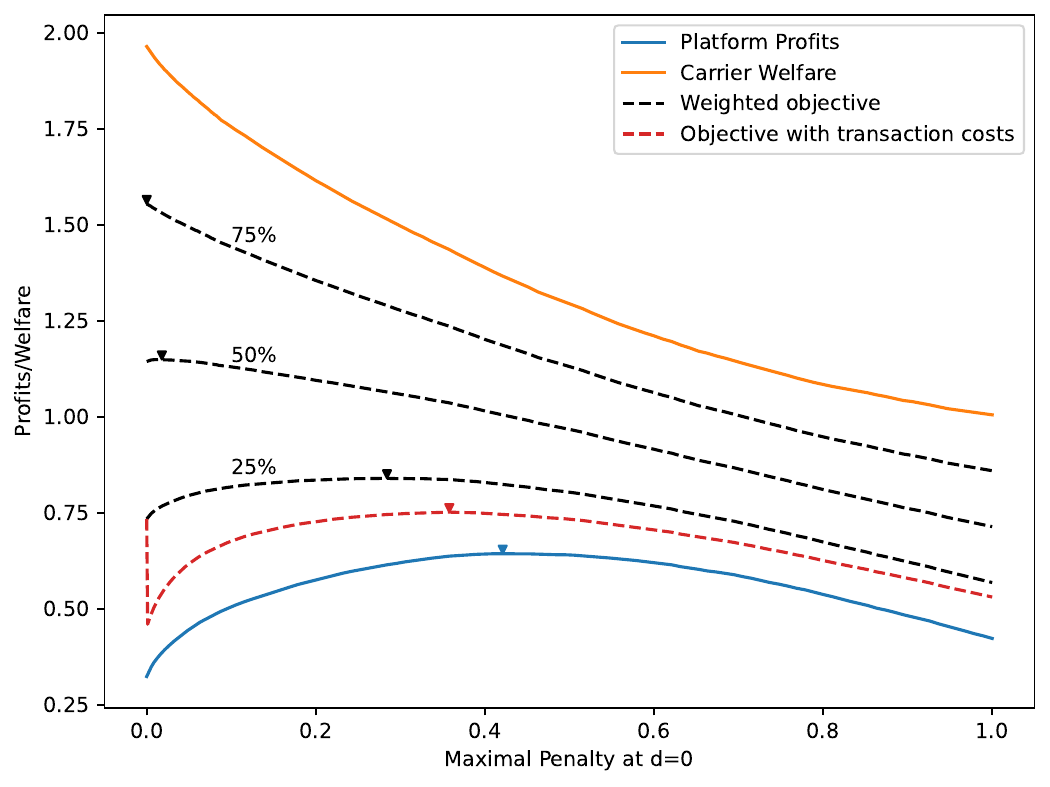}
        \caption{Illustration}
        \label{fig:monetary_caveats1}
    \end{subfigure}
    \hfill
    \begin{subfigure}{0.48\textwidth}
        \centering
        \includegraphics[width=\textwidth]{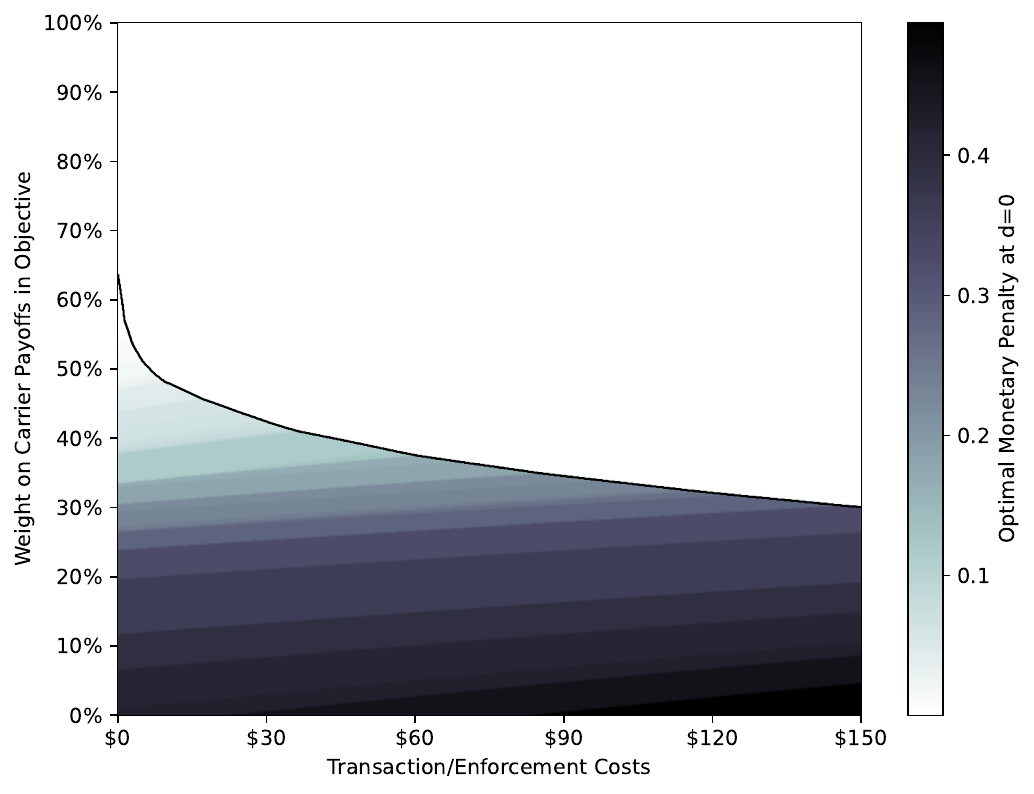}
        \caption{Grid: weights and transaction costs}
        \label{fig:monetary_caveats2}
    \end{subfigure}
    \end{center}

    {\footnotesize \textsc{Note}: Profits, losses, and penalties are scaled by average lane market price of a shipment. Panel \subref{fig:monetary_caveats1} illustrates the effect of a weighted objective and transaction costs on platform profits and total welfare, with objective maximizing penalties highlighted with corresponding triangles. Panel \subref{fig:monetary_caveats2} presents the grid of weights and transaction costs considered, with the corresponding optimal penalty level shown as a color (darker=higher).}
\end{figure}

An immediate question that arises is why the firm does not use monetary penalties in practice. To rationalize these, I can explore two caveats to the baseline results. Firstly, the model does not account for carrier's decisions to join or leave the platform. As there may be some set-up costs and time investment to continuously monitor the platform, the number of active carriers in the market may be sensitive to the average welfare of carriers. Secondly, I have so far assumed that charging monetary fees for cancellations is costless. In practice, there may be significant transaction fees associated with collecting these fees, particularly due to the enforcement costs associated with pursuing carriers often based in other states. Below I provide some quantification of these concerns. 

To approximate the effect of long-term carrier acquisition and retention on the platform, I treat the platform as maximizing a weighted sum of its own short-term profits and the welfare of carriers. Different weighted objectives are illustrated in Figure \ref{fig:monetary_caveats1}. As the weight on carrier welfare increases, the optimal penalty level decreases. Secondly, I can suppose that the platform bears some transaction cost for each monetary cancellation fee that it collects. As illustrated by the dashed red line in Figure \ref{fig:monetary_caveats1}, this transaction cost leads to a discontinuity in the objective function when moving from zero penalties to positive penalties. For a large enough transaction cost, the optimal penalty level is zero, even for a lower weight on carrier welfare. 

The joint effect of the weight on carrier welfare and transaction costs is presented in Figure \ref{fig:monetary_caveats2}. A weight of approximately \cfZeroPenaltyWeight on carrier welfare is sufficient to rationalize why the firm would not use monetary penalties at all. Transaction costs on their own are not sufficient to explain the absence of monetary penalties---even for a high transaction cost of \$150, some weight on carrier welfare is still necessary to rationalize the avoidance of monetary penalties.\footnote{While \$150 is large for a transaction cost, consider that monetary fees for shipper cancellations (the aforementioned TONU fee) range from \$150 to \$300. The absence of such monetary fees for carrier cancellations in the industry would a priori suggest transaction or enforcement costs of a similar order of magnitude.} Furthermore, for a given weight on carrier welfare below the threshold, the optimal penalty level is increasing in the transaction cost, as the firm is able to pass on part of the transaction cost to the carriers. 

Taken together, the results suggest that the firm avoids monetary penalties due to strong concerns about the long-term growth of the platform, which are not captured by the short-term profit-maximizing model. The choice of reputational or monetary penalties will thus depend on the degree of lock-in that the platform has over carriers, which is likely to be low in the highly fragmented trucking industry.

\subsection{Timing of the Penalty}

Although the primary focus of the counterfactual analysis has been on the level and type of the penalties, the rich dynamic model also permits an analysis of different timings of the penalties---that is, not only how \emph{much} to charge, but also \emph{when} to charge a given amount. This can also be thought of as examining different shapes of the penalty schedule. To illustrate this, I compare the stair-shaped increasing penalty schedule of the status quo to a flat penalty schedule, where the penalty is the same for all days of cancellation.

\begin{figure}
    \caption{Comparison of flat and increasing penalties.}
    \label{fig:flat_increasing_comparison}
    \begin{center}
    \begin{subfigure}{0.45\textwidth}
        \centering
        \includegraphics[width=\textwidth]{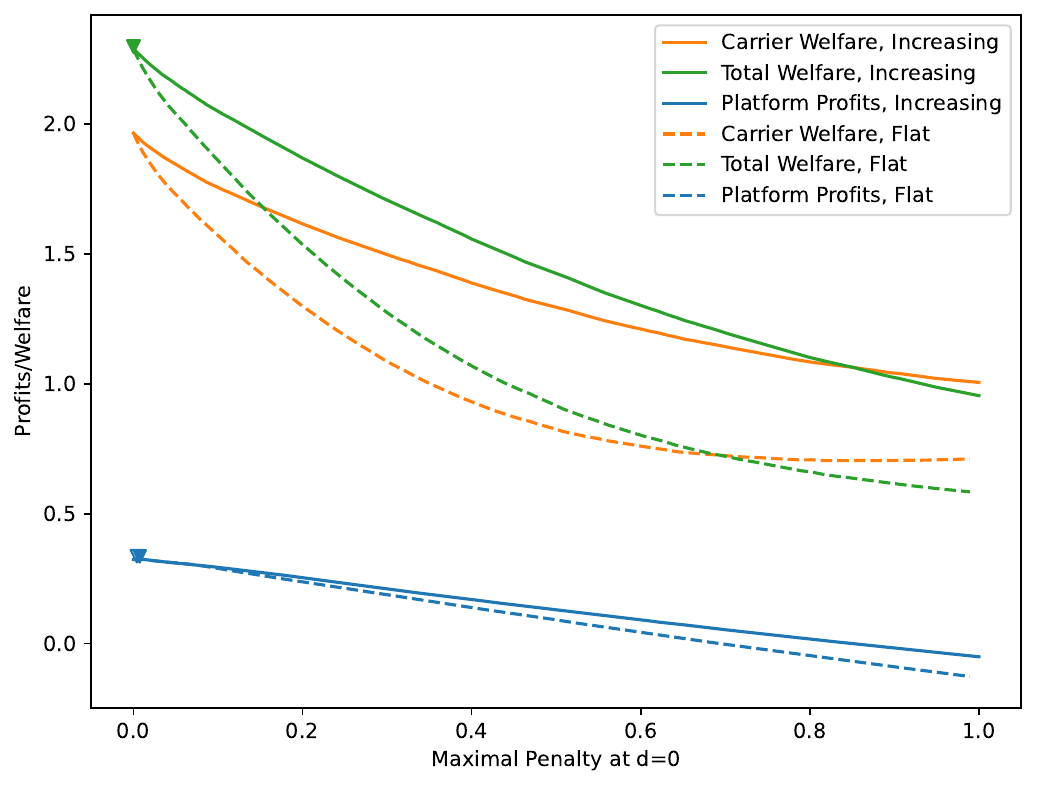}
        \caption{Reputational penalties}
        \label{fig:flat_increasing_comparison_reputational}
    \end{subfigure}
    \hfill
    \begin{subfigure}{0.45\textwidth}
        \centering
        \includegraphics[width=\textwidth]{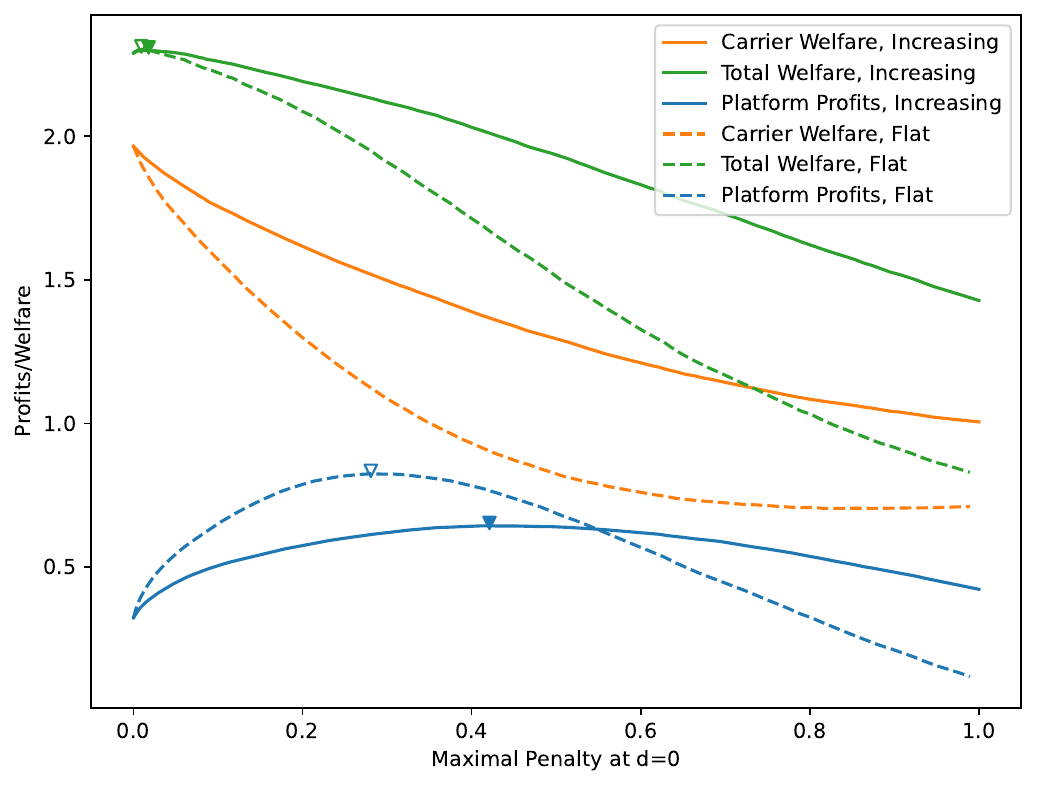}
        \caption{Monetary penalties}
        \label{fig:flat_increasing_comparison_monetary}
    \end{subfigure}
\end{center}
\end{figure}

Figure \ref{fig:flat_increasing_comparison} presents the results of the comparison between flat and increasing penalties. For reputational penalties, the shape of the profit function is slightly steeper under flat penalties, but welfare decreases much faster as the penalty level increases. Intuitively, carriers are exposed to more outside offers in the earlier days before pickup (see Figure \ref{fig:platform_features}), so being subject to higher penalties on these days is very costly in terms of opportunity costs. Monetary penalties, on the other hand, are able to extract more rents from carriers on these early days, leading to a steeper increase in platform profits. This analysis demonstrates that profit-seeking firms may reduce overall welfare not only by setting penalties too \emph{high}, but also by setting them too \emph{early}.

\subsection{Full commitment and late-clearing policies}

Contracts with limited commitments are pervasive in the trucking industry and extend beyond cancellation policies. For example, long-term contracts between shippers and carriers or brokers typically only specify a price and a general service level, but do not contain strict guarantees on the number of shipments to be executed. Prior research on this contracting process has focused on the negative aspects of these weak commitments. \textcite{caplice2006combinatorial} highlight the difficulty of awarding long-term contracts through combinatorial auctions: potential complementarities between different lanes are not guaranteed to be exercised in practice as both carriers and shippers are free to partner with other parties at any time. \textcite{harris2022long} show how price fluctuations in the spot market increase carriers' propensity to reject tenders under long-term contracts, which is detrimental to the concerned shipper. In contrast, my model explicitly highlights the \emph{value} of contractual flexibility in enabling market participants to act on new opportunities they encounter through the search process.

To fully illustrate the value of flexibility, consider a full-commitment policy with infinite cancellation penalties. This policy is the extreme opposite of the profit- and welfare-maximizing reputational penalty, where carriers are free to cancel at any time. Table \ref{tab:cf_results} summarizes the results of multiple different counterfactual policies, including the full-commitment policy with infinite penalties. The full-commitment policy not only drastically reduces carrier welfare but also the platform's profits via a steep dropoff in the match rate, highlighting the firm incentives to avoid binding commitments even when individual instances of reneging are costly.

\begin{table}[htbp]
    \begin{center}
        \caption{Comparison of Counterfactual Policies}
        \resizebox{\textwidth}{!}{
            \begin{tabular}{l|cccccc}
\toprule
{} &    Status Quo & Max Welfare & Max Profit & Zero Penalty & Infinite Penalty & Late Clearing \\
Metric            &               &             &            &              &                  &               \\
\midrule
Penalty Type      &  Reputational &   Pecuniary &  Pecuniary &            - &                - &             - \\
Total Revenues    &          7.46 &        7.48 &       7.20 &         7.42 &             3.96 &          2.00 \\
                  &               &      (0.3\%) &    (-3.5\%) &      (-0.6\%) &         (-47.0\%) &      (-73.2\%) \\
Total Welfare     &          2.25 &        2.30 &       1.95 &         2.29 &             0.19 &         -0.53 \\
                  &               &      (2.5\%) &   (-13.3\%) &       (1.8\%) &         (-91.7\%) &     (-123.5\%) \\
Platform Profit   &          0.32 &        0.40 &       0.82 &         0.32 &            -0.68 &         -0.71 \\
                  &               &     (23.7\%) &   (155.2\%) &       (0.1\%) &        (-310.5\%) &     (-318.2\%) \\
Carrier Welfare   &          1.92 &        1.90 &       1.12 &         1.97 &             0.87 &          0.18 \\
                  &               &     (-1.1\%) &   (-41.6\%) &       (2.1\%) &         (-55.0\%) &      (-90.8\%) \\
Matches           &          7.33 &        7.35 &       7.05 &         7.30 &             3.87 &          1.88 \\
                  &               &      (0.2\%) &    (-3.8\%) &      (-0.5\%) &         (-47.3\%) &      (-74.4\%) \\
Match Rate        &          0.54 &        0.54 &       0.52 &         0.54 &             0.30 &          0.14 \\
                  &               &      (0.2\%) &    (-3.5\%) &      (-0.4\%) &         (-45.7\%) &      (-74.2\%) \\
Match Welfare     &          0.40 &        0.40 &       0.42 &         0.40 &             0.49 &          0.46 \\
                  &               &     (-0.1\%) &     (2.9\%) &      (-0.0\%) &          (20.5\%) &       (14.9\%) \\
Transaction Price &          0.88 &        0.88 &       0.90 &         0.88 &             0.94 &          0.80 \\
                  &               &      (0.1\%) &     (1.2\%) &      (-0.2\%) &           (6.0\%) &       (-9.0\%) \\
Average Bid       &          1.07 &        1.07 &       1.10 &         1.07 &             1.24 &          2.14 \\
                  &               &     (-0.0\%) &     (2.1\%) &      (-0.1\%) &          (15.4\%) &       (99.2\%) \\
\bottomrule
\end{tabular}

            }
            \label{tab:cf_results}
        \end{center}
        {\footnotesize \textsc{Note}: The table presents counterfactual outcomes under different penalty schedules. Change relative to status quo in parentheses. Absolute numbers are multiples of average lane market price of a shipment. Infinite penalty is approximated by a penalty of $\kappa = 10^6$. Match welfare is average over final matches on platform.}
\end{table}

Finally, recall that the driving force behind the firm's incentives are the opportunity costs of commitment that are passed through into carriers' bids. Presumably, an alternative means of reducing these opportunity costs would be to delay the matching time, so that carriers have more information about their outside options at the time of bidding. We know that when the \emph{entire market} is centralized at a later matching time, timing frictions become less significant \parencite{roth1994jumping}. Instead, the \emph{unilateral} late clearing scenario, included in Table \ref{tab:cf_results}, exhibits severe negative effects on both total welfare and platform profits, with reductions of \cfLateClearWelfarePct and \cfLateClearProfitPct respectively. The massive reduction in platform profits can be attributed to the large drop in the number of matches, resulting in large losses for the platform due to penalties for unmatched shipments. This loss occurs despite an improvement in the per-shipment margin, with a much lower average transaction price, which is likely due to most matches occurring with late arriving carriers, who have lower opportunity costs at the time of bidding.

The model suggests a mechanism to explain why the platform's unilateral shift to a later matching time proves detrimental to all parties. Carriers remain pressured to accept or reject outside offers before knowing the platform auction outcome, leading to high attrition rates. Matching strong bidders earlier allows them to reject outside offers with greater confidence (reflected in higher reservation prices), thereby reducing attrition and ultimately fostering a more efficient allocation of shipments. Consequently, in the absence of centralized matching, early matching with non-binding agreements emerges as the most effective market design, both from a firm and social planner's perspective.

\section{Conclusion}
\label{sec:conclusion}

I provide the first comprehensive empirical analysis of pricing the right to renege in matching markets, using a novel dataset from a digital brokerage platform with detailed information on reneged matches. In the stylized model, I first show the importance of considering both the downstream effect of penalties on the propensity to renege and the upstream effects on the initial matching probability and transaction price. In particular, I show that even when the penalty itself is non-pecuniary, the value of the right to renege can be priced into the final transaction price, such that the firm has a sufficiently strong incentive to offer the flexibility.

By extending the model to a dynamic setting and structurally estimating it, I can  simulate counterfactual policies and evaluate the welfare implications of different penalty schedules. I find that the current near-zero reputational penalties are nearly optimal for both social welfare and firm profits. In contrast, moving to a monetary penalty, which intuitively should enable more efficient trade, instead distorts the platform's incentives by allowing it to extract more rents from carriers, at the cost of overall welfare. Barriers to the implementation of monetary penalties, whether by regulation or through transaction costs, are thus counter-intuitively welfare-improving in this context.

\newpage
\printbibliography

\newpage

\appendix
\appendixpage

\setcounter{table}{0}
\renewcommand{\thetable}{A.\arabic{table}}
\setcounter{figure}{0}
\renewcommand{\thefigure}{A.\arabic{figure}}
\setcounter{equation}{0}
\renewcommand{\theequation}{A.\arabic{equation}}

\section{Appendix}

\subsection{Data Construction}
\label{sec:data_construction}

\newcommand{\dcNCompletedAN}{190,555\xspace}
\newcommand{\dcNCancelledAN}{60,725\xspace}
\newcommand{\dcNTotalAN}{251,280\xspace}
\newcommand{\dcNBidSharingRemoved}{26,473\xspace}
\newcommand{\dcPctBidSharingRemoved}{0.1\%\xspace}
\newcommand{\dcNOriginalAccepted}{1,265,348\xspace}
\newcommand{\dcNImputeMethodOne}{11,083\xspace}
\newcommand{\dcNImputeMethodTwo}{6,181\xspace}
\newcommand{\dcNImputeMethodThree}{233,055\xspace}
\newcommand{\dcNFinalImputed}{1,515,667\xspace}
\newcommand{\dcNDuplicateIntents}{88,590\xspace}
\newcommand{\dcPctDuplicateIntents}{1.2\%\xspace}
\newcommand{\dcNViewTimeReplaced}{6,126,701\xspace}
\newcommand{\dcNANPriceFromIntent}{3,364,463\xspace}
\newcommand{\dcNANPriceFallback}{199,203\xspace}
\newcommand{\dcNMissingANDropped}{4,515\xspace}
\newcommand{\dcNPostFalloffDropped}{11,973\xspace}
\newcommand{\dcNTTTImputedPerMile}{0\xspace}
\newcommand{\dcNTTTImputedTTM}{0\xspace}
\newcommand{\dcPctTTTMissing}{11.9\%\xspace}

\newcommand{\esNInitial}{87,824\xspace}
\newcommand{\esNMissingPrices}{2\xspace}
\newcommand{\esNWinsorized}{430\xspace}
\newcommand{\esNANPriceFilter}{604\xspace}
\newcommand{\esNAcceptedAfterPickup}{82\xspace}
\newcommand{\esNEarlyAcceptances}{327\xspace}
\newcommand{\esNOutsideWindow}{3,088\xspace}
\newcommand{\esNLateCancellations}{78\xspace}
\newcommand{\esNFinal}{83,213\xspace}
\newcommand{\esNShipments}{8,822\xspace}
\newcommand{\esNAcceptedTotal}{9,747\xspace}
\newcommand{\esNExcludedCurrentMatch}{2,325\xspace}
\newcommand{\esNExcludedEverReneged}{1,550\xspace}
\newcommand{\esNAcceptedAlt}{5,872\xspace}

This section details the construction of the estimation sample from the platform's raw data records. The process involves reconstructing incomplete records, imputing missing timestamps, and applying sample restrictions motivated by the structural model.

\subsubsection{Accept-Now Reconstruction}

The platform's dedicated logging of accept-now transactions is incomplete---some completed shipments are known to have been matched via the accept-now option but lack explicit records of the acceptance itself. To address this, I reconstruct accept-now records from two sources. Completed accept-nows are identified as shipments processed through the platform's automatic matching pipeline where the carrier had a prior detailed view of the shipment listing, yielding \dcNCompletedAN records. Cancelled accept-nows are identified as carrier-initiated falloffs where the carrier had viewed the shipment but placed no bid, implying that the only way they were matched is via the accept-now option. This yields \dcNCancelledAN records. Both identification strategies are validated against the 2022 records, where the raw data coverage appears most complete. In total, \dcNTotalAN accept-now records are reconstructed.

\subsubsection{Bid Acceptance Imputation}

Similarly to accept-now transactions, many bids that were accepted by the platform lack explicit acceptance timestamps in the raw data. I impute the remaining using three methods, applied sequentially.

First, I match bids to completed shipment records via carrier identity, shipment identity, and exact price agreement, recovering both acceptance and confirmation timestamps for \dcNImputeMethodOne bids. Second, I infer bid acceptances from carrier cancellation records. In such cases I infer the acceptance time as the close of the earliest auction round in which the reserve price exceeded the bid price (\dcNImputeMethodTwo bids). Third, I apply a dominance argument: if a bid with a strictly lower quality-adjusted price and a lower raw price than an already-accepted bid was placed before that acceptance, it must have been accepted no later than the worse bid (\dcNImputeMethodThree bids). This third method identifies accepted-but-unconfirmed bids---cases where the platform selected the bid but the carrier did not respond within the confirmation window---which are classified as reneged bids in the structural model. These imputation methods bring the total number of bid acceptances from \dcNOriginalAccepted in the raw data to \dcNFinalImputed in the processed data.

\subsubsection{Sample Filters}

Several additional filters are applied to construct the estimation sample. Bids placed through the platform's bid-sharing feature, where a partner broker submits bids on behalf of carriers, are excluded to focus on direct carrier behavior (\dcNBidSharingRemoved bids, \dcPctBidSharingRemoved of total). For carriers who place a bid or use accept-now, the view timestamp is replaced with the time of the carrier's action so that the timing of the decision, rather than the initial browsing event, determines the number of days until pickup used in estimation. 
Observations where a carrier's bid is accepted after they have already canceled the same shipment before are removed as data inconsistencies (\dcNPostFalloffDropped observations).

The estimation sample is restricted to the Seattle to San Francisco lane over 2021--2022 (\esNInitial observations after lane and year selection). Bids with normalized prices outside $[0.7, 2.0]$ are winsorized as outliers (\esNWinsorized removed), and observations where the accept-now price exceeds 1.2 times the spot rate are excluded (\esNANPriceFilter removed). Observations with bids accepted after the scheduled pickup time (\esNAcceptedAfterPickup), acceptances earlier than the auction clearing window (\esNEarlyAcceptances), views outside the auction window (\esNOutsideWindow), and cancellations recorded after pickup (\esNLateCancellations) are also removed. The final estimation sample contains \esNFinal observations across \esNShipments shipments.

\subsubsection{Cancellation Sample Restrictions}

For the confirmation and cancellation components of the likelihood, I impose two additional restrictions to avoid modeling within-platform substitution, which would require tracking carriers' full portfolio of platform matches and integrating over unobserved external offer arrivals.

First, a bid's acceptance is only included in the confirmation likelihood if the carrier does not already hold a confirmed match on the platform for the same pickup date and has not already reneged on a previously accepted bid that day. This excludes \esNExcludedCurrentMatch bids due to an existing match and \esNExcludedEverReneged due to prior reneging, retaining \esNAcceptedAlt of \esNAcceptedTotal accepted bids. The implicit assumption is that a carrier who has already cancelled in favor of an off-platform opportunity will not accept a different platform shipment for the same date. Second, as described in \autoref{sec:appendix_estimation_details}, confirmation and cancellation data is limited to the first accepted bid of each carrier-day pair to avoid correlated cancellation decisions across shipments.

\subsection{A Repeated Game Model of Reputational Penalties}
\label{sec:appendix_repeated_model}

This section presents a repeated game model of reputational penalties, which serves as a theoretical foundation for the modeling choices of the model in the main text. Recall that the platform penalizes past carrier cancellations by penalizing them in the bid scoring rule. Whereas in the main text, we model this penalty via a simple reduced-form parameter $\kappa$ (see \autoref{sec:toy_model}), here we microfound the penalty in a repeated game setting and show how it maps to the reduced-form parameter $\kappa$ in the main text. The analysis presented here also establishes some theoretical implications for a carrier's optimal bidding behavior in response to a reputational penalty, which we can show are at odds with the empirical evidence presented in \autoref{sec:background}. 

\subsubsection{Carriers' Optimal Response to a Bid Penalty}

Consider a carrier bidding on a shipment. The carrier's chooses a bid $b$, while facing a cost $c$ and a quality adjustment score $q \geq 0$ that is set by the platform in accordance with the carrier's past behavior. The penalty is added to the bid in order to determine the carrier's effective bid in the auction, which we denote as $x \equiv b+q$. The carrier's conditional win probability is $\gamma(x) \in (0,1)$ with $\gamma'(x) < 0 \, \forall x$. The carrier's expected profit from bidding $b$ is given by:
\begin{equation}
    \mathbb{E}[\text{Profit}] = (b - c) \cdot \gamma(x)
\end{equation}

We can use the first-order condition to derive the optimal bid $b^*(c,q)$:
\begin{equation}
    \pdv{\mathbb{E}[\text{Profit}]}{b} = \gamma(x) + (b - c) \cdot \gamma'(x) = 0
\end{equation}

Which can be re-arranged to obtain $b^*(c,q) = c - \frac{\gamma(x)}{\gamma'(x)}$. Denote $R(x) \equiv -\frac{\gamma(x)}{\gamma'(x)}$ as the markup over cost that the carrier applies when bidding in the auction and is always positive given that $\gamma'(x) < 0$. 

To ensure a maximum, we can check the second-order condition:
\begin{equation}
    \pdv[2]{\mathbb{E}[\text{Profit}]}{b} = 2 \gamma'(x) + (b - c) \cdot \gamma''(x) < 0
\end{equation}

Substituting in the FOC, we have:
\begin{equation}
    \label{eq:penalty_bidding_soc}
    \pdv[2]{\mathbb{E}[\text{Profit}]}{b} = 2 \gamma'(x) - \frac{\gamma(x)}{\gamma'(x)} \gamma''(x) < 0
\end{equation}  

This imposes necessary conditions on the win probability function $\gamma(x)$ to guarantee a well-defined optimal bidding strategy. This condition also restricts the derivative of the markup $R(x)$, which is:
\begin{equation}
    R'(x) = \pdv{}{x} \left(-\frac{\gamma(x)}{\gamma'(x)}\right) = \frac{\gamma(x) \gamma''(x) - \gamma'(x)^2}{\gamma'(x)^2} 
\end{equation}

Multiplying the second-order condition \eqref{eq:penalty_bidding_soc} by $\gamma'(x)<0$ gives us:
\begin{align}
    2\left(\gamma'(x)\right)^2 - \gamma(x) \gamma''(x) & > 0 \notag \\
    \gamma(x) \gamma''(x) - 2\left(\gamma'(x)\right)^2 &< 0 \notag \\
    \gamma(x) \gamma''(x) - \gamma'(x)^2 &< \gamma'(x)^2 \notag \\
    \frac{\gamma(x) \gamma''(x) - \left(\gamma'(x)\right)^2}{\left(\gamma'(x)\right)^2} = R'(x) &< 1 
\end{align}

Which implies that the optimal markup $R(x)$ may be increasing in the index $x$ (the sum of the bid and the penalty), but it must be increasing at a rate that is less than 1.

\subsubsection{Empirical Implications for Bidding Behavior}

The model has immediate predictions for a carrier's reaction to a change in the reputational penalty, which can be empirically tested. Differentiating the implicit equation $b^*(c,q) = c - \frac{\gamma(b^*(c,q) + q)}{\gamma'(b^*(c,q) + q)}$ with respect to $q$ gives us:
\begin{align*}
    \frac{\partial b^*(c,q)}{\partial q} = R'(b^*(c,q) + q) \left( \frac{\partial b^*(c,q)}{\partial q} +1 \right) \\
    \frac{\partial b^*(c,q)}{\partial q} \left( 1 - R'(b^*(c,q) + q) \right) = R'(b^*(c,q) + q) \\
    \frac{\partial b^*(c,q)}{\partial q} = \frac{R'(b^*(c,q) + q)}{1 - R'(b^*(c,q) + q)}
\end{align*}

Note that we have established that $R'(b^*(c,q) + q) < 1$, which implies that the denominator is always positive. 

There are then two possibilities for the sign of the derivative $\frac{\partial b^*(c,q)}{\partial q}$. The optimal bid can be increasing in the penalty if $R'(b^*(c,q) + q) > 0$. Alternatively, the optimal bid can be decreasing in the penalty if $R'(b^*(c,q) + q) < 0$, effectively compensating for the penalty by reducing the bid. But note that the optimal bid can never overcompensate for the penalty. To see this we can re-arrange the following:
\begin{align*}
    \frac{\partial b^*(c,q)}{\partial q} &< -1 \\
    \frac{R'(b^*(c,q) + q)}{1 - R'(b^*(c,q) + q)} &< -1 \\
    R'(b^*(c,q) + q) &<  R'(b^*(c,q) + q) - 1 \\
    0 &< -1
\end{align*}

This is a contradiction, which implies that the optimal bid can never overcompensate for the penalty. But as shown in \autoref{tab:penalty_effects}, the empirical evidence suggests that carriers are overcompensating for the penalty. A reasonable explanation for this discrepancy is the vague statement of the platform's penalty policy, which does not give carriers any quantitative information about the penalty. It is thus likely that carriers have overestimated the strength of the penalty, which is one of the reasons for estimating the reduced-form parameter $\kappa$ as the \emph{subjective} penalty that carriers perceive rather than deriving it in a repeated-game setting on the basis of the platform's true policy.

\subsubsection{Impact of a Bid Penalty on the Carrier's Continuation Value}

We can also use this analysis to microfound the reduced-form parameter $\kappa$ in the main model. The key insight is that the reputational penalty $q$ reduces the carrier's flow profit, which in turn reduces the carrier's continuation value on the platform.

We start by analyzing the impact of a change in the score $q$ on the carrier's flow profit. Fix an initial score $q_0 \ge 0$ and a cost draw $c$.   Then let
\[
\pi(c,q)
   \;=\;\bigl(b^{*}(c,q)-c\bigr)\,
          \gamma\!\bigl(b^{*}(c,q_0)+q_0\bigr)
\]
denote the indirect per-period profit conditional on the cost draw \(c\) and an initial score \(q_0\).

We can analyze the impact of an increase in the score \(q\) on the flow profit through the envelope theorem.
\begin{align}
\frac{\partial \pi(c,q)}{\partial q}
     &= \underbrace{\partial_q\Pi(b^{*},c,q_0)}_{\text{direct}} 
        + \underbrace{\partial_b\Pi(b^{*},c,q)\,
                      \frac{\partial b^{*}}{\partial q}}_{\text{indirect}} \notag\\
     &= (b^{*}\!-\!c)\,\gamma'(b^{*}\!+\!q) \;+\; 0 \quad(\text{FOC}) \notag\\
     &= -\,\gamma\!\bigl(b^{*}(c,q)+q\bigr)  < 0.
     \label{eq:envelope_derivative}
\end{align}

Where the second term vanishes because the FOC implies that $\partial_b\Pi(b^{*},c,q) = 0$ at the optimal bid $b^{*}(c,q)$. Equation \eqref{eq:envelope_derivative} is the classic \emph{envelope theorem}: once the bid is optimal, a marginal change in $q$ harms profits solely through its explicit appearance, with the loss equal to the win probability, which is non-zero by assumption. The take-away from this is that an increase in the score $q$ always reduces the carrier's flow profit.

Now suppose that a cancellation is penalized by a permanent increase in the score by $\Delta q > 0$. Define the ex-ante per-period profit as $u(q) \equiv \mathbb{E}_{c}[\pi(c,q)]$ and the present discounted value of a carrier's participation on the platform as $V(q) \equiv \sum_{t=0}^{\infty} \delta^t u(q) = \frac{u(q)}{1-\delta}$, where $\delta \in (0,1)$ is the discount factor. The change in continuation value due to a cancellation is then given by:

\begin{equation}
    \kappa(\Delta q) = V(q_0) - V(q_0 + \Delta q) = \frac{u(q_0) - u(q_0 + \Delta q)}{1 - \delta}.
\end{equation}

This maps the penalty $\Delta q$ into a change in the carrier's continuation value, which becomes the reduced-form parameter $\kappa$ in the main model. 

There are two key reasons for this approach. First, the previous section established that in a rational full-information repeated game, the carriers will never over-compensate for the penalty, which is at odds with the empirical evidence presented in \autoref{sec:background}. 

Second, adding the full repeated game dynamics to the main model would drastically increase the computational complexity of the model. It would require expanding the state space of the carriers' decision problem with an additional dimension that tracks the carrier's current score. It would also introduce a complex dependence between the stage auction game and the repeated game. The optimal bidding behavior in the stage game would depend on the effective cancellation penalty, which itself will depend on the present discounted value of all the future stage-game payoffs. This would thus create an additional level of nesting in the model, making it computationally intractable.

\subsection{Screening Motives of Reputational Penalties}
\label{sec:appendix_screening}

In the main text, the model focuses on the incentive effects of reputational penalties, assuming they reduce the present discounted value of a carrier's future profits on the platform without directly affecting the firm's profits. However, this assumption may not hold if reputational penalties also help the firm by \textit{screening} out low-quality carriers. Specifically, if reputational penalties lead to the exclusion of carriers with high cancellation rates, the firm might improve its overall performance by reducing cancellations. A counterpoint to this argument is that excluding carriers with high cancellation rates may also reduce the platform's margins, as these carriers may also be offering lower bids (recall the negative relationship between cancellation rates and bid amounts in Figure \ref{fig:confirm_cancel_prob_price}).

To evaluate this possibility, the following reduced-form exercise---with minimal modelling assumptions---approximates the potential benefits of screening out frequent cancellers:

\begin{enumerate}
\item For each carrier with at least 10 bookings, compute the mean cancellation rate and average margins directly from the data.
\item Exclude the top $X$\% of carriers ranked by their cancellation rate.
\item Evaluate the impact of this exclusion on the overall cancellation rate and margins.
\end{enumerate}

We focus on margins here under the assumption that the platform's remaining carriers will pick up the slack from the excluded carriers. While unrealistic, this assumption is biased in favor of screening, as it assumes that the platform can fully replace the excluded carriers with other carriers and does not suffer any other negative consequences from the reduced carrier base (in particular, reduced competition in auctions). Let $b$ denote the booking price of a shipment booking and let $v$ denote the value of the shipment to the platform. We focus on two types of margins:

\begin{itemize}
    \item \textbf{Booking Margin} $= \tfrac{v - b}{v}$ \\ (assumes the booking is kept)
    \item \textbf{Final Margin} $= \mathbbm{1}[\text{No cancel}]\tfrac{v - b}{v} - \tfrac{1}{10}\mathbbm{1}[\text{Cancel}]$ \\ (accounts for the impact of cancellations)
\end{itemize}

The final margin is the relevant profit margin we will focus on. It assumes the worst-case scenario in case of a cancellation: that a cancelled booking will fail to be matched in time and will incur the indirect penalty of 10\% of the shipment value.\footnote{Recall from Section \ref{sec:platform_params_est} that this heuristic was used internally by the firm as a proxy for the impact of failed shipments on their clients' contract renewal probabilities.} The final margin is a more accurate measure of the platform's profitability, as it accounts for the impact of cancellations on the platform's revenue, but is again biased in favor of screening, as it assumes a worse impact of cancellations than the platform actually experiences.

\begin{figure}[h]
\begin{center}
    \caption{Effect of Screening Frequent Cancellers on Cancellation Rates and Margins}
    \label{fig:screening_approximation}    
    \includegraphics[width=0.8\textwidth]{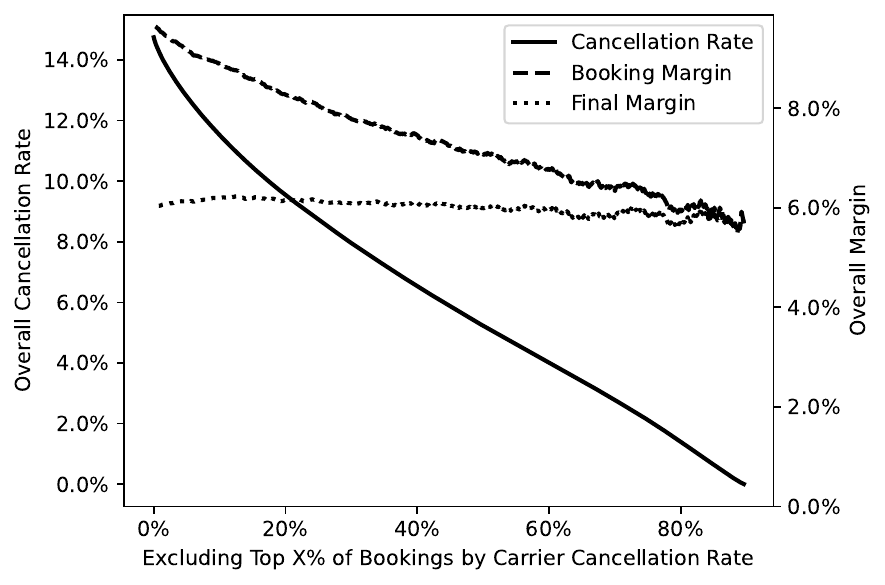}
\end{center}
{\footnotesize \textsc{Note}: Figure plots the effect of excluding the top $X$\% (as a fraction of bookings) of carriers based on their average cancellation rate across all confirmed bookings. Sample restricted to carriers with at least 10 bookings. Left axis represents effect on cancellation rate and right axis represents effect on margins. Booking margin is defined as the difference between the shipment value and the booking price, divided by the shipment value. Final margin is defined as the booking margin, adjusted for the impact of cancellations (assuming a cancellation incures a 10\% penalty for a failed shipment).}
\end{figure}

Figure \ref{fig:screening_approximation} shows the impact of excluding carriers with the highest cancellation rates on the overall cancellation rate and margins. The figure shows that excluding carriers with the highest cancellation rates reduces the overall cancellation rate but also reduces the booking margins, implying that more frequently cancelling carriers also tend to bid lower. The overall impact of both forces almost exactly cancel each other out in terms of the final margin, suggesting that the net benefit of screening out high-cancellation-rate carriers is minimal.

In conclusion, while reputational penalties might have a screening effect by discouraging or excluding frequent cancellers, the net benefit to the firm is minimal under conservative assumptions. The loss of low-cost carriers offsets the gains from reduced cancellations. Moreover, reducing the carrier base may adversely affect the platform’s ability to meet demand. Therefore, the incentive effects of reputational penalties, as discussed in the main text, remain the primary channel through which these penalties influence carrier behavior and platform performance.

\subsection{Special case of the envelope theorem}

\begin{lemma}
    \label{lemma:derivative}
    Given a continuous and differentiable function $f(x)$ and continuous and differentiable distribution $G(x)$, define:
    $$h(x) = \int \max \{f(x),y + g(x) + c\}dG(y) = G(f(x)-g(x)-c)f(x) + \int_{f(x)-g(x)-c}^\infty (y + g(x) + c) dG(y)$$

    Then the derivative of $h(x)$ is $h'(x) =  G(f(x) - g(x) - c)f'(x) + (1-G(f(x) - g(x) - c))g'(x)$
\end{lemma}

\begin{proof}
    \begin{align*}
        h'(x) = \frac{\partial}{\partial x} G(f(x)-g(x)-c)f(x) + \frac{\partial}{\partial x} \int_{f(x)-g(x)-c}^\infty y dG(y) &= \\
        G(f(x) - g(x) - c)f'(x) + \textcolor{red}{G'(f(x) - g(x) - c)(f'(x) - g'(x))f(x)}& \\
        - \textcolor{red}{G'(f(x) - g(x) - c)(f'(x) - g'(x))f(x)} + (1-G(f(x) - g(x) - c))g'(x)& \\
        =  G(f(x) - g(x) - c)f'(x) + (1-G(f(x) - g(x) - c))g'(x)
    \end{align*}
\end{proof}

Note that Lemma \ref{lemma:derivative} is really just a special case of the envelope theorem in a form more readily applicable to the model being studied in this paper. It says that the derivative with respect to the expectation over the maximum of a binary choice with a random payoff is equivalent to the derivative with respect to the fixed utility component of each choice, weighted by the respective choice probabilities.

\subsection{Value Functions in Continuous Time Markov Models}
\label{sec:value_derivation}

As described in Section \ref{sec:model}, the model takes place in continuous time, but all payoffs and decisions occur through Poisson arrival processes. This gives rise to a continuous-time Markov chain, where states proceed from one to another in a discrete way. Furthermore, there is no discounting over time in this paper, due to the short time horizons involved, which also simplifies the derivation of the value functions.

We start with a general derivation. Suppose we have states $s \in \mathcal{S}$, and $\lambda_{s'|s}$ be the Poisson arrival rates of a state $s'$ given that the current state is $s$. Because Poisson arrival rates are distributed exponentially, we can also describe the transition process as a combined arrival rate $\bar{\lambda}_s = \sum_{s'} \lambda_{s'|s}$, and a probability distribution over the states $P(s'|s) = \tfrac{\lambda_{s'|s}}{\bar{\lambda}_s}$ which gives the probability of transitioning from $s$ to $s'$ \emph{conditional} on transitioning at all. Furthermore, suppose we receive an instantaneous flow payoff $\phi(s'|s)$ when transitioning from $s$ to $s'$. If we assume $\phi(s|s) = 0$, then we can ignore transitions from a state to itself without loss of generality, and let $\lambda_{s|s}=0$.

Now, because there is no discounting nor any other time-varying component, the value function of a state $s$ is simply given by the probabilities of all successor states $s'$, weighted by the flow payoff and value function of that state:
\begin{equation}
    U_s = \sum_{s'} P(s'|s) \Big[ \phi(s'|s) + U_{s'} \Big]
\end{equation}
We can now apply this to Equation \ref{eq:unmatched_value}. The implicit state space here is $(d,Unmatched)$, where the latter variable indicates that carrier $i$ is currently unmatched with shipment $j$. Two events can shift a carrier out of this state. Firstly, they may receive an outside offer at rate $\lambda_d$, which they will take if it is attractive enough, thereby exiting the market, or refuse if they would rather keep searching. This event thus gives payoff
\begin{equation}
    \int \max\{U_{d,\bar{d}_i}(c_{ij}),\tilde{\pi}_{ij'} - c_{ij} + \rho(c_{ij}) \} dG_d(\tilde{\pi}_{ij'})
\end{equation}
Secondly, they may transition to the next day $d-1$, which occurs at exogenous rate $\eta$, which gives payoff $\bar{U}_{d-1}(c_{ij})$. Thus, the value function of a matched carrier is given by:
\begin{equation}
    U_{d,\bar{d}_i}(c_{ij}) = \frac{1}{\eta +\lambda_{d}} \Big[ \lambda_{d}  \Big(\int \max\{U_{d,\bar{d}_i}(c_{ij}),\tilde{\pi}_{ij'} - c_{ij} + \rho(c_{ij}) \} dG_d(\tilde{\pi}_{ij'}) \Big)  + \eta U_{d-1}(c_{ij}) \Big]
\end{equation}
Eventually, if the carrier hasn't found any worthwhile shipment, they will end up unmatched in the absorbing state with payoff $U_{-1}(c_{ij}) = 0$, which makes the value function finite.

Notice that this value function involves some probability of remaining in the current state, which occurs when an outside offer is not attractive enough to induce a cancellation. This occurs at the cutoff $R^U_{d,\bar{d}_i}(c_{ij}) = U_{d,\bar{d}_i}(c_{ij})+c_{ij}-\rho(c_{ij})$. As stated earlier, we can ignore any transition from a state to itself, so we can rewrite the value function as follows:
\begin{align*}
    U_{d,\bar{d}_i}(c_{ij}) = \frac{1}{\eta +\lambda_{di}} \Big[ &\lambda_{di}  \Big( \int_{R^U_{d,\bar{d}_i}(c_{ij})} \tilde{\pi}_{ij'} - c_{ij} + \rho(c_{ij})  dG_d(\tilde{\pi}_{ij'}) \\
    &+ G(R^U_{d,\bar{d}_i}(c_{ij}))U_{d,\bar{d}_i}(c_{ij}) \Big)  + \eta \bar{U}_{d-1}(c_{ij}) \Big]
\end{align*}
Which is the form used in the main text. For computational purposes we can further re-arrange this equation to:
\begin{equation*}
    U_{d,\bar{d}_i}(c_{ij}) = \frac{1}{\eta +\lambda_{di}(1-G(R^U_{d,\bar{d}_i}(c_{ij})))} \Big[ \lambda_{di}  \Big( \int_{R^U_{d,\bar{d}_i}(c_{ij})} \tilde{\pi}_{ij'} - c_{ij} + \rho(c_{ij})  dG_d(\tilde{\pi}_{ij'}) \Big)  + \eta \bar{U}_{d-1}(c_{ij}) \Big]
\end{equation*}
Benchmarks of the value function iteration have shown this re-arranged form to converge faster (when inserting the current iteration into the LHS to obtain the next iteration from the RHS). Intuitively, the original puts greater weight on the previous iteration of the value, and thus dampens the iteration, which generally reduces speed but may be more robust to oscillations in the iteration. Note that in both forms, the value function we are solving for shows up in both sides of the equation (implicitly in $R^U_{d,\bar{d}_i}(c_{ij})$), so solving the problem still requires a fixed point iteration even though we have a finite and discrete time index $d$.

We can similarly re-arrange the value function of a carrier with a matched shipment in Equation \ref{eq:matched_value}. We first define $R^M_{d,\bar{d}_i}(b_{ij},c_{ij}) = V_{d,\bar{d}_i}(b_{ij},c_{ij})+c_{ij} + \kappa_d -\rho(c_{ij})$, so that the value of a pending bid is now:
\begin{multline*}
    V_{d,\bar{d}_i}(b_{ij},c_{ij}) = \frac{1}{\eta + \lambda_d(1-G_d(R^M_{d,\bar{d}_i}(b_{ij},c_{ij})))} \Big[ \lambda_d  \Big(\int_{R^M_{d,\bar{d}_i}(b_{ij},c_{ij})}^\infty (\tilde{\pi}_{ij'} - c_{ij} - \kappa_d + \rho(c_{ij})) dG_d(\tilde{\pi}_{ij'}) \Big) \\ + \eta \bar{V}_{d-1}(b_{ij},c_{ij}) \Big]
\end{multline*}
Finally, we can do the same with the value function of a carrier with a pending bid in Equation \ref{eq:pending_bid_value}. We first define $R^P_{d,\bar{d}_i}(b_{ij},c_{ij}) = W_{d,\bar{d}_i}(b_{ij},c_{ij})+c_{ij} - \rho(c_{ij})$, so that the value of a pending bid is now:
\begin{multline*}
    W_{d,\bar{d}_i}(b_{ij},c_{ij}) = \frac{1}{\eta + \gamma_d(b_{ij}) + \lambda_d(1-G_d(R^P_{d,\bar{d}_i}(b_{ij},c_{ij})))} \Big[ \lambda_d  \Big(\int_{R^P_{d,\bar{d}_i}(b_{ij},c_{ij})}^\infty (\tilde{\pi}_{ij'} - c_{ij} + \rho(c_{ij})) dG_d(\tilde{\pi}_{ij'}) \Big) \\
    + \gamma_d(b_{ij})(V_{d,\bar{d}_i}(b_{ij},c_{ij}) ) + \eta \bar{W}_{d-1}(b_{ij},c_{ij}) \Big]
\end{multline*}

\subsection{Derivative of bidding first order condition}
\label{sec:bid_derivative}

To solve the bidding problem, we solve for the first order condition.

Similarly to the value functions themselves, the derivatives can be represented in recursive form.

We begin with the derivative of $W_{d,\bar{d}_i}(b_{ij},c_{ij})$ with respect to $b_{ij}$. Making use of Lemma \ref{lemma:derivative}, we have

\begin{multline*}
     (\eta+\gamma_d(b_{ij}) + \lambda_d) \pdv{W_{d,\bar{d}_i}(b_{ij},c_{ij})}{b_{ij}} + \gamma_d'(b_{ij})W_{d,\bar{d}_i}(b_{ij},c_{ij}) = \\
     \notag \Big[\lambda_d G_d(R^P_{d,\bar{d}_i}(b_{ij},c_{ij}))\pdv{W_{d,\bar{d}_i}(b_{ij},c_{ij})}{b_{ij}} + \gamma_d'(b_{ij})\big(\alpha_d V_{d,\bar{d}_i}(b_{ij},c_{ij}) + (1-\alpha_d)U_{d,\bar{d}_i}(c_{ij}) \big) \\
     \notag + \alpha_d\gamma_d(b_{ij})\pdv{V_{d,\bar{d}_i}(b_{ij},c_{ij})}{b_{ij}} + \eta \pdv{W_{d-1,\bar{d}_i}(b_{ij},c_{ij})}{b_{ij}} \Big] \\
\end{multline*}

Re-arranging yields the following recursive expression:

\begin{multline*}
     \notag \Rightarrow \pdv{W_{d,\bar{d}_i}(b_{ij},c_{ij})}{b_{ij}} = \frac{1}{\eta + \gamma_d(b_{ij}) +\lambda_d\big(1 - G_d(R^P_{d,\bar{d}_i}(b_{ij},c_{ij})) \big)} \\ \Big[
     \notag \gamma_d'(b_{ij}) \Big( \alpha_d V_{d,\bar{d}_i}(b_{ij},c_{ij}) + (1-\alpha_d)U_{d,\bar{d}_i}(c_{ij}) - W_{d,\bar{d}_i}(b_{ij},c_{ij}) + \frac{\alpha_d\gamma_d(b_{ij})}{\gamma_d'(b_{ij})}\pdv{V_{d,\bar{d}_i}(b_{ij},c_{ij})}{b_{ij}} \Big) \\
    + \eta \pdv{W_{d-1,\bar{d}_i}(b_{ij},c_{ij})}{b_{ij}} \Big]
\end{multline*}

We then need the derivative for the matched value ($V_{d,\bar{d}_i}$):

\begin{align*}
    \pdv{V_{d,\bar{d}_i}(b_{ij},c_{ij})}{b_{ij}} &= \frac{\eta}{\eta + \lambda(1-G_d(R^M_{d,\bar{d}_i}(b_{ij},c_{ij})))} \pdv{V_{d-1,\bar{d}_i}(b_{ij},c_{ij})}{b_{ij}} \\
    \pdv{V_{-1,\bar{d}_i}(b_{ij},c_{ij})}{b_{ij}} &= 1 \\
    \Rightarrow \pdv{V_{d,\bar{d}_i}(b_{ij},c_{ij})}{b_{ij}} &= \prod_{k=0}^d \frac{\eta}{\eta + \lambda(1-G_k(R^M_{k,\bar{d}_i}(b_{ij},c_{ij})))} \equiv P_{d,\bar{d}_i}(\text{no cancel}|b_{ij},c_{ij})
\end{align*}

We can then write the FOC as a weighted sum:

\begin{multline*}
     \pdv{W_{d,\bar{d}_i}(b_{ij},c_{ij})}{b_{ij}} = \sum_{k=0}^{d^b_{ij}} \overbrace{\Big[ \prod_{\ell=k}^{d^b_{ij}} \frac{\eta}{\eta + \gamma_\ell(b_{ij}) +\lambda_\ell\big(1 - G_\ell(R^P_{\ell,\bar{d}_i}(b_{ij},c_{ij})) \big)} \Big] \eta^{-1}}^{\text{Probability of bid surviving from $d$ to $k$}} \\
     \gamma_k'(b_{ij}) \Big( \underbrace{\alpha_k V_{k,\bar{d}_i}(b_{ij},c_{ij}) + (1-\alpha_k)U_{k,\bar{d}_i}(c_{ij}) - W_{k,\bar{d}_i}(b_{ij},c_{ij})}_{\text{Benefit of winning net of opportunity costs}} + \underbrace{\frac{\alpha_k\gamma_k(b_{ij})}{\gamma_k'(b_{ij})} P_{k,\bar{d}_i}(\text{no cancel}|b_{ij},c_{ij})}_{\text{Markup}} \Big) \\
      = 0
\end{multline*}

While superficially complicated, this FOC is essentially a weighted sum of day-specific FOCs, each of which captures the marginal benefit of increasing the bid on the given day. The terms highlighted in braces (a day-specific derivative) take the familiar form observed in the standard theory of first-price auctions. These day-specific derivatives are then weighted according to the probability of surviving to that day.

\section{Likelihood Derivation and Estimation Procedure}
\label{sec:appendix_likelihood}

This appendix provides the full derivation of the carrier model likelihood and the formal description of the block coordinate estimation procedure summarized in Section \ref{sec:estimation}.

\subsection*{Likelihood Derivation}

The data used to estimate the carrier model consists of the following:

\begin{enumerate}[noitemsep,topsep=0pt]
    \item \textsc{First stage decisions}: For each carrier $i$ and shipment $j$, I observe the choice $x_{ij} \in \{B,A,I\}$, corresponding to the decision to bid, Accept-now, or ignore the shipment. For carrier $i$, let $K^B_i \subseteq K_i$ denote the set of bids, $K^A_i \subseteq K_i$ the set of Accept-Now matches, and $K^I_i \subseteq K_i$ the set of ignored shipments. 
    \item \textsc{Bids}: For shipments $j \in K^B_i$, I also observe the carrier's bids $b_{ij} \in \mathbb{R}_+$.
    \item \textsc{Confirmations}: A subset of bids $\bar{K}^B_i \subseteq K^B_i$ are winning bids, giving the carrier the opportunity to confirm their bid. I thus observe $y_{ij} \in \{1,0\}$ for $j \in \bar{K}^B_i$, where $1$ denotes a confirmation. I let $\tilde{K}^B_i \subseteq \bar{K}^B_i$ denote the set of shipments that a carrier confirms.
    \item \textsc{Cancellation decisions}: For shipments matched by confirming a bid or using Accept-Now ($\tilde{K}^B_i \cup K^A_i$), I also observe, for each valid day before pickup $d$, whether the carrier canceled on that day or not, so I have cancellation decisions $z_{ijd} \in \{1,0\}$ for $j \in \tilde{K}^B_i \cup K^A_i$, where $1$ denotes a cancellation. Let $z_{ij}$ denote the vector of cancellation decisions.
\end{enumerate}
In addition, the likelihood needs to account for the unobservable costs. This includes both the mean carrier cost $\bar{c}_i$ and the shipment-specific costs $c_{ij} = \bar{c}_i + \epsilon_{ij}$. To start, I assume that $\bar{c}_i$ is also observed, so that I can condition the likelihood on it, before eventually integrating it out. 

I start by forming the likelihood of the observed bids. Recall that I can use the first-order condition of the carrier's bidding problem to invert the bid into their cost, so the likelihood of bids is obtained from the distribution of costs that I aim to estimate. I denote this inversion as $c_{ij} = b_{\bar{d}_i}^{*-1}(b)$. Recall that $c_{ij} = \bar{c}_i + \epsilon_{ij}, \epsilon_{ij} \sim N(0,\sigma_\epsilon)$. I can then write the likelihood of a bid conditional on $\bar{c}_i$ and parameters $\theta$ as: 
\begin{equation}
    \label{eq:bids_density}
    \mathcal{L}(b_{ij}|\bar{c}_i, \theta) = \phi((b_{\bar{d}_i}^{*-1}(b) - \bar{c}_i)/\sigma_\epsilon)\Big|\pdv{b_{\bar{d}_i}^{*-1}(b)}{b_{ij}} \Big|
\end{equation}
where $\phi(.)$ denotes the density function of a standard Normal distribution. The derivative term inside the absolute value is needed to properly account for the transformation of the random variable.

Below, I derive the remaining component of the likelihood by first assuming that I know the shipment specific cost $c_{ij}$. This is true for shipments a carrier has bid on, as I can invert the bid to obtain the cost. For the remaining shipments, I integrate out the unobservable cost.

The first stage choices are obtained straightforwardly from Equation \ref{eq:first_stage_prob}. I denote their conditional likelihood as $P(x_{ij}|c_{ij},\theta)$. 

Next, the bid confirmation probabilities are obtained from Equation \ref{eq:confirm_prob}. These are only relevant for observations with bids, and can thus all be conditioned on the bid and the corresponding inverted cost $c_{ij} = b^{*-1}(b_{ij})$
\begin{equation}
    P(y_{ij}|b_{ij},\theta) = P_{\bar{d}_i}(confirm|b_{ij},c_{ij},d^a_{ij},d^b_{ij})^{y_{ij}} \left(1-P_{\bar{d}_i}(confirm|b_{ij},c_{ij},d^a_{ij},d^b_{ij})\right)^{(1-y_{ij})}
\end{equation}
where $y_{ij}$ is shorthand for the decision to confirm a bid, conditional on winning. Finally, I obtain the cancellation probabilities from Equation \ref{eq:cancel_prob}. These cover every day before pickup $d$ during which a shipment was matched, which rules out multiple cancellations for one shipment.
\begin{equation}
    P(z_{ij}|b_{ij},c_{ij},\theta) = \prod_{d=\underline{d}_{ij}}^{\overline{d}_{ij}} P_d(cancel|b_{ij},c_{ij})^{z_{ijd}} (1-P_d(cancel|b_{ij},c_{ij}))^{(1-z_{ijd})}
\end{equation}
where $z_{ij}$ is shorthand for the vector of cancellation decisions and $z_{ijd}$ is the decision for day $d$. Computing the full likelihood for a single carrier index $i$ requires careful consideration of which  likelihoods are jointly integrated.

The panel structure of the data requires that the unobservable mean cost $\bar{c}_i$ be jointly integrated out over the whole set of decisions of a single carrier. Formally, suppose a carrier bids on every shipment they see. Then the joint probability of their actions is:
\begin{equation*}
    \int \prod_{j \in K^B_i}  [ \mathcal{L}(b_{ij}|\bar{c}_i, \theta)  \underbrace{ P(x_{ij} = B|c_{ij}) ]\prod_{j \in \bar{K}^B_i} \left[ P(y_{ij}|b_{ij},\theta) \right] \prod_{j \in \tilde{K}^B_i} \left[ P(z_{ij}|b_{ij},c_{ij},\theta)\right]}_{\textrm{Independent of variable of integration $\bar{c}_i$}} dF(\bar{c}_i)
\end{equation*}
where $x_{ij}$ and $z_{ij}$ are shorthand for the first-stage choice and cancellation decisions, respectively. Since the terms relating to decisions on bidding, confirmation, and cancellation are fully determined by the observed bid $b_{ij}$ and the inverted cost $c_{ij}$, these can be moved outside the integral. This simplification is directly attributable to Assumption \ref{assumption:carrier_cost_knowledge}, which requires that carriers only make use of $c_{ij}$ in the computation of their optimal policies. For carriers that also use the Accept-Now option or ignore some viewed shipments, additional terms are added to the integral, as described in the following.

For any shipments where a carrier used Accept-Now, no bid is observed. I first write the likelihood conditional on $\bar{c}_i$, and integrate out the idiosyncratic component $\epsilon_{ij}$ jointly over the first-stage choice and cancellation decisions, as follows:
\begin{equation}
    P^A(x_{ij},z_{ij}|\bar{c}_i, b^A_{ij}, \theta) = \int P(x_{ij}=A|\bar{c}_i+\epsilon,\theta)P(z_{ij}|b^A_{ij},\bar{c}_i+\epsilon,\theta) dF(\epsilon)
\end{equation}

For shipments that were ignored by a carrier, I can simply integrate the likelihood of the first stage choice:
\begin{equation}
    P^{I}(x_{ij}|\bar{c}_i,\theta) = \int P(x_{ij}=Ignore|\bar{c}_i+\epsilon,\theta)dF(\epsilon)
\end{equation}
Now, I can gather all the shipments of carrier $i$, and write the likelihood of all terms to be integrated over the unobserved $\bar{c}_i$:
\begin{equation}
    \mathcal{L}_i^{int}(\theta) = \int \Big[ \prod_{j \in K^B_i} \mathcal{L}(b_{ij}|\bar{c}_i, \theta) \prod_{j \in K^A_i} P^A(x_{ij},z_{ij}|\bar{c}_i, b^A_{ij}, \theta) \prod_{j \in K^I_i} P^{I}(x_{ij}|\bar{c}_i,\theta) \Big] dF(\bar{c}_i)
\end{equation}
And thus, the full likelihood of the model is:
\begin{equation}
    \mathcal{L}(\theta) = \prod_{i=1}^N \Big[ \mathcal{L}_i^{int}(\theta) \underbrace{\prod_{j \in K^B_i} P(x_{ij}=B|c_{ij})}_{\text{Choice to bid}} \underbrace{\prod_{j\in \bar{K}^B_i} P(y_{ij}|b_{ij},\theta)}_{\text{Confirmations}} \underbrace{\prod_{j \in \tilde{K}^B_i} P(z_{ij}|b_{ij},b^{*-1}(b_{ij}),\theta)}_{\text{Cancellations}}  \Big] \equiv \prod_{i=1}^N \mathcal{L}_i(\theta)
\end{equation}

For convenience, we will define the log-likelihood of a single carrier as $\ell_i(\theta) = \log \mathcal{L}_i(\theta)$.

\subsection{Block Coordinate Estimation}

While in principle it is possible to use the full likelihood to jointly estimate all parameters of the model, it is computationally expensive to do so and it is more sensitive to model misspecification. For example, if the model of first-stage choices---which is not a primary focus of this paper---is misspecified, the parameters relating to the outside offer process and cancellation penalties may be biased to create a better fit. This is a particular concern because there are many more observations of first-stage choices than of confirmations and cancellations. For these reasons, I adopt an alternating optimization approach. In a first step, the parameters governing cancellation behavior are estimated using only the confirmation and cancellation decisions, taking some cost parameters as fixed. The parameters governing the first stage choices are then chosen to maximize the full likelihood, taking the existing parameters as given. This alternation is repeated until both sets of parameters converge. 

More formally, let $\theta_1 = (\kappa,\lambda,\mu^\pi, \sigma^\pi)$ denote the parameters of the outside offer process and the cancellation penalties, and $\theta_2 = (\mu_c, \sigma_c, \epsilon_c, c_{bid}, c_{AN}, \sigma_{choice})$ denote the parameters of the cost distribution and the first-stage choices. 

We then define the two following objectives:
\begin{align}
    \tilde{Q}^{(1)}(\theta_1,\theta_2) &\equiv \sum_{i=1}^N \Big[ \sum_{j\in \bar{K}^B_i} \log P(y_{ij}|b_{ij},\theta_1,\theta_2) + \sum_{j \in \tilde{K}^B_i} \log P(z_{ij}|b_{ij},b^{*-1}(b_{ij}),\theta_1,\theta_2)  \Big] \label{eq:Q1}\\
    Q^{(2)}(\theta_1,\theta_2) &\equiv \sum_{i=1}^N \ell_i(\theta_1,\theta_2) \label{eq:Q2}
\end{align}

The first objective function is further augmented with a penalty on negative inverted costs. This penalty is economically motivated because negative marginal costs are not sensible in this context. This penalty also plays a key role in the identification argument that follows shortly below. Thus we have: 
\begin{equation}
Q^{1}(\theta_1,\theta_2) = \tilde{Q}^{(1)}(\theta_1,\theta_2)  + \psi \left[\sum_{i=1}^N \sum_{j=1}^{k_i^B} \min\{c_{ij}(b_{ij},\theta_1,\theta_2),0\} \right]
\end{equation}

We then define the blockwise score vectors:
\[
q^{(1)}(\theta)\equiv\nabla_{\theta_1}Q^{(1)}(\theta_1;\theta_2),
\qquad
q^{(2)}(\theta)\equiv\nabla_{\theta_2}Q^{(2)}(\theta_2;\theta_1),
\]
and stack $g(\theta)\equiv\big(q^{(1)}(\theta)^\prime,\;q^{(2)}(\theta)^\prime\big)^\prime$.
Our estimator $\hat\theta$ is defined as any solution to the stacked score system
\begin{equation}
g(\hat\theta)=0.
\label{eq:stacked-root}
\end{equation}

Although this estimation strategy relies on the model's likelihood, it is formally interpreted as a Generalized Method of Moments (GMM) estimator. In this framework, the stacked score vector $g(\theta)$ defines the moment conditions, with the identity matrix serving as the weighting matrix. For further discussion on the general equivalence between maximum likelihood estimation and GMM, see \textcite{newey1994large}.

Some additional parametric restrictions are imposed on the model to reduce the computational burden and stabilize the estimation process. These are detailed in Appendix \ref{sec:appendix_estimation_details}. The offline estimation of the conditional win probabilities is detailed in Appendix \ref{sec:win_rate}. For details on the estimation of the asymptotic variances of the estimates, the reader is referred to Appendix \ref{sec:struct_est_var}.

\subsection{Implementation Details}
\label{sec:appendix_estimation_details}

Here I discuss additional restrictions imposed on the model for estimation purposes, as well as the implementation of an economically motivated assumption of non-negative costs, which aid in identification.

To start, while the model is derived by normalizing utilities into monetary equivalents, they are not estimated as dollar amounts. Instead, they are estimated as a proportion of the current market rate, as in Section \ref{sec:background}. The trucking industry experiences regular seasonal fluctuations, in addition to cost variations due to macroeconomic factors (such as the fuel price shock caused by the Russian invasion of Ukraine in 2022). To remove the noise from these fluctuations and pool the data across periods, all price variables in the data are first divided by a third-party estimate of the prevailing spot price for shipments along the lane (see \autoref{sec:data_construction} for details on the imputation of missing rates). This will be referred to as the \emph{market rate}, and all relevant structural parameters will be expressed as a fraction of this market rate.

I then impose monotonicity and non-negativity assumptions on $\kappa_d$, so that $\kappa_{d-1} \geq \kappa_{d} \geq 0,\quad \forall d$ . If cancellation penalties were not monotonically increasing as the pickup time draws closer, it would be in the interest of a carrier to hold off on cancelling even when they already have a better offer in hand. To discipline the time path of the parameters governing the outside offer distribution ($\mu_d^\pi, \sigma_d^\pi, \lambda_d$), I impose a second-order polynomial structure in the number of days until pickup. Furthermore, I assume that all day-specific parameters are shared between $d=4,5,6,7$ as no bids are accepted by the platform before then, so there is limited information on cancellation and bid reneging behavior.

Finally, as the model treats the confirmation and cancellation decisions of a carrier on different shipments as separate, I filter the data to only include confirmations and cancellations on the \emph{first} bid of each carrier-day pair that is accepted by the platform. This complements the sample restriction described in \autoref{sec:data_construction}, which excludes carriers who already hold a confirmed match or have previously reneged. Together, these restrictions sidestep the issue of correlated confirmation and cancellation decisions, which would otherwise necessitate integrating over the arrivals of unobserved external offers in the likelihood calculation, adding a significant computational burden.

\subsection{Offline estimation of win rates}
\label{sec:win_rate}

Before estimating the parameters of the structural model, I first estimate the conditional win probabilities of the auction at the lane level. The FOC in Equation \ref{eq:bid_foc} involves the day-specific win rates $\gamma_d(b)$ and their derivatives $\gamma_d'(b)$. While these can be solved for through market equilibrium conditions (and are, in the counterfactual simulations), doing so for every evaluation of the likelihood is computationally expensive, as it involves solving for a fixed point in the space of bidding strategies (the optimal bid is a function of the win probabilities and the win probabilities are a function of the optimal bidding strategies). To avoid this computational burden, I estimate the win probabilities off-line from the data directly, in the spirit of \textcite{guerre2000optimal}.\footnote{In standard auction settings as studied by \textcite{guerre2000optimal}, the win probability is a straightforward function of the bid distribution (which can be efficiently nonparametrically estimated) and the number of opponents. The complex auction format studied here does not permit such a closed-form relationship, so the conditional win probability is estimated directly.}

The FOC in Equation \ref{eq:bid_foc} involves the day-specific win rates $\gamma_d(b)$ and their derivatives $\gamma_d'(b)$. While in principle these can be derived through market equilibrium conditions, doing so is computationally expensive, as it involves solving for a fixed point in the space of bidding strategies (the optimal bid is a function of the win probabilities and the win probabilities are a function of the optimal bidding strategies). To avoid this computational burden, I simply estimate the win probabilities off-line from the data directly.

As with all other price variables in the data, I normalize all bids by the contemporaneous lane spot market rate for the respective shipment's pickup date, in order to pool data over the two-year period.

I fit a modified logistic function to the hazard rate, i.e., the probability that an auction is won on day $d$ conditional on the bid still existing on day $d$:
\begin{equation}
    \frac{\gamma(b;\theta_\gamma)}{\eta + \gamma(b;\theta_\gamma)} = \frac{\theta_{\gamma 1}}{1 + \exp((\log(b)-\theta_{\gamma 2})/\theta_{\gamma 3})}
\end{equation}

This yields the following expressions for $\gamma$ and $\gamma'$:
\begin{align}
    \gamma(b,\theta_\gamma) &= \frac{\eta \theta_{\gamma 1}}{1+exp((\log(b)-\theta_{\gamma 2 })/\theta_{\gamma 3})-\theta_{\gamma 1}} \\
    \gamma'(b,\theta_\gamma) &= -\frac{\eta \theta_{\gamma 1} exp((\log(b)-\theta_{\gamma 2 })/\theta_{\gamma 3})}{b\theta_{\gamma 3}\big( 1+exp((\log(b)-\theta_{\gamma 2 })/\theta_{\gamma 3})-\theta_{\gamma 1}\big)^2}
\end{align}

The form of this function is chosen for its reasonable fit. In addition, with $\theta_{\gamma 1} \geq 0$ and $\theta_{\gamma 3} \geq 0$, the function is guaranteed to be monotonically decreasing.

The estimated win probability function is denoted by $\hat{\gamma}(b)$ and is fitted via maximum likelihood on the full bid data. The fit of the function is presented below in Figure \ref{fig:win_prob_estimation}

\begin{figure}[htbp]
    \centering
    \caption{Non-parametric vs. fitted estimates of win probabilities (Seattle-San Francisco)}
    \label{fig:win_prob_estimation}
    \subfloat[Kernel regression of win probabilities]{\includegraphics[width=0.45\textwidth]{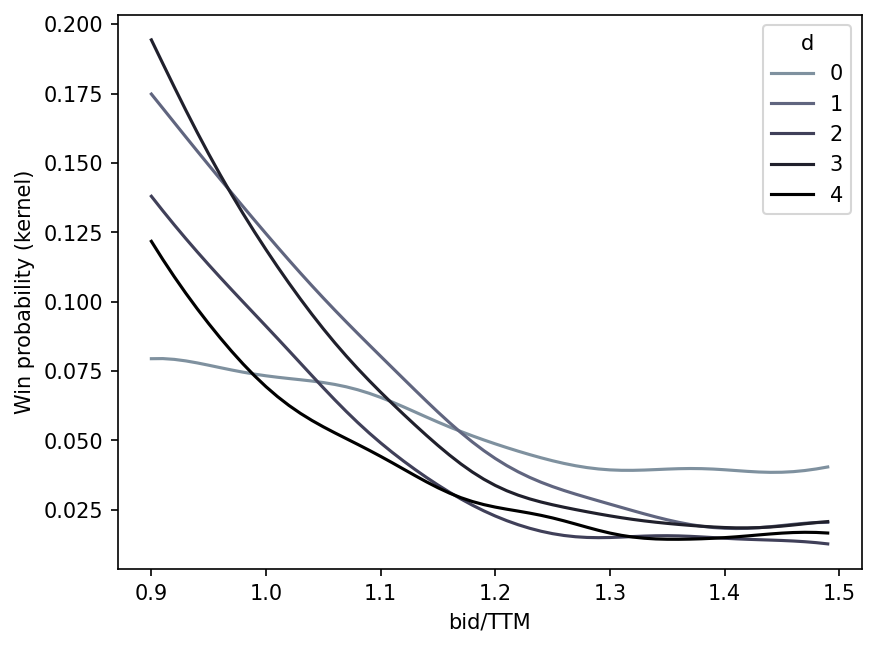}}
    \hfill
    \subfloat[Fitted probabilities]{\includegraphics[width=0.45\textwidth]{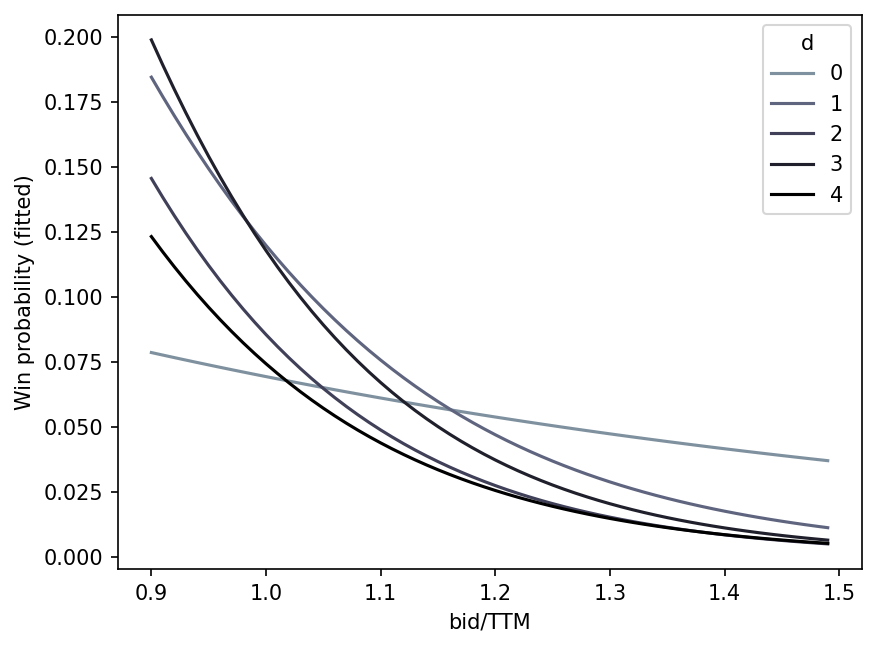}}
\end{figure}

\subsection{Carrier Model: Variance and Raw Parameter Estimates}
\label{sec:struct_est_var}

The structural parameters $\theta$ are estimated using an alternating block coordinate descent algorithm. The parameter vector is partitioned into two blocks, $\theta = (\theta_1', \theta_2')'$, corresponding to the parameters of the outside offer process and cancellation penalties ($\theta_1$) and the parameters of the cost distribution and first-stage choices ($\theta_2$). The estimation procedure alternates between maximizing two distinct objective functions until convergence:
\begin{align}
    \hat{\theta}_1 &= \arg\max_{\theta_1} Q^{(1)}(\theta_1, \hat{\theta}_2) \\
    \hat{\theta}_2 &= \arg\max_{\theta_2} Q^{(2)}(\hat{\theta}_1, \theta_2)
\end{align}
where $Q^{(1)}$ is the likelihood of confirmation and cancellation decisions augmented by the penalty term for negative inverted costs, and $Q^{(2)}$ is the full likelihood of the model.

Upon convergence, the estimator $\hat{\theta}$ satisfies the first-order conditions of both optimization problems simultaneously. We formally interpret this converged estimator as a Generalized Method of Moments (GMM) estimator. Let $m_i(\theta)$ denote the stacked vector of score contributions for carrier $i$:
\begin{equation}
    m_i(\theta) = \begin{pmatrix}
        \nabla_{\theta_1} \ell_i^{(1)}(\theta_1, \theta_2) + \nabla_{\theta_1} \psi \min\{c_{ij}, 0\} \\
        \nabla_{\theta_2} \ell_i^{(2)}(\theta_1, \theta_2)
    \end{pmatrix}
\end{equation}
where $\ell_i^{(1)}$ and $\ell_i^{(2)}$ are the observation-level log-likelihood components for the two steps, and $\psi \min\{c_{ij}, 0\}$ is the contribution of observation $i$ to the penalty term. The sample moment condition is given by $g_n(\theta) = \frac{1}{N} \sum_{i=1}^N m_i(\theta)$. The estimator $\hat{\theta}$ solves the system $g_n(\hat{\theta}) = 0$.

Because the penalty term $\psi \min\{c_{ij}, 0\}$ introduces kinks in the objective function at points where the inverted cost is exactly zero, the objective function is not continuously differentiable everywhere. Consequently, standard asymptotic arguments based on Taylor expansions are not directly applicable. Instead, we rely on Theorem 7.2 of \textcite{newey1994large}, which establishes the asymptotic normality of minimum distance and GMM estimators with non-smooth objective functions.

Under the regularity conditions specified in \textcite{newey1994large}, the estimator is asymptotically normal:
\begin{equation}
    \sqrt{N}(\hat{\theta} - \theta_0) \xrightarrow{d} \mathcal{N}(0, V)
\end{equation}
The asymptotic variance matrix $V$ is given by the standard sandwich formula. Since the number of moment conditions equals the number of parameters (exact identification), the weighting matrix is effectively the identity, and the variance simplifies to:
\begin{equation}
    V = G^{-1} \Omega (G^{-1})'
\end{equation}
where:
\begin{itemize}
    \item $G = E[\nabla_\theta m(z, \theta_0)]$ is the expected Jacobian of the moment vector. This matrix captures the cross-dependencies between the two estimation steps, including the off-diagonal blocks $\nabla_{\theta_1 \theta_2} Q^{(1)}$ and $\nabla_{\theta_2 \theta_1} Q^{(2)}$.
    \item $\Omega = E[m(z, \theta_0) m(z, \theta_0)']$ is the covariance matrix of the scores.
\end{itemize}

We estimate the asymptotic variance $\hat{V}$ using the sample analogs of these matrices evaluated at the estimated parameters $\hat{\theta}$:
\begin{align}
    \hat{\Omega} &= \frac{1}{N} \sum_{i=1}^N m_i(\hat{\theta}) m_i(\hat{\theta})' \\
    \hat{G} &= \nabla_\theta g_n(\hat{\theta})
\end{align}
Due to the non-smoothness of the objective function and the complexity of the analytical derivatives, we compute the Jacobian $\hat{G}$ using numerical differentiation with a step size calibrated to the scale of the parameters. This approach is consistent with the results in Theorem 7.4 of \textcite{newey1994large}, which validates the use of numerical derivatives for variance estimation in non-smooth contexts provided the step size shrinks at an appropriate rate relative to the sample size.

Table \ref{tab:raw_params_block1} reports the raw structural parameter estimates from the first block and their asymptotic standard errors. Table \ref{tab:raw_params_block2} reports the raw structural parameter estimates from the second block and their asymptotic standard errors.

\begin{table}[htbp]
    \begin{center}
        \caption{Raw structural parameter estimates: first block}
        \label{tab:raw_params_block1}
        \begin{tabular}{lrr}
\toprule
                    Parameter &  Estimate &  Std. Error \\
\midrule
  $\log(\kappa_0-\kappa_{1})$ &    -5.342 &   2.418e-06 \\
  $\log(\kappa_1-\kappa_{2})$ &    -6.128 &   5.212e-05 \\
  $\log(\kappa_2-\kappa_{3})$ &    -8.075 &       0.027 \\
  $\log(\kappa_3-\kappa_{4})$ &    -5.046 &   1.410e-05 \\
           $\log(\kappa_{4})$ &    -8.789 &       0.069 \\
$\log((1-\alpha_0)/\alpha_0)$ &     0.204 &       0.609 \\
$\log((1-\alpha_1)/\alpha_1)$ &    -0.221 &       0.246 \\
$\log((1-\alpha_2)/\alpha_2)$ &    -0.497 &       0.028 \\
$\log((1-\alpha_3)/\alpha_3)$ &    -0.837 &       0.488 \\
$\log((1-\alpha_4)/\alpha_4)$ &    -1.176 &       1.815 \\
                 $\mu^{\pi0}$ &     1.118 &       0.349 \\
                 $\mu^{\pi1}$ &    -0.085 &       0.325 \\
                 $\mu^{\pi2}$ &     0.051 &       0.018 \\
              $\sigma^{\pi0}$ &     2.137 &       0.129 \\
              $\sigma^{\pi1}$ &     0.298 &       0.093 \\
              $\sigma^{\pi2}$ &     0.058 &       0.103 \\
            $\log(\lambda^0)$ &     0.124 &       0.022 \\
            $\log(\lambda^1)$ &    -0.045 &       0.001 \\
            $\log(\lambda^2)$ &     0.017 &       0.010 \\
\bottomrule
\end{tabular}

    \end{center}
    
    {\footnotesize \textsc{Note:} As explained in Appendix \ref{sec:appendix_estimation_details}, additional structure is placed on the parameters. Firstly, the cancellation penalty schedule is constrained to be non-decreasing. I enforce this by estimating the log of the difference in the penalty from one day to the next. The attention probabilities are strictly bounded between 0 and 1 by estimating the logistic transformation of the raw attention probabilities. The parameters of the outside offer distribution are constrained to be second order polynomials of the number of days until pickup. }
\end{table}

\begin{table}[htbp]

    \begin{center}
    \caption{Raw structural parameter estimates: second block}
    \label{tab:raw_params_block2}
    \begin{tabular}{lrr}
\toprule
        Parameter &  Estimate &  Std. Error \\
\midrule
        $\mu^c_0$ &     0.758 &       0.018 \\
        $\mu^c_1$ &     0.858 &       0.010 \\
        $\mu^c_2$ &     0.860 &       0.007 \\
        $\mu^c_3$ &     0.865 &       0.007 \\
        $\mu^c_4$ &     0.839 &       0.036 \\
     $\sigma^c_0$ &     0.233 &       0.008 \\
     $\sigma^c_1$ &     0.192 &       0.009 \\
     $\sigma^c_2$ &     0.192 &       0.015 \\
     $\sigma^c_3$ &     0.195 &       0.016 \\
     $\sigma^c_4$ &     0.223 &       0.588 \\
$\sigma^\epsilon$ &     0.082 &       0.031 \\
$\sigma^{choice}$ &     0.158 &       0.063 \\
        $c_{bid}$ &     0.019 &       0.021 \\
         $c_{AN}$ &    -0.521 &       0.144 \\
\bottomrule
\end{tabular}

    \end{center}
    
    {\footnotesize \textsc{Note:} Table presents parameters estimated from the second step. Upper half reports day-specific parameters, lower half reports parameters that are common across days. $\mu^c_d$ and $\sigma^c_d$ are the mean and standard deviation of the distribution of the mean cost $\bar{c}_i$ and $\sigma^\epsilon$ is the standard deviation of the carrier-shipment idiosyncratic cost $\epsilon_{ij}$. $c_{bid}$ and $c_{AN}$ are the hassle costs of bidding and using the Accept-Now option, respectively. $\sigma^{choice}$ is the standard deviation of the idiosyncratic term in the first-stage choice model.}
\end{table}

\newcommand{\mcGammaLoc}{1.2\xspace}
\newcommand{\mcGammaScale}{0.2\xspace}
\newcommand{\mcReps}{500\xspace}
\newcommand{\mcSampleSize}{10,000\xspace}
\newcommand{\mcAvgWins}{5,672\xspace}
\newcommand{\mcAvgConfirmed}{1,615\xspace}
\newcommand{\mcAvgCancelled}{269\xspace}

\subsection{Monte Carlo Simulation}
\label{sec:monte_carlo}

For the Monte-Carlo exercise, I specify $G(.)$ as a Normal $N(\mu_G, \sigma_G)$, while the win probability $\gamma(b)$ is specified as a logistic distribution with location \mcGammaLoc and scale \mcGammaScale. This version of the model also adds an attention probability $\alpha$ to be estimated, as well as an additional component of the likelihood from the confirmation probability.

Table \ref{tab:monte_carlo} shows the results of the Monte-Carlo simulation, both with a penalty on negative inverted costs ($\psi=50$) and without. The penalty clearly helps with the separate identification of $\mu_G$ and $\kappa$, bringing down their standard errors significantly. Furthermore, all parameter point estimates are within one standard deviation of their true values, meaning that the true parameters are well within the confidence intervals of the estimates. Lastly, the Monte-Carlo exercise replicates a key feature of the data, which is the progressively declining sample size: of \mcSampleSize bidders, only \mcAvgWins win, \mcAvgConfirmed confirmed bids, and \mcAvgCancelled cancellations on average across the Monte-Carlo runs.

\begin{table}[htbp]
    \begin{center}
    \caption{Results of Monte Carlo Simulation with \mcReps runs of \mcSampleSize bidders each}
    \label{tab:monte_carlo}
    \begin{tabular}{lrrrrr}
\toprule
 &  & \multicolumn{2}{c}{$\psi = 50$} & \multicolumn{2}{c}{$\psi = 0$} \\
 & True & MC Mean & MC S.D. & MC Mean & MC S.D. \\
\midrule
$\mu_G$ & 1.000 & 0.969 & 0.075 & 1.318 & 0.680 \\
$\sigma_G$ & 0.500 & 0.517 & 0.106 & 0.533 & 0.119 \\
$\kappa$ & 0.100 & 0.066 & 0.078 & 0.425 & 0.702 \\
$\alpha$ & 0.500 & 0.497 & 0.032 & 0.497 & 0.028 \\
\bottomrule
\end{tabular}

    \end{center}
    {\footnotesize \textsc{Note}: The parameter $\psi$ is the penalty applied to negative inverted costs. On average, \mcAvgWins winning bids, \mcAvgConfirmed confirmed bids, \mcAvgCancelled cancellations.}
\end{table}

\subsection{Platform Parameters Estimation}
\label{sec:platform_params_est}

This section describes the estimation of the remaining platform parameters, namely:
\begin{itemize}
    \item The arrival rates of carriers, shipments, views, shipper cancellations, and auction clearing rounds.
    \item The distribution of shipment values to the platform.
    \item The conditional reserve and Accept-Now prices.
\end{itemize}

For the purposes of this section, let $m$ denote each separate \emph{market}, defined as a unique combination of origin and destination, and pickup date.

\paragraph{Arrival Rates}

We begin by estimating the exponential rates of the three event types which can be simply counted: carriers, shipments, and auction clearing rounds. For each of these events $e\in\{carrier,shipment,clear\}$, let $N^e_{m,d}$ denote the number of events of type $e$ in market $m \in 1,\dots,M$ on day $d$ from departure. As the events are assumed to follow a Poisson process, the count of events follow a Poisson distribution, so that:
\begin{equation*}
    \mathbb{E}[N^e_{m,d}] = \lambda^e_d.
\end{equation*}
Thus, a simple method-of-moments estimator for each of these arrival rates is the sample mean of the counts:
\begin{equation*}
    \hat{\lambda}^e_d = \frac{1}{M} \sum_{m=1}^M N^e_{m,d} 
\end{equation*}
with the asymptotic variance given by the sample variance of the counts divided by $M$.

Shipment cancellations and views have rates that are specified on a per-shipment and per-pair basis, and can only occur once, respectively. Let $Y^{shipcancel}_{j,m,d}\in\{0,1\}$ denote the event that shipment $j$ in market $m$ on day $d$ is cancelled, and $Y^{view}_{ij,m,d}\in\{0,1\}$ denote the event that shipment $j$ in market $m$ on day $d$ is viewed by carrier $i$. Given the properties of the Poisson process, the probabilities of these events are
\begin{align*}
    P(Y^{shipcancel}_{j,m,d}=1) &= \frac{\lambda^{shipcancel}_d}{\eta + \lambda^{shipcancel}_d} \equiv p^{shipcancel}_d  \\
    P(Y^{view}_{ij,m,d}=1) &= \frac{\lambda^{view}_d}{\eta + \lambda^{view}_d} \equiv p^{view}_d
\end{align*}
where $\eta$ is the rate of transition between days, normalized to 1. The sample analogues of these probabilities are the sample proportions of cancellations and views for each market and day, among the set of \emph{uncancelled} shipments and \emph{unmatched} pairs, respectively. These are used to estimate the arrival rates of cancellations and views:
\begin{align*}
    \hat{\lambda}^{shipcancel}_d &= \frac{\hat{p}^{shipcancel}_d}{1-\hat{p}^{shipcancel}_d} \\
    \hat{\lambda}^{view}_d &= \frac{\hat{p}^{view}_d}{1-\hat{p}^{view}_d}
\end{align*}
with the asymptotic variances of the estimates obtained by multiplying the variances of the sample proportions by the square of the derivative of the transformation function.

Due to the large number of arrival rate parameters, the estimates and their confidence intervals are reported graphically in Figure \ref{fig:lambda_platform} in the main text.

\paragraph{Shipment Values}

The shipment values to the platform are directly observed. I fit a log-normal distribution to the data by simply taking the sample mean $\hat{\mu}$ and sample variance $\hat{\sigma}^2$ of the log-transformed shipment values, reported in Table \ref{tab:shipment_value_dist}. In addition, the cost of not matching a shipment is calibrated at 10\% of the shipment value, which is the same heuristic used by the platform in its development of the newer auction format introduced in 2023.

\begin{table}[htbp]
    \begin{center}
    \caption{Shipment value distribution parameters}
    \label{tab:shipment_value_dist}
    \begin{tabular}{lcc}
        \toprule
        Parameter & Estimate & Standard Error \\
        \midrule
        $\mu$ & -0.008 & 0.002 \\
        $\sigma^2$ & 0.021 & 0.0003 \\
        \bottomrule
    \end{tabular}
    \end{center}
\end{table}

\paragraph{Reserve and Accept-Now Prices}

The conditional Accept-Now price function $p^{AN}(v_j)$ and conditional reserve price functions $r_d(v_j)$ and are fitted through second-order polynomials. 

The Accept-Now price function is estimated through a simple linear regression:
\begin{equation*}
    p^{AN}(v_j) = \theta^{AN}_{0} + \theta^{AN}_{1} v_j + \theta^{AN}_{2} v_j^2 + \epsilon_{j}
\end{equation*}

Reserve prices up until the last 24 hours before pickup, namely $r_d(v_j)$ for $d \in \{1,2,3,4\}$, are estimated through a similar regression:
\begin{equation*}
    r_d(v_j) = \theta^R_{0d} + \theta^R_{1} v_j + \theta^R_{2} v_j^2 + \epsilon_{j}
\end{equation*}
where the intercept term $\theta^R_{0d}$ is specific to the number of remaining days until departure $d$. This allows for the reserve price to increase as the departure date approaches, and avoids any potential ``crossing'' of the reserve price functions across different days.

The effective reserve price for the last 24 hours at $d=0$, $r_0(v_j)$, is less straightforward to estimate as there is no unique observed reserve price during this period. We utilize a moment-inequality approach. Let $\mathcal{J}$ denote the set of shipments unmatched at the start of the last 24 hours, and $\mathcal{J}^U \subseteq \mathcal{J}$ those still unmatched at the end. For a shipment $j$, the theoretical reserve price must be bounded below by the highest accepted bid (if any) and above by the lowest rejected bid (if the shipment remained unmatched). We define these bounds as:
\[
\underline b_j \equiv \max_{i\in\mathcal I_j^A} b_{ij}
\quad\text{and}\quad
\overline b_j \equiv \min_{i\in\mathcal I_j\setminus \mathcal I_j^A} b_{ij},
\]
where $\underline b_j=-\infty$ if no bid was accepted and $\overline b_j=+\infty$ if no rejected bid exists to form the upper bound. The model implies the moment inequalities $r_0(v_j) \geq \underline b_j$ for all $j \in \mathcal{J}$ and $r_0(v_j) \leq \overline b_j$ for all $j \in \mathcal{J}^U$.

To estimate $\theta^{R_0}$, we minimize a positive-part quadratic objective function that penalizes violations of these bounds:
\begin{equation*}
    \mathcal{Q}_N^{\text{ineq}}(\theta^{R_0}) = \frac{1}{N}\sum_{j\in\mathcal J} \left[ r_0(v_j,\theta^{R_0}) - \underline b_j \right]_{-}^2 + \frac{1}{N}\sum_{j\in\mathcal J^U} \left[ \overline b_j - r_0(v_j,\theta^{R_0}) \right]_{-}^2
\end{equation*}
where $[x]_{-}\equiv \min\{x,0\}$. Standard errors are computed by treating the estimator as first-order equivalent to a just-identified GMM estimator formed by the set of binding inequalities. The variance-covariance matrix is estimated using a cluster-robust plug-in estimator, allowing for correlations within market-day clusters.

The results of the OLS regressions ($d \geq 1$) and the inequality estimator ($d=0$) are reported together in Table \ref{tab:price_regressions}.

\begin{table}
    \begin{center}
    \caption{Reserve and Accept-Now price estimates}
    \label{tab:price_regressions}
    \begin{tabular}{lccc}
\toprule
{} & Reserve ($d=0$) & Reserve ($d \geq 1$) &  Accept Now \\
               &                 &                      &             \\
\midrule
$v_j$          &         -0.3325 &            1.1939*** &   1.2870*** \\
               &        (1.4522) &             (0.0823) &    (0.1854) \\
$v_j^2$        &          0.1779 &           -0.4189*** &  -0.3420*** \\
               &        (0.7151) &             (0.0388) &    (0.0877) \\
$d = 1$        &                 &            0.1672*** &             \\
               &                 &             (0.0432) &             \\
$d = 2$        &                 &            0.1596*** &             \\
               &                 &             (0.0432) &             \\
$d = 3$        &                 &            0.1586*** &             \\
               &                 &             (0.0431) &             \\
$d = 4$        &                 &            0.1308*** &             \\
               &                 &             (0.0432) &             \\
Constant       &         1.2169* &                      &    -0.1707* \\
               &        (0.7330) &                      &    (0.0973) \\
R-squared      &                 &               0.1554 &      0.3532 \\
R-squared Adj. &                 &               0.1552 &      0.3530 \\
N              &                 &                18250 &        6841 \\
\bottomrule
\end{tabular}

    \end{center}
\end{table}

\newpage
\section{Additional Figures}
\renewcommand{\thefigure}{A\arabic{figure}}
\setcounter{figure}{0}

\begin{figure}[htbp]
    \caption{Comparison of lane frequencies on the platform and across the U.S.}
    \label{fig:map_shipment_freq}
    \centering
    \begin{subfigure}{0.6\textwidth}
        \includegraphics[width=\textwidth]{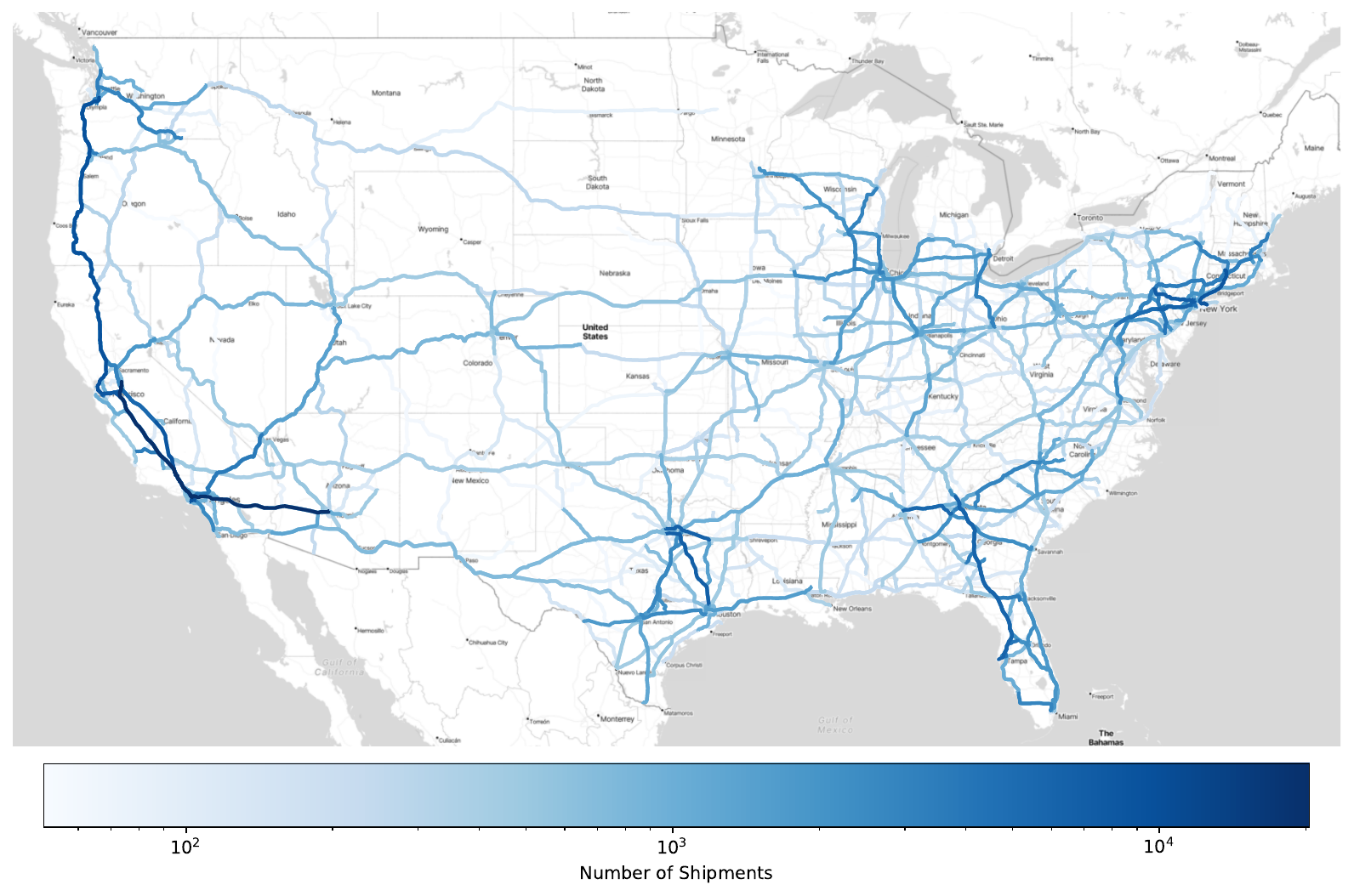}
        \caption{Platform Lane Frequencies}
        \label{fig:map_shipment_freq_platform}
    \end{subfigure}
    \hfill
    \begin{subfigure}{0.6\textwidth}
        \includegraphics[width=\textwidth]{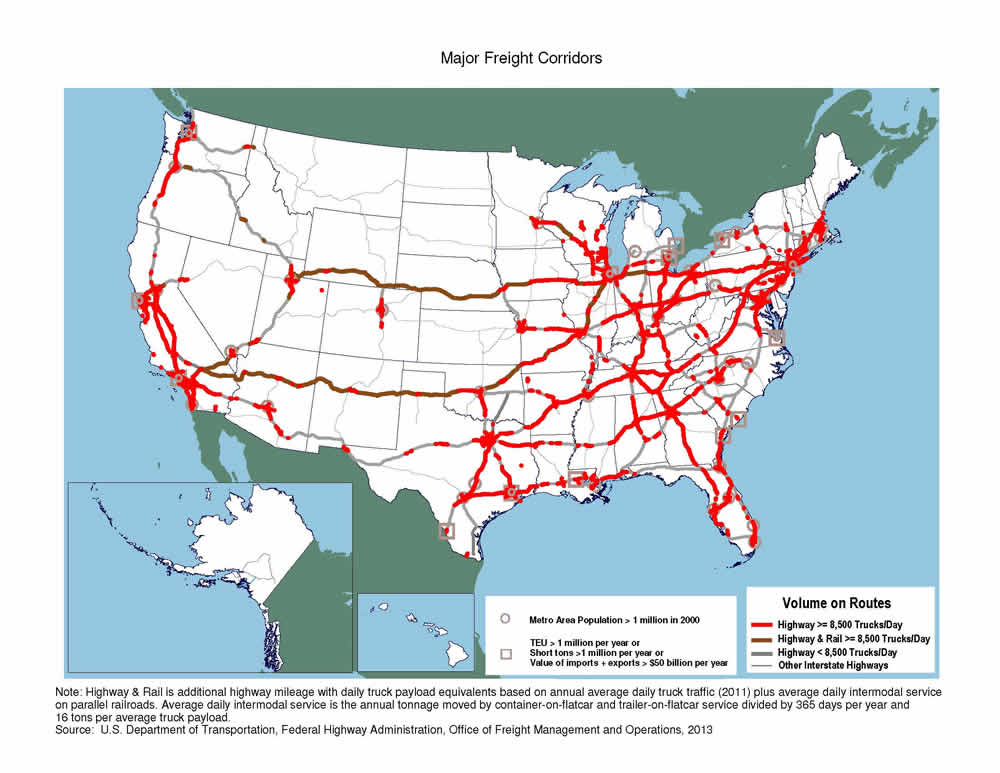}
        \caption{Lane Frequencies Across the U.S.}
        \label{fig:map_shipment_freq_us}
    \end{subfigure}
\end{figure}

\begin{figure}[htbp]
    \centering
    \caption{Distribution of carrier miles driven in a year}
    \label{fig:CDF_miles_driven}
    \includegraphics[width=0.6\textwidth]{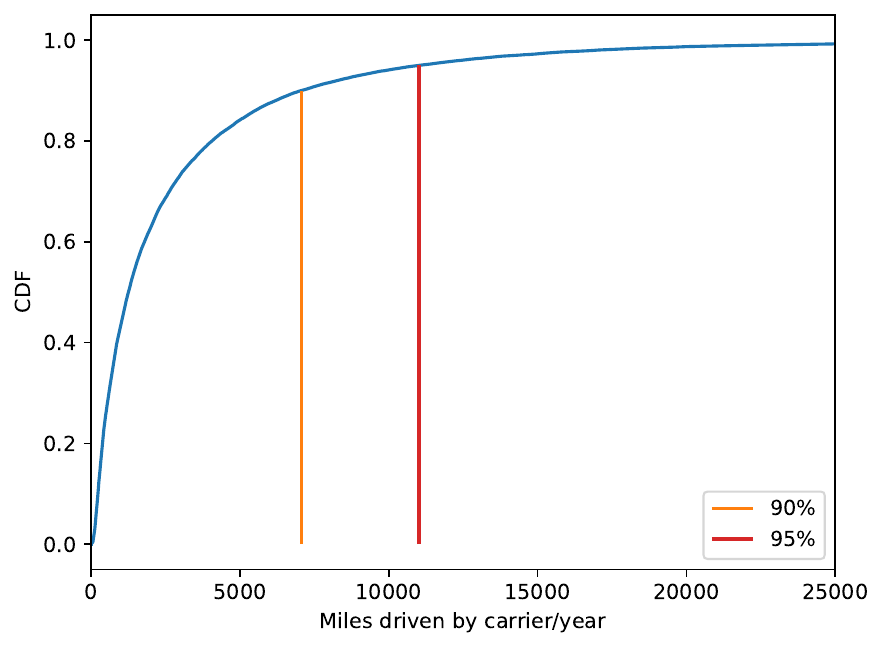}
\end{figure}

\begin{figure}[htbp]
    \begin{center}
    \caption{Confirmation probability vs. bid or expected value}
    \label{fig:confirm_cancel_prob_amounts_expected}
    \subfloat[Bid amount]{\includegraphics[width=0.45\textwidth]{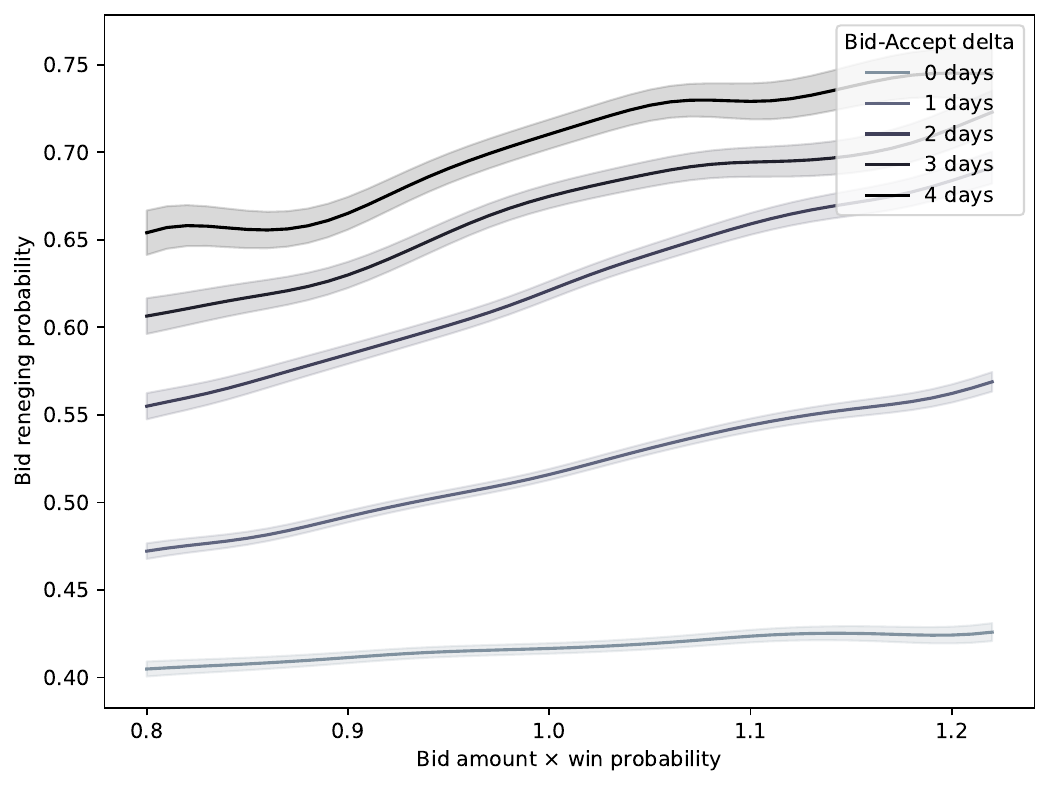}}
    \hfill
    \subfloat[Expected Payoff]{\includegraphics[width=0.45\textwidth]{figs/confirm_prob_expected_days.pdf}}
    \end{center}
    {\footnotesize{\textsc{Note}: Figures are obtained through a two-dimensional kernel regression, conditioning both on different time horizons for attrition, and measures of payoffs. Kernel bandwidth chosen in accordance with Silverman's rule of thumb. Win probability in panel (b) is also computed with a one-dimensional kernel regression of bid acceptance on bid amount.}}
\end{figure}

\end{document}